\documentclass{aa}  

\usepackage{graphicx}
\usepackage{comment}

\usepackage{xcolor}
\usepackage[normalem]{ulem} %for strikeout /sout
\usepackage{txfonts}
\usepackage[colorlinks,allcolors=blue]{hyperref}
\newcommand{\kmpers}      {\mbox{\rm km~s$^{-1}$}}
\newcommand{\kms} {\kmpers}

\newcommand{\lvmap}{$l\varv$-map}
\newcommand{\flva}{F$l\varv$A}
\newcommand{\cxya}{C$xy$A}

\begin{document}

   \title{The SEDIGISM survey: The influence of spiral arms \\on the molecular gas distribution of the inner Milky Way}

   \subtitle{}

   \author{D. Colombo\inst{1}\thanks{dcolombo@mpifr-bonn.mpg.de},
          A. Duarte-Cabral\inst{2},
          A. R. Pettitt\inst{3},
          J.\,S. Urquhart\inst{4},
          F. Wyrowski\inst{1},
          T. Csengeri\inst{5},\\
          K. R. Neralwar\inst{1},
          F. Schuller\inst{1,6},
          K.\,M. Menten\inst{1},
          L. Anderson\inst{7,8,9},
          P. Barnes\inst{10,11,12},
          H. Beuther\inst{13},\\
          L. Bronfman\inst{14},
          D. Eden\inst{15},
          A. Ginsburg\inst{16},
          T. Henning\inst{13},
          C. K\"onig\inst{1},
          M.-Y. Lee\inst{17},
          M. Mattern\inst{18},\\
          S. Medina\inst{1},
          S.\,E. Ragan\inst{2},
          A. J. Rigby\inst{2},
          \'A. S\'anchez-Monge\inst{19},
          A. Traficante\inst{20},
          A. Y. Yang\inst{1},
          M. Wienen\inst{1}
          }

   \institute{Max-Planck-Institut f\"ur Radioastronomie, Auf dem H\"ugel 69, 53121 Bonn, Germany
   \and
   School of Physics \& Astronomy, Cardiff University, Queen’s Building, The Parade, Cardiff CF24 3AA, UK
   \and
   Department of Physics, Faculty of Science, Hokkaido University, Sapporo 060-0810, Japan
   \and
   Centre for Astrophysics and Planetary Science, University of Kent, Canterbury, CT2\,7NH, UK
   \and
   Laboratoire d’astrophysique de Bordeaux, CNRS, Univ. Bordeaux, B18N, allée Geoffroy Saint-Hilaire, F-33615 Pessac, France
   \and
   Leibniz-Institut für Astrophysik Potsdam (AIP), An der Sternwarte 16, D-14482 Potsdam, Germany
   \and
   Department of Physics and Astronomy, West Virginia University, Morgantown, WV 26506
   \and
   Green Bank Observatory, P.O. Box 2, Green Bank, WV 24944
   \and
   Center for Gravitational Waves and Cosmology, West Virginia University, Chestnut Ridge Research Building, Morgantown, WV 26505
   \and
   Department of Astronomy, University of Florida, 211 Bryant Space Science Center, Gainesville, FL 32611, USA
   \and
   Space Science Institute, 4765 Walnut St Suite B, Boulder, CO 80301, USA
   \and
   School of Science and Technology, University of New England, Armidale NSW 2351, Australia
   \and
   Max-Planck-Institut f\"ur Astronomie, K\"onigstuhl 17, D-69117 Heidelberg, Germany
   \and
   Astronomy Department, Universidad de Chile, Casilla 36-D, Santiago, Chile
   \and
   Astrophysics Research Institute, Liverpool John Moores University, IC2, Liverpool Science Park, 146 Brownlow Hill, Liverpool, L3 5RF, UK
   \and
   University of Florida Department of Astronomy,
   Bryant Space Science Center Gainesville, FL 32611, US
   \and
   Korea Astronomy and Space Science Institute, 776 Daedeok-daero, Yuseong-gu, Daejeon 34055, Republic of Korea
   \and
   Commissariat à l’énergie atomique et aux énergies alternatives, 91191 Gif-sur-Yvette, Saclay, France
   \and
   I.\ Physikalisches Institut, Universität zu K\"oln, Z\"ulpicher Strasse 77, 50937 Cologne, Germany
   \and
   IAPS-INAF, Via Fosso del Cavaliere, 100, 00133 Rome, Italy
   }    
   \date{Received XXX; accepted XXX}

% \abstract{}{}{}{}{} 
% 5 {} token are mandatory
 
  \abstract
  {The morphology of the Milky Way is still a matter of debate. In order to shed light on uncertainties surrounding the structure of the Galaxy, in this paper, we study the imprint of spiral arms on the distribution
and properties of its molecular gas. To do so, we take full advantage of the SEDIGISM (Structure, Excitation, and Dynamics of the Inner Galactic Interstellar Medium) survey that observed a large area of the inner Galaxy in the $^{13}$CO\,(2-1) line at an angular resolution of 28". We analyse the influences of the spiral arms by considering the features of the molecular gas emission as a whole across the longitude--velocity map built from the full survey. Additionally, we examine the properties of the molecular clouds in the spiral arms compared to the properties of their counterparts in the inter-arm regions. Through flux and luminosity probability distribution functions, we find that the molecular gas emission associated with the spiral arms does not differ significantly from the emission between the arms. On average, spiral arms show masses per unit length of $\sim10^5-10^6$\,M$_{\odot}\,$kpc$^{-1}$. 
  This is similar to values inferred from data sets in which emission distributions were segmented into molecular clouds. By examining the cloud distribution across the Galactic plane, we infer that the molecular mass in the spiral arms is a factor of 1.5 higher than that of the inter-arm medium, similar to what is found for other spiral galaxies in the local Universe. We observe that only the distributions of cloud mass surface densities and aspect ratio in the spiral arms show significant differences compared to those of the inter-arm medium; other observed differences appear instead to be driven by a distance bias. By comparing our results with simulations and observations of nearby galaxies, we conclude that the measured quantities would classify the Milky Way as a flocculent spiral galaxy, rather than as a grand-design one.}


   \titlerunning{SEDIGISM spiral arms}
   \authorrunning{D. Colombo, A. Duarte-Cabral, A. R. Pettitt et al.}
   \maketitle
%%%%%%%%%%%%%%%%%%%%%%%%%%%%%%%%%%%%%%%%%%%
\section{Introduction}
\label{S:introduction}
%%%%%%%%%%%%%%%%%%%%%%%%%%%%%%%%%%%%%%%%%%%

Spiral galaxies dominate the star formation budget of the local Universe. Understanding how spiral arms and, in general, the dynamical environment influence star formation and the properties of the cold, dense progenitor gas has become of significant importance in recent years because of the advent of observational surveys that are beginning to probe the interstellar medium (ISM) in nearby galaxies on parsec scales \citep[e.g. ][]{sun2018}.

Spiral arms possess a variety of different shapes and extents, and their possible origin mechanism is, as yet, not entirely clear \citep{dobbs_baba2014}. Historically, the tightness of the arms around galactic centres (i.e. the pitch angle) has been one of the primary criteria used to classify galactic morphology (e.g. \citealt{hubble1926}, \citealt{de_vaucouleurs1959}). Spiral galaxies have also been categorised based on the visual distinctiveness of their arms or the number thereof (\citealt{elmegreen1990}). Grand-design galaxies (such as M51 or NGC\,628) are characterised by two long and fairly symmetric arms while flocculent galaxies (such as NGC\,7793 or NGC\,7331) have multiple, fragmented, and generally shorter arms. This second classification appears to be directly connected with the physical mechanisms that create arms, which leave imprints on the distribution of the various galactic components. The arms of grand-design spirals coincide with a real gravitational potential depth that is  noticeable in infrared images as an excess of old stars, while this is not present in flocculent spirals, where the arms are primarily composed of patches of gas and young stars. Flocculent spirals are supposedly generated by local disc instabilities, while the grand-design character is associated with large-scale quasi-stationary density waves or tidal interactions, or with the presence of a bar \citep{dobbs_baba2014}. These two arm classes are not mutually exclusive. The M51 galaxy, which is often put forward as an archetypal example of a grand-design galaxy, shows flocculent-type arms in its outer region, which can no longer be associated with a density wave \citep{meidt13,colombo2014b}.

Cold gas appears to be strongly influenced by the spiral-arm perturbation. Evidence of this can be acquired by simple inspection of CO images of nearby galaxies: molecular gas emission within the spiral arms is much brighter than in the space between them (the inter-arm regions) at both high (e.g., $\sim$pc scale \citealt{koda2009}, \citealt{gratier12}, \citealt{donovan_meyer2013}, \citealt{schinnerer2013}, \citealt{druard2014}, \citealt{pan2017}, \citealt{leroy2017}, \citealt{elmegreen2017}, \citealt{gallagher2018}) and low resolution (e.g., $\sim$kpc scale \citealt{helfer03}, \citealt{leroy2008},  \citealt{wilson2009}, \citealt{rahman2012}, \citealt{bolatto2017}, \citealt{sorai2019}). For comparison, the arms appear much fainter in images of the old stellar population \citep{elmegreen2011}.
This could be due to the collisional nature of the cold gas, which reacts strongly to any small perturbation in the stellar distribution. With the transit through the spiral shock induced by arms, the gas undergoes compression and a series of effects that leave an imprint on its distribution, properties, and structure (see \citealt{dobbs_baba2014} for a review). The rate of cloud--cloud collisions is enhanced within spiral arms due to orbit crowding, generating large molecular gas complexes (e.g. \citealt{tasker09,dobbs_pdc2015}). Most of the star-formation regions are located in the spiral arms, meaning that stellar feedback and supernova explosions are more frequent there, increasing turbulence, and possibly enabling the formation of large cloud complexes on the interacting surfaces of expanding shells \citep{inutsuka2015}. At the same time, spiral arm streaming motions might reduce the environmental pressure on the surface of the clouds which increases their stable mass and generates a population of unbound objects within the arms \citep{meidt13}. Upon leaving the arms, the large clouds that originated within the spiral perturbation feel the elevated shear of the differentially rotating galactic disc which results in their transition into elongated structures such as spurs, feathers, and branches, as predicted by simulations (e.g. \citealt{dobbs_bonnell2006}, \citealt{dobbs_pringle2013}, \citealt{duarte-cabral2017}) and clearly visible in high-resolution CO maps \citep{schinnerer2017}. Gas itself can dampen the prominence of spiral arms \citep{dobbs_baba2014}, while the gravitational instability of the gas disc might be one of the dominant mechanisms producing more flocculent-like spiral features (as in the case of M33, \citealt{dobbs2018}).

Over the years, quantification of the effects of the spiral arm perturbation on the distribution and the properties of the molecular gas has been pursued in various ways. For instance, the distribution of the CO flux in different regions of nearby galaxies has been studied through probability distribution functions (PDFs). \cite{hughes2013a} find clear differences between the PDFs drawn from the dynamical environments of M51: inter-arm PDFs are narrower than spiral-arm and galaxy-centre PDFs which show departures from a pure log-normal shape. The authors interpret those departures as the signature of a combination of effects acting within the spiral arms, such as streaming motions, shocks, and stellar feedback, together with self-gravity of the gas within the clouds. Similarly, the integrated intensity PDFs from the bar, centre, and arm regions of  the galaxy M83 show differences in the tails and the overall shape \citep{egusa2018}, possibly due to the higher velocity dispersion of the gas in the central region compared to the spiral arms.

High-resolution observations of nearby galaxies show that even over-densities of the molecular ISM (i.e. giant molecular clouds; GMCs) feel the perturbing forces of spiral arms.  The GMCs located in the spiral arms and central region of M51 are (on average) brighter and have larger velocity dispersions compared to similarly sized objects in the inter-arm regions. Moreover, their mass spectra show shapes that reflect the passage of the clouds through the different dynamical environments (\citealt{colombo14a}; see also \citealt{koda2009}), with spectra for the spiral arm  that extend to larger masses than those of the inter-arm regions.  Additionally, the recent work of \cite{braine2020} shows that the non-axisymmetric potential of M51 caused by spiral arms generates an elevated number of spiral arm GMCs with retrograde rotation compared to clouds in the inter-arm regions. This is also seen in simulation works that show that large GMCs forming from agglomerations of smaller clouds show a large degree of retrograde rotation compared to the galactic disc \citep{dobbs08}. In the barred spiral galaxy M100, the situation is similar: central clouds are more massive, denser, and have higher velocity dispersions than objects in the inter-arm regions \citep{pan2017}. Nevertheless, GMC properties in several other spiral galaxies do not differ signficantly from spiral arms to inter-arm regions \citep{donovan_meyer2013}, even if, more recently, \cite{rosolowsky2021} observed some slightly  higher surface densities and lower virial parameters in clouds in the spiral arms compared to objects in inter-arm regions.

Differing cloud properties  across various dynamical environments are also seen in high-resolution galactic disc simulations. For example, \cite{fujimoto2014} perform a simulation of a galaxy similar to M83 and observe that the distributions of cloud properties extend to different maxima depending on whether they are located in the bar, spiral arms, or disc regions. These authors find that a large fraction of massive clouds are formed by agglomeration. They also observe a population of unbound objects that are typically observed as a product of cloud interactions in dense filamentary structures. Similarly, \cite{nguyen2018} find that simulations that include spiral perturbation tend to generate a high degree of agglomeration of clouds within the spiral arms, with a general decrease in the number of  small and medium-sized objects (with masses $<10^6$\,$M_{\odot}$) in the disc, together with an incremental increase in the size  of the unbound population. Cloud properties do not appear to be strongly influenced by the kind of spiral perturbation (flocculent, grand-design, or perturbed by a companion iteration), as shown by \cite{pettitt2020}, who nevertheless found that the cloud mass spectra and contrast between arms and inter-arm regions differ depending on the type of spiral arms.

Our position within the disc of the Galaxy renders observations of the kind described above much more complicated for the Milky Way, although we are able to probe much smaller physical scales and gather relatively large samples for statistical analyses. Several works have attempted to unveil possible environmental differences in the Milky Way's gas distribution. Molecular gas features in the Milky Way's longitude--velocity map (\lvmap{}) can be associated with spiral arms (\citealt{dame01}, \citealt{reid2014}, \citealt{rigby2016}), even if substantial amounts of inter-arm gas is observed. The earlier work of \cite{dame86} revealed that GMCs are identified almost exclusively along the spiral arms. Similarly, \cite{stark06} observe that large complexes are mostly associated with the arms, while the distribution of smaller clouds is more random. This result was later confirmed by \cite{roman_duval2009} who concluded that the absence of large GMCs in the Milky Way's inter-arm regions implies that clouds form in the spiral arms and that these clouds must be short-lived (with lifetimes $<10^7$\,yr). Nevertheless, later studies of the first Galactic quadrant, using higher resolution data and more advanced molecular cloud identification techniques \citep{colombo2019}, did not find a significant difference in terms of the mass and the size of the clouds between the spiral arms and the inter-arm regions. Molecular cloud catalogues of the complete Galactic plane based on the $^{12}$CO(1-0) data of \cite{dame01} have been presented independently by \cite{rice2016} and \cite{miville_deschenes2017}. The two catalogues differ in a number of ways, but the denser and larger complexes appear to describe some spiral structure in the top-down view of the Milky Way. \cite{rigby2019} used $^{13}$CO\,(3-2) data to identify thousands of molecular gas clumps (somewhat smaller than the clouds segmented in the previous studies), and find that clumps in the spiral arms appear to have larger line widths, virial parameters, and excitation temperatures than objects in the inter-arm regions\footnote{The studies listed in this paragraph used  models from various sources and tracers to define the spiral arms, such as 21\,cm line data \citep{shane1972}, compilations \citep{vallee2008}, and masers \citep{reid2014}}.

Dynamical environments can also play an important role in shaping the morphology of molecular clouds. In recent years, a number of highly elongated clouds have been observed in the Milky Way \citep{jackson2010,goodman2014,ragan2014,wang2015,zucker2015,abreu-vicente2016,mattern2018a}. An attempt to uniformly categorise these elongated clouds into at least two broad classes (giant molecular filaments and Milky Way Bones) based on their aspect ratio and density was made by \cite{zucker2018}. Studying their locations, \cite{zucker2018} \citep[see also ][]{abreu-vicente2016} conclude that only 35\% of those large-scale filaments can be associated with the spiral arms. The formation and origin of these highly elongated clouds is a matter of debate. Sub-parsec-resolution simulations of \cite{smith2014} that try to reproduce the four-arm spiral structure of the Milky Way find that large-scale filaments tend to form preferentially in spiral arms. However, the \cite{smith2014} simulations are likely hindered in their predictive power by the fact that they do not include stellar feedback of any kind or gas self-gravity. Indeed, more recently, \cite{smith2020} found that supernovae can randomise the alignment between these filaments and spiral arms. Using lower resolution simulations, which in turn take into account a wider range of physical processes, \cite{duarte-cabral2016} observed that the most elongated clouds can be found exclusively in the inter-arm medium (or close to the spiral arm entry point) and that their morphology is due to the intense shear in this region, beyond the protection of low-shear spiral arms. Additionally, the arm region of their simulation appears to harbour the complexes with the largest sizes and highest velocity dispersions. Further, \cite{duarte-cabral2017} suggest that those elongated features merge with each other to become GMC complexes in the spiral arm, and so it is unlikely that large-scale filaments in the Milky Way actually trace the spiral arms.

Aiming to shed new light on the nature of the Milky Way's spiral structure, in this paper, we study the influence of the dynamical environment generated by spiral arms on the distribution and properties of the molecular gas in the inner Milky Way. To do so, we use $^{13}$CO\,(2-1) data from the Structure, Excitation, and Dynamics of the Inner Galactic Interstellar Medium (SEDIGISM; \citealt{schuller2017, schuller2021}) survey described in Section~\ref{S:data}. We consider the four-armed spiral model of the Milky Way provided by \cite{taylor_cordes1993}. This model is described in Section~\ref{S:sparm_models}. Two methodologies are used for the analyses described here (Section~\ref{S:methods}): we first  study the full distribution of molecular gas in longitude--velocity space (Section~\ref{SS:flva}) and then we use the locations of discretised molecular clouds extracted from the SEDIGISM survey (Section~\ref{SS:cxya}). In Section~\ref{S:results}, we present the results of our analysis: the cumulative distribution of the flux with respect to the velocity offset from the spiral arms (Section~\ref{SS:cumdist_voff}), the PDFs of the gas associated with the spiral arms, inter-arm, and Galactic centre (Section~\ref{SS:pdf}), the cloud numbers, and molecular gas mass per unit length and unit area values for each spiral arm (Section~\ref{SS:sparm_linmass}), and the properties of the clouds in the spiral arms and the inter-arm regions (Section~\ref{SS:gmc_props}). We conclude with a discussion of the nature of the spiral features of the  Milky Way, comparing our findings with observations of nearby galaxies and numerical works (Section~\ref{S:disc}).

\begin{figure*}
    \centering
        \includegraphics[width=0.88\textwidth]{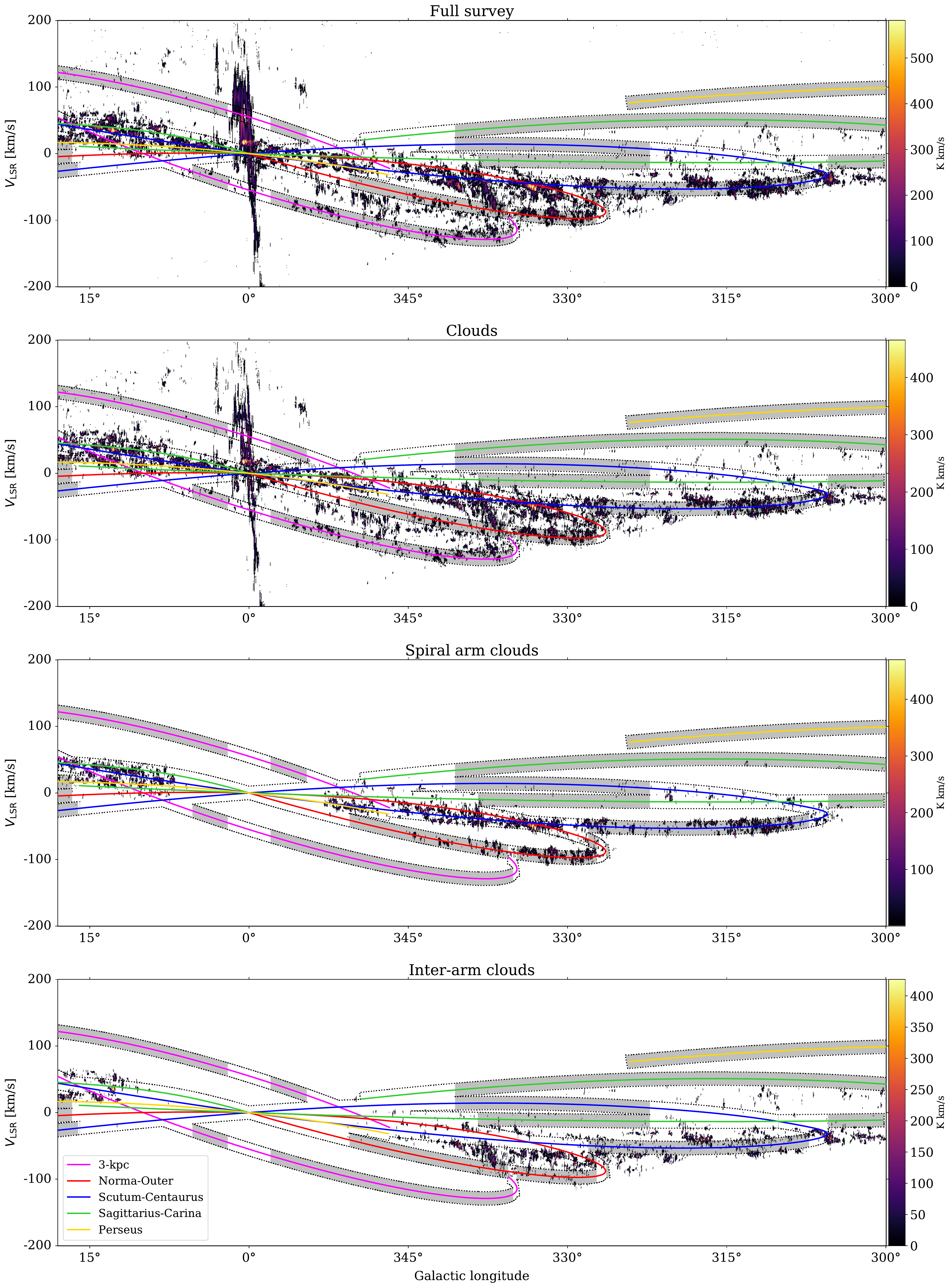}
   \caption{\emph{From top to bottom:}  Longitude--velocity map (\lvmap{}) from the trunk-masked full survey data cube (see Section~\ref{S:data}), the cloud data cube, the data cube obtained for the clouds located in the spiral arms, and the data cube obtained for the clouds located in the inter-arms (following the \cxya{} method described in Section~\ref{SSS:gmc_match}). In each panel, the spiral arm tracks defined by \cite{taylor_cordes1993} are overlaid: the 3\,kpc arms in magenta, the Norma-Outer arm in red, the Scutum-Centaurus arm in blue, the Sagittarius-Carina arm in green, and the Perseus arm in yellow. Dotted black lines surround areas of the $lv-$map that have a velocity offset with respect to the spiral arms within 10\,\kms. Shaded patches indicate regions along the spiral arms where the velocity difference between adjacent arms (along the velocity axis) is greater than 20\,\kms. In the last two panels, the clouds associated with the 3~kpc arms and the clouds with unreliable distance are excluded as they cannot be unequivocally associated with a spiral arm or inter-arm region (see Section~\ref{SSS:gmc_match}). 
   }
    \label{F:sparm_lvmaps}
\end{figure*}

%%%%%%%%%%%%%%%%%%%%%%%%%%%%%%%%%%%%%%%%%%%
\section{Data}
\label{S:data}
%%%%%%%%%%%%%%%%%%%%%%%%%%%%%%%%%%%%%%%%%%%
For the analysis presented in this paper, we use the $^{13}$CO\,(2-1) data from the SEDIGISM survey obtained with the Atacama Pathfinder Experiment 12m submillimeter telescope (APEX, \citealt{guesten2006}). The SEDIGISM project is fully described in \cite{schuller2017} and \cite{schuller2021}; here we provide only a brief summary. We utilise the full contiguous survey data as per the DR1\footnote{\url{https://sedigism.mpifr-bonn.mpg.de/cgi-bin-seg/SEDIGISM_DATABASE.cgi}}  that covers $-60^{\circ}\leq l \leq 18^{\circ}$ and $|b|\leq 0.5^{\circ}$, plus small latitude extensions towards particular regions. The DR1 $^{13}$CO\,(2-1) data have an average  1$\sigma_{\rm RMS}$ of 0.8-1\,K (in $T_{\rm mb}$) per 0.25 km\,s$^{-1}$ channel width, and an angular resolution $\theta_{\rm FWHM, mb}=28''$.

The survey data are provided as tiles of roughly $2^{\circ}\times1^{\circ}$ for a velocity range from $-200$ to 200 km\,s$^{-1}$. The noise across the SEDIGISM survey data is not uniform, and in order to retain only the significant emission, we need to mask the data cubes. Each tile is masked using the mask cubes provided by the dendrogram trunk. \cite{duarte-cabral2021} (hereafter DC21) used the dendrogram technique \citep[see ][]{rosolowsky08} to generate the basic data structure necessary for the cloud identification algorithm (see Section~\ref{SSS:gmc_cat}). In the dendrogram, the trunk is defined by all the connected regions within the masked data cube in position--position--velocity. The masking is generally obtained by considering a few times the local (line-of-sight-wise) or global signal-to-noise ratio (see DC21 for further details). The dendrogram trunk by construction contains all the significant emission in a data cube. We therefore use these trunk-masked cubes, which provide a good flux recovery whilst also minimising the inclusion of the noisy regions. 

We build integrated intensity maps from masked data cubes and $lv-$maps from each tile and we combine them using the STARLINK \citep{currie2014} KAPPA package, using the WCSMOSAIC task\footnote{\url{http://starlink.eao.hawaii.edu/docs/sun95.htx/sun95ss204.html\#Q1-231-838}} to build the maps from the full survey (Fig.~\ref{F:sparm_lvmaps}). A similar procedure is used to generate the maps from molecular cloud datacubes. 

\begin{figure*}
    \centering
        \includegraphics[width=\textwidth]{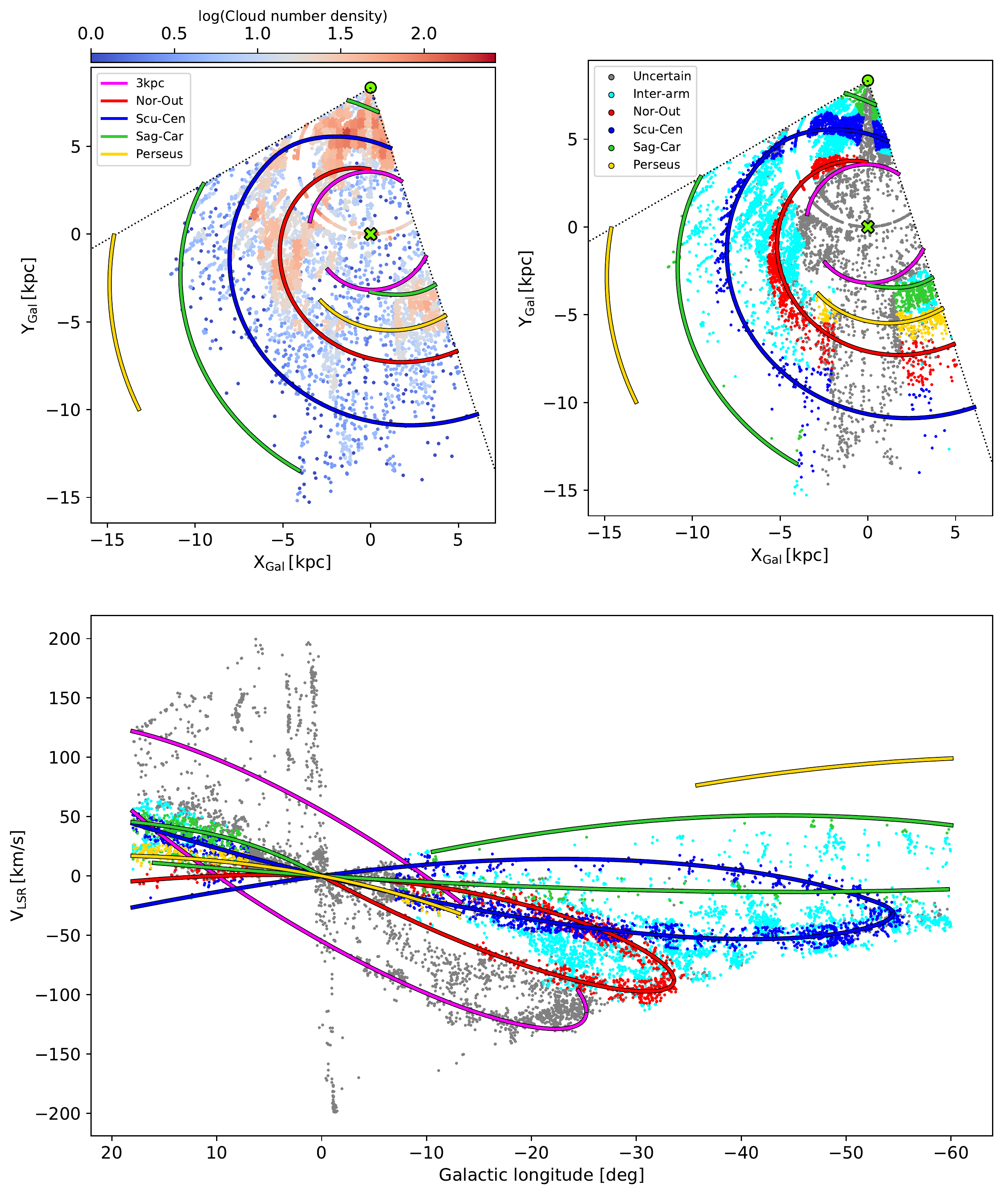}
   \caption{Face-on view of the Galactic region surveyed by SEDIGISM (confined within the dotted lines; top panels) overlaid with spiral arm tracks defined by TC93: the 3\,kpc arms in magenta, the Norma-Outer arm in red, the Scutum-Centaurus arm in blue, the Sagittarius-Carina arm in green, and the Perseus arm in yellow (the solid lines merely trace the bottom of the potential, and do not correspond to a real `thickness'). The position of the Sun is indicated with a green circle, while the Galactic centre is shown with a green `X'. In the top-left panel, coloured dots represent the number density distribution of all molecular clouds identified within the SEDIGISM field by DC21. The top-right and bottom panels show the distribution of the clouds in the full sample with respect to the spiral arms in $xy$  and $lv$ space, respectively. Clouds are colour-encoded by the attributed spiral arm. Objects in the inter-arm region are in cyan and clouds with uncertain allocation are shown in grey. The latter consist of clouds associated with the 3~kpc arms and those with unreliable distances.}
   \label{F:sparm_gmc_match}
\end{figure*}

%%%%%%%%%%%%%%%%%%%%%%%%%%%%%%%%%%%%%%%%%%%
\section{Spiral arm models}
\label{S:sparm_models}
%%%%%%%%%%%%%%%%%%%%%%%%%%%%%%%%%%%%%%%%%%%
We use the models from \cite{taylor_cordes1993} (hereafter TC93, see also \citealt{cordes2004}) to draw spiral arms across the Milky Way disc mapped by SEDIGISM. These models follow the distribution of known HII regions and improve on earlier models of \cite{georgelin_georgelin1976}. \cite{georgelin_georgelin1976} models have also been independently verified by \cite{russeil2003}, who examined the position of the arms using also H$\alpha$, CO, radio continuum, and absorption data. The arms in this model are polynomial perturbations to single log-spiral structures, and as such do not trace out a single pitch angle with radius, seen most notably towards Carina. These tracks are the same as those used in \citet{schuller2021} and \citet{urquhart2021} and include additional near and far 3~kpc arm features that have non-zero radial velocities. The four primary arms are projected into longitude-velocity space using modern values for the local standard of rest and rotation curve as a function of radius, assuming purely circular orbits (see Sect. 3.2 of \citealt{schuller2021} for details). Top-down positions and longitude-velocity tracks of the spirals are shown in Fig.~\ref{F:sparm_gmc_match}. Spiral arms in this model are named 3~kpc, Norma-Outer (or Nor-Out), Sagittarius-Carina (or Sag-Car), Scutum-Centaurus (or Scu-Cen), and Perseus.

We choose the TC93 model set for consistency with previous SEDIGISM studies and also because, in the Galactic region spanned by SEDIGISM, it is the only one fitted independently of the CO emission, which avoids possible configuration biases. Another more recent and already widely used spiral arm model is presented in \cite{reid2019}. However, that model is only well defined in the I and II Milky Way quadrants, and in the remaining quadrants, the tracks on the \lvmap\ follow the $^{12}$CO\,(1-0) emission peaks of \cite{dame01} survey data, which make them prone to the aforementioned biases.
We instead use the \cite{reid2019} model set to benchmark the robustness of our results regarding model choices in Appendix~\ref{A:reid19}. 

There have been several other attempts to model the Milky Way spiral structure. For example, \cite{hou-han2014} used HII regions, GMCs, and masers positions to define the Milky Way morphology using `polynomial' log-spirals. However, their arm tracks do not match the local material particularly well, and the non-log-spiral Sagittarius-Carina arm in the model makes it hard to globally fit the spiral tracks on the \lvmap{} (see quadrant I in Fig~26 of \citealt{pettitt2014}). \cite{koo2017} extrapolated a map of the 21~cm line emission and derived the position of the spiral arms. However, their models have a large gap in the IV Galactic quadrant where most of the SEDIGISM data are collected.

In order to adapt the TC93 models to the SEDIGISM data, we interpolate $V_{\rm lsr}$, $R$, and $d$ of each arm onto the survey cube coordinates (whose pixel size is 9.5$^{\prime\prime}$) considering the full extent of the arm using the {\sc scipy interp1d\footnote{\url{https://docs.scipy.org/doc/scipy/reference/generated/scipy.interpolate.interp1d.html}}} method and a spline, cubic interpolation. 

In our analyses, we assume the spiral arms to be confined to the Galactic plane, i.e. we do not consider the Galactic latitude distribution of the diffuse emission and clouds to perform the spiral arm association. This assumption is motivated by the fact that the Milky Way warp starts at a Galactocentric distance of  $\sim10$\,kpc \citep[e.g. ][]{levine2006}, which almost coincides  with the margin of our sensitivity limit. \cite{dame-thaddeus2011} identified a spiral arm at high latitude 
in the 21~cm line emission. However, this arm is observed in the Milky Way I quadrant and at a distance of $\sim21$\,kpc, a region not probed by our survey. 

%%%%%%%%%%%%%%%%%%%%%%%%%%%%%%%%%%%%%%%%%%%%%%%%%%%%%%%%%%%%%%%%%%%%%%%%%%%%%%%%%%%%%
\section{Methods}
\label{S:methods}
%%%%%%%%%%%%%%%%%%%%%%%%%%%%%%%%%%%%%%%%%%%%%%%%%%%%%%%%%%%%%%%%%%%%%%%%%%%%%%%%%%%%%
To compare the distribution of the molecular gas associated with spiral arms to that of the molecular gas outside of the arms we made use of two methods: the first one involves the integrated flux of the trunk-masked data across the full \lvmap\ of the survey; the second uses the molecular clouds identified within the SEDIGISM field (from DC21) and associates clouds with the spiral arms considering their position in $x_{\rm Gal}$, $y_{\rm Gal}$, and $V_{\rm lsr}$ space. We used these two methods as they are highly complementary, and in an attempt to compensate for the endemic biases that affect each of the methods. Working on the \lvmap{} allows us to associate all the significant emission to spiral arms or inter-arm regions. Due to our position within the Galactic disc, this method suffers from projection effects, meaning that inter-arm gas is potentially attributed to spiral arms and vice versa. For the other method, using the SEDIGISM field, segmenting the gas into discrete elements (clouds) allows the assignment of these elements to particular positions across the Galactic disc, removing part of the projection effect bias. However, it is difficult to find distances of sources in the inner Galaxy, and many SEDIGISM clouds have `unreliable' distances (as defined by DC21; see also \ref{SSS:gmc_cat}). Additionally, clouds constitute only the denser parts of the molecular ISM; in other words, the cloud segmentation filters out the most diffuse $^{13}$CO emission. Therefore, we can consider that, to first order, the first (full $l\varv$ assignment, hereafter \flva{}) method provides to an `upper limit' of the spiral arm/inter-arm association, while the second (cloud $xy$ assignment, hereafter \cxya{}) method provides more of a `lower limit' of the spiral arm/inter-arm association.

%%%%%%%%%%%%%%%%%%%%%%%%%%%%%%%%%%%%%%%%%%%%%%%%%%%%%%%%%%%%%%%%%%%%%%%%%%%%%%%%%%%%%
\subsection{\flva\ method: molecular gas distribution with respect to the spiral arms in $lv$ space}
\label{SS:flva}
%%%%%%%%%%%%%%%%%%%%%%%%%%%%%%%%%%%%%%%%%%%%%%%%%%%%%%%%%%%%%%%%%%%%%%%%%%%%%%%%%%%%%

To define the $^{13}$CO emission associated with spiral arms, we calculate the Euclidean distance between each \lvmap\ pixel and every spiral arm point. The minimisation of this distance identifies the spiral arm point closest to a given pixel in the \lvmap . In this way, each \lvmap\ pixel is uniquely associated to a spiral arm point defined by the model tracks.  Because of the interpolation described in section~\ref{S:sparm_models}, each pixel assumes the properties of its associated spiral arm point, in particular the heliocentric distance that we use below to obtain physical properties of the spiral arm gas. The absolute velocity difference or offset, $\Delta V$, between the \lvmap\ pixel and the associated spiral arm point is then calculated. We consider as part of the spiral arms all \lvmap\ regions where $\Delta V < 10$\kms, which is of the order of the amplitude of streaming motions around the spiral arms of the Milky Way and is generally used to define velocity offsets corresponding to material within the spiral arms \citep{reid2014,grosbol_carraro2018,ramon-fox_bonnell2018,xu2018,wang2020_gaia}. %($\pm7$\kms\ as measured by \citealt{reid2014}).

This kind of analysis is affected by a number of potential issues, and Fig.~\ref{F:sparm_lvmaps} clearly illustrates how the spiral structure in the inner Milky Way is tightly convoluted, making it difficult to properly separate one arm from another. In particular, towards the Galactic centre, several arm tracks converge, making it hard to disentangle the emission from each arm using the $l\varv-$maps alone \citep[but see ][]{mertsch-vittino2020}. 

%%%%%%%%%%%%%%%%%%%%%%%%%%%%%%%%%%%%%%%%%%%%%%%%%%%%%%%%%%%%%%%%%%%%%%
\subsection{\cxya\ method: molecular cloud distribution with respect to the spiral arms in $xy$ space}
\label{SS:cxya}
%%%%%%%%%%%%%%%%%%%%%%%%%%%%%%%%%%%%%%%%%%%%%%%%%%%%%%%%%%%%%%%%%%%%%%%%%%%%%%%%%%%%%

%%%%%%%%%%%%%%%%%%%%%%%%%%%%%%%%%%%%%%%%%%%%%%%%%%
\subsubsection{The SEDIGISM molecular cloud catalogue}
\label{SSS:gmc_cat}
%%%%%%%%%%%%%%%%%%%%%%%%%%%%%%%%%%%%%%%%%%%%%%%%%%
The molecular cloud catalogue from the full SEDIGISM data is fully described in DC21. The catalogue was built using the Spectral Clustering for Molecular Emission Segmentation (SCIMES) algorithm (\citealt{colombo15}, see also \citealt{colombo2019} for a description of the updated version). This applies a spectral clustering method to identify discrete objects (i.e. molecular clouds) from a dendrogram of emission features \citep{rosolowsky08} without the need for preceding data smoothing. Considering that $\sim30\%-50\%$ of the flux appears to be diffuse or with low S/N, this method allows a good separation of clouds across the same line of sight. Indeed, $\sim82\%$ of sightlines are assigned to a single cloud, $\sim16\%$ to two clouds, $\sim2\%$ to three clouds, and less than 1\% to more than three clouds (see DC21, their Section 3.1.3 for further details). In total, 10663 molecular clouds have been decomposed from the SEDIGISM data. The heliocentric distance to clouds was calculated assuming the rotation model of \cite{reid2019}. To solve the kinematic distance ambiguity (KDA), a set of robust distance indicators was used that includes masers, dark clouds, HI self-absorption, dust clumps, the size--line width relation, and 3D extinction mapping  (see DC21, their Section 4.2 for full details). 

The distribution of the clouds in the SEDIGISM coverage is shown in a top-down view of the Milky Way in Fig.~\ref{F:sparm_gmc_match}. Qualitatively, it seems that more objects are located across the near arm sections of the Norma-Outer, Scutum-Centaurus, and Sagittarius-Carina arms, except for the Scutum-Centaurus and Norma-Outer inter-arm region (around $x_{\rm Gal},y_{\rm Gal} \approx -7, 0$\,kpc). 

We consider a set of subsamples for the analyses in this paper. The entire sample contains all 10300 clouds of the SEDIGISM catalogue. The sample with reliable distances, referred to hereafter as the distance reliable sample, contains all clouds with a good distance estimation ($d_{\rm reliable}=1$ in the catalogue as explained in DC21, their section 4). This sample is useful for the study of cloud cumulative quantities (such as fluxes and masses) that do not require closed contours to be reliable. 

Following the name convention of DC21, the science sample is constituted by all objects that have a reliable distance ($d_{\rm reliable}=1$), do not touch the upper and lower edges of the survey datacubes (edge$=0$, in the catalogue), and are well resolved (cloud area in arcsec $>3\Omega_{\rm beam}$, where the beam size $\Omega_{\rm beam}\sim888$\,arcsec$^2$). This sample is used here to analyse all the properties of the clouds, especially those that require well-resolved clouds and closed contours.

We also use complete distance-limited samples that share the same set of criteria as the science sample, but only include clouds with distances in the range of $2.5-5$\,kpc, and with masses above $3.1\times10^2$\,M$_{\rm mol}$ and effective radii above 1\,pc (see DC21, their Appendix C, for further details). This sample is useful for assessing whether trends observed in the science sample are robust against distance biases. 

In addition, we analyse subsamples of clouds that contain certain star-formation signposts. The ATLASGAL sample indicates clouds that contain at least one source identified within the APEX Telescope Large Area Survey of the Galaxy data (\citealt{schuller2009}, see also \citealt{urquhart2021}); the HMSF-sample includes objects that show signs of high-mass star formation in various tracers (see DC21, Sections 4.3 and 5.1 for further details).

\begin{table*}
\centering
\caption{Results of the cloud association to spiral arms for the TC93 model.}
\begin{tabular}{ccccccc}
\hline
Cloud ID & Cloud Name & Closest spiral arm & Minimum distance & $\Delta V$ & $p_{\rm val}$ & Location \\
\hline
 &  &  & $\mathrm{kpc}$ & $\mathrm{km\,s^{-1}}$ & & \\
\hline
1 & SDG300.618-0.2307 & Scu-Cen & 0.80 & 1.82 & 0.00 & IA \\
2 & SDG300.555-0.2366 & Scu-Cen & 0.81 & 1.80 & 0.00 & IA \\
... & ... & ... & ... & ... & ... & ... \\
194 & SDG304.498-0.0151 & Sag-Car & 0.12 & 7.41 & 0.97 & SA \\
195 & SDG303.246-0.2143 & Sag-Car & 0.32 & 9.52 & 0.89 & SA \\
... & ... & ... & ... & ... & ... & ... \\
10663 & SDG016.163+0.1113 & 3kpc & 2.15 & 9.62 & 0.00 & IA \\
\hline
\end{tabular}
\tablefoot{Only a few lines are shown, the full table is provided online. From left to right: cloud identification number; cloud name; closest spiral arm label; minimum distance to the closest spiral arm in kpc; velocity offset with respect to the closest spiral arm in \kms\ , $p-$value from the $\chi^2$ test described in Section~\ref{SSS:gmc_match} and equation~\ref{E:chi2}, location with respect to the spiral arm; `SA' indicates clouds associated to a spiral-arm
region, `IA' indicates clouds in the inter-arm region.}
\label{T:spiral_arm_association}
\end{table*}

\begin{figure}
    \centering
        \includegraphics[width=0.45\textwidth]{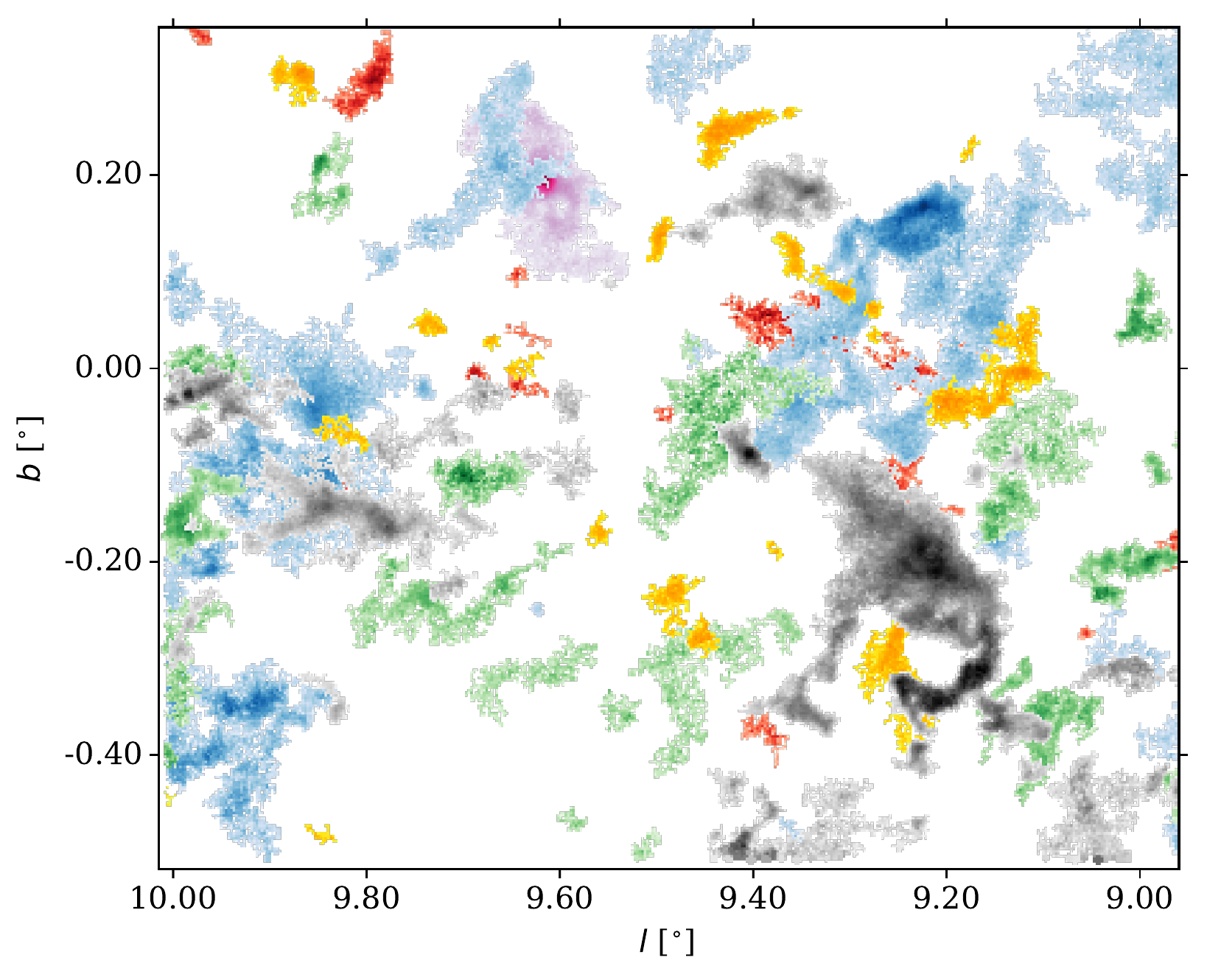}
   \caption{Multi-coloured integrated intensity maps of the $^{13}$CO\,(2-1) emission in an inset of the `G009' SEDIGISM field. Clouds attributed to a given spiral arm are colour-coded: magenta (3~kpc), red (Norma-Outer), blue (Scutum-Centaurus), green (Sagittarius-Carina), yellow (Perseus), grey (inter-arm or unreliable distance clouds). For visualisation purposes, in these images, we include also clouds attributed to the 3~kpc arms and with uncertain location, even if they are not used in further analyses. The same images for the full survey are collected in Appendix~\ref{A:multi_colored}.}
    \label{F:g009_example}
\end{figure}

%%%%%%%%%%%%%%%%%%%%%%%%%%%%%%%%%%%%%%%%%%%%%%%%%%%%%%%%%%%%%%%%%%%%%%%%%%%%%%%%%%%%%%%%%
\subsubsection{Association of molecular clouds to spiral arms and distance redefinition}
\label{SSS:gmc_match}
%%%%%%%%%%%%%%%%%%%%%%%%%%%%%%%%%%%%%%%%%%%%%%%%%%%%%%%%%%%%%%%%%%%%%%%%%%%%%%%%%%%%%%%%%

Segmented clouds can be considered as single discrete objects that have well-defined extents, locations in the Galactic disc, and velocities. With respect to the pixels in the \lvmap,\, clouds have a distance and associated uncertainty derived independently from the spiral arm model (see DC21 for a full description of the cloud distance assignment methods). Therefore, to match a given cloud with its closest spiral arm, we use a simple $\chi^2$ test on the distance--longitude plane (equivalent to the $xy$ plane):
\begin{equation}\label{E:chi2}
\chi^2 = \frac{(d_{\rm cloud}-d_{\rm arm})^2}{\sigma^2_{ d}} + \frac{(l_{\rm cloud}-l_{\rm arm})^2}{\sigma^2_{ l}},    
\end{equation}
where $d_{\rm cloud}$ and $d_{\rm arm}$ are the distance to the cloud and a given point on the spiral arm, respectively; $l_{\rm cloud}$ and $l_{\rm arm}$ is the longitude of the cloud centroid and to a given point on the spiral arm, respectively. The distance uncertainty of a cloud, $\sigma_{\rm d}$, is calculated in DC21 and we refer the reader to that work for details. For an uncertainty on the cloud longitude, we assume the following:
\begin{equation}\label{E:sigma_l}
\sigma_{ l} = \sigma_{\rm maj}\cos(pa),     
\end{equation}
where the cloud semi-major axis ($\sigma_{\rm maj}$) is measured via a moment method (see \citealt{rl06} for further details) applying a principal component analysis of the cloud projection onto the plane of the sky, while the position angle ($pa$) is given by the orientation of the cloud major axis with respect to the $x-$direction in the datacube (which for our case is Galactic longitude). We use equation~\ref{E:chi2} to calculate the $\chi^2$ between  position of a cloud and the location of each spiral arm point from our adopted model. We then associate a cloud to the closest spiral arm point by minimising the $\chi^2$. We assume that a cloud is within the spiral arms if the $p-$value from the $\chi^2$ test satisfies $p_{\rm val}>0.05$, and if $\Delta V < 10$\,\kms{}. We assessed how our chosen velocity offset influences the properties of the clouds in the spiral arms versus the inter-arm region in Appendix~\ref{A:voff}. However, this method does not perform well for the clouds near the spiral arm tangent points, as a $\Delta V$ in velocity space does not correspond to a fixed width on the $xy$ plane. In this case, we also consider a cloud to be part of a given spiral arm if the minimum offset between cloud and arm on the $xy$ plane is < 200\,pc, assuming a conservative 400\,pc width \citep{vallee2008}.  

Using this method, we also attempt to redefine the distance of the clouds with unreliable distance attribution as per DC21 (2913 objects). We solve the kinematic distance ambiguity (KDA) whenever the near distance puts a given cloud within the associated arm, while the far distance locates the object in the inter-arm region, or vice versa. In those cases, we favour the distance solution that places a cloud within an arm. If both distance solutions attribute a cloud to spiral arms or inter-arm regions, we keep the original distance of the object as defined by DC21, with the unreliable flag. This method implicitly assumes that the presence of the spiral arms favours cloud formation \citep[see e.g. ][]{wang2020_thor}, or simply that spiral-arm regions contain more clouds than inter-arm regions, as observed in nearby galaxies \citep[e.g. ][]{colombo14a}. By applying this spiral-arm criterion, we solve the KDA for 139 more clouds, that is: 139 objects that originally had an unreliable distance attribution are now assigned to a spiral-arm region. DC21 assumed 12 distance flags (called $d_{\rm flag}$) that identify the method used to find the distance of a given cloud (see their Table 1 and Figure 6), and the objects for which the KDA has been resolved with the spiral-arm method assume $d_{\rm flag}=13$. With these new distances, we recalculate the properties of those clouds as described in DC21, their Section 3.2. 

Considering this, we now include 7889 objects with reliable distances in the SEDIGISM cloud catalogue (distance reliable sample). With the addition of these new reliable cloud distances, the science sample (see Section~\ref{SSS:gmc_cat}) now constitutes 6782 objects (as opposed to the 6664 clouds in the original science sample of DC21) and the complete distance-limited sample has 981 clouds. We use our updated catalogue for the remainder of the paper. The inclusion of the additional clouds does not significantly change the results and main conclusions of the paper. More details are provided in Appendix~\ref{A:dflag13}. A representative part of the updated catalogue (with the additional spiral arm information) is shown in Table~\ref{T:spiral_arm_association}. The updated catalogue will be available online as part of the SEDIGISM database.

The association between clouds and spiral arms is ambiguous in certain Galactic regions. Figure~\ref{F:sparm_gmc_match} shows the result of the cloud matching to the spiral arms on the $xy$ plane (top right) and $lv$ plane (bottom). On the $xy$ plane, the clouds appear slightly downstream with respect to the position of the spiral arms. Some objects associated to the 3~kpc arms are actually far away from them on the $xy$ plane, but have a distance uncertainty that crosses the arms. Other objects that appear quite close to the 3~kpc arms are assigned as inter-arm clouds, because their velocity offset is larger than 10~\kms.

The reason behind the mismatch between $xy-$ and $lv-$planes across the 3~kpc region could arise from the strong non-circular motions that are not accounted for in the derivation of the heliocentric distances of the clouds. We therefore consider the clouds associated to the 3~kpc arms to have an uncertain allocation. Together with the clouds that have an unreliable distance attribution, the 3~kpc-associated objects constitute the uncertain location sample of (2210) clouds and we do not consider them in any of the analyses performed with the \cxya{} method. 

Some clouds in the upstream region of the Scutum-Centaurus and Norma-Outer arms appear very close to the arms on the $xy$ plane, but they have a velocity offset $>10$\,\kms\ . Those objects end up to be attributed to the inter-arm region. Additionally, Fig.~\ref{F:sparm_gmc_match} (left panel) shows that many clouds that appear to be attributed to the inter-arm regions are actually quite close to the spiral arms in $lv-$space, but are further away from them in the $xy$ plane. This is best illustrated in Fig.~\ref{F:sparm_lvmaps} (bottom panels). In particular, the region of the \lvmap\ where $\Delta V<10$\,\kms\ contains $\sim$95\% of the flux of the clouds associated to the spiral arms with the CxyA method, but also $\sim$40\% of the flux from the inter-arm clouds (considering only objects with reliable distances and not associated with the 3~kpc arms). Indeed, as the allocation of the clouds to a specific region is performed considering the position of the cloud centroid, the clouds themselves can extend within the spiral arm or inter-arm region on the \lvmap. Additionally, some objects of the inter-arm region  are far from every spiral arm on the $xy$ plane ($p-$value < 0.05 or with a $xy$ offset > 200\,pc, considering the requirements of \cxya{}), but are in the $\Delta V<10$\,\kms\ area of the \lvmap\ due to projection effects.

\begin{figure*}
        \includegraphics[width=\textwidth]{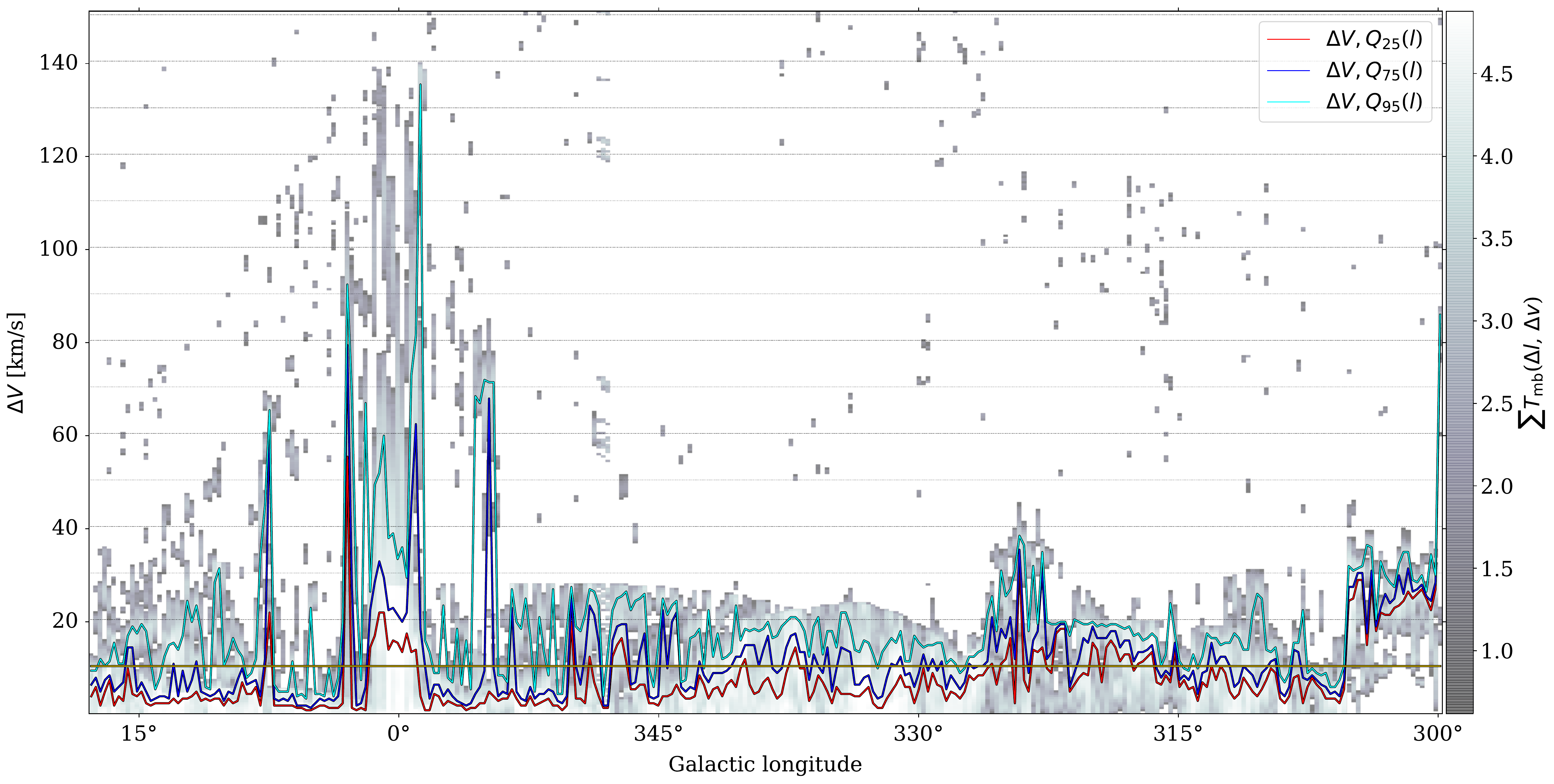}
   \caption{Velocity offset with respect to the closest spiral arm versus Galactic longitude, integrated across the whole \lvmap. The colour of each pixel shows the value of the integrated flux in a bin of $\Delta l \sim0.5^{\circ}$ and $\Delta V = 0.5$\,\kms{}. Red, blue, and cyan lines mark the $\Delta V$ values that encompass the 25\%, 75\%, and 95\% quartile values (respectively) of the flux distribution in a given longitudinal bin. The yellow horizontal line marks the position of a velocity offset $\Delta V = 10$\,\kms{}, which indicates our spiral-arm velocity threshold.}
    \label{F:lvoff_map}
\end{figure*}

In Fig.~\ref{F:g009_example}, every cloud evident in the integrated intensity maps is shaded with a colour that represents its association to a given spiral arm. The full survey data displayed in this way are shown in Appendix~\ref{A:multi_colored}. Such visualisation is useful to explore the complexity of inner Galactic structure across the line of sight (see e.g. Fig.~\ref{F:sed_sparms_337}, lower panel). In addition, it is interesting to see how many bright clouds appear to be located within the inter-arm region. This visualisation might suggest that some redefinition of the spiral arm models in the Milky Way fourth quadrant is needed. For instance, the G305 complex shown in the upper panel of Fig.~\ref{F:sed_sparms_301} is usually considered
as part of the Scutum-Centaurus arm \citep{clark_porter2004}, but seems to be a  mostly inter-arm region. 

\begin{figure*}
    \centering
        \includegraphics[width=0.85\textwidth]{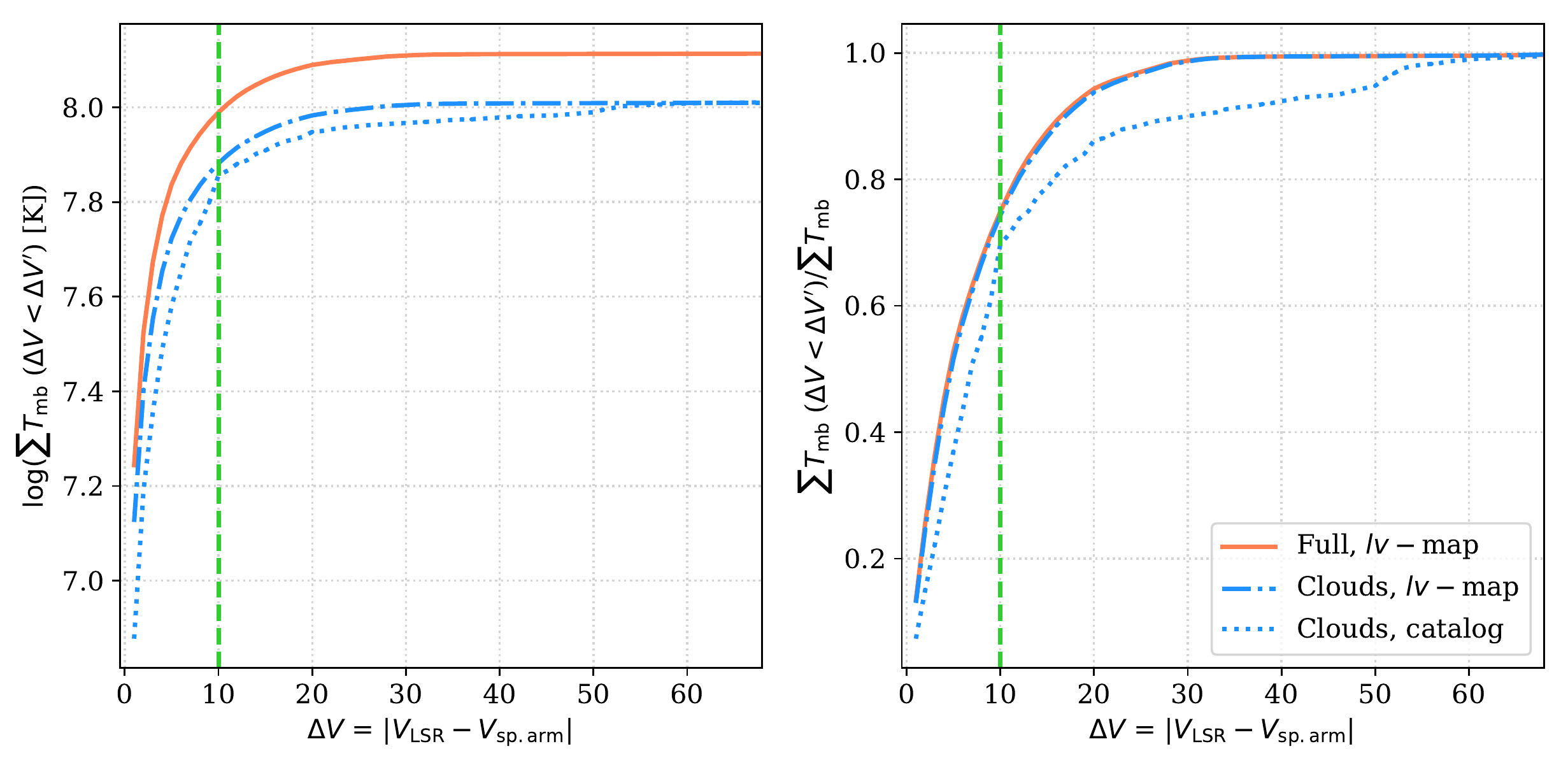}
   \caption{Cumulative distribution of the integrated intensity of the velocity offsets with respect to spiral arms ($\Delta V$; left) and cumulative distribution normalised by the total flux in their respective datasets (right). In the panels, we show cumulative distribution from the different methods described in the text: direct integration of the original $lv-$map data cube (orange full line), integration of the $lv-$map generated by cloud masks (blue dotted-dashed line), integration of the cloud fluxes as per their cloud velocity centroid association to the closest spiral arm (blue dotted lines). The dashed green line shows the 10\kms{} width used to define arm association.}
    \label{F:cumdistr}
\end{figure*}

%%%%%%%%%%%%%%%%%%%%%%%%%%%%%%%%%%%%%%%%%%%%%%%%%%%%%%%%%%%%%%%%%%%%%%
\section{Results}
\label{S:results}
%%%%%%%%%%%%%%%%%%%%%%%%%%%%%%%%%%%%%%%%%%%%%%%%%%%%%%%%%%%%%%%%%%%%%%
We compare the properties of the molecular gas within the spiral arms and the inter-arm region across the SEDIGISM field. In particular, we calculate the cumulative distribution of the flux with respect to the velocity offset from the spiral arms (Section~\ref{SS:cumdist_voff}). In Section~\ref{SS:pdf}, we derive the flux and luminosity probability distribution functions (PDFs) of the gas associated with the spiral arms, the inter-arm regions, and the Galactic centre. We measure the cloud linear mass (i.e. mass per unit length), surface density, and number density for each spiral arm (Section~\ref{SS:sparm_linmass}), and finally we examine whether the properties of the clouds in the spiral arms differ from those of the inter-arm regions (Section~\ref{SS:gmc_props}).

%%%%%%%%%%%%%%%%%%%%%%%%%%%%%%%%%%%%%%%%%%%%%%%%%%%%%%%%%%%%%%%%%%%%%%%%%%%%
\subsection{Flux cumulative distribution with respect to spiral arm velocity offsets}
\label{SS:cumdist_voff}
%%%%%%%%%%%%%%%%%%%%%%%%%%%%%%%%%%%%%%%%%%%%%%%%%%%%%%%%%%%%%%%%%%%%%%%%%%%%

Excluding the Galactic centre, the distribution of flux within the SEDIGISM fields is generally closely concentrated around the spiral arm loci. This is visible from the $lv-$maps in Fig.~\ref{F:sparm_lvmaps}. Nevertheless, a non-negligible fraction of emission and clouds are observed in the inter-arm regions. Figure~\ref{F:lvoff_map} provides an alternative visualisation of a longitude--velocity map where, instead of the $V_{\rm lsr}$, we use the velocity offset with respect to its closest spiral arm, $\Delta V$, on the $y$-axis. It is interesting to note that, except for a few cases, the emission is concentrated within $\Delta V<30$\,\kms. This result agrees with the analysis of \cite{urquhart2021} (see their Figs.~7 and 8). In addition, the offset that we use to define a spiral arm, $\Delta V=10$\,\kms{}, contains more than $\sim50\%$ of the emission in a given longitudinal bin. The exceptions include the longitudes towards the Galactic centre (approximately $-2\leq l \leq 2$ deg); and the regions $300\leq l \leq 305$ deg and around $l \sim 320$ deg, where a large amount of emission is significantly offset from the spiral arm loci in the \lvmap{}, which is possibly due to mismatches between the data and the model used (see Section~\ref{SSS:gmc_match}, but also Appendix~\ref{A:reid19}). Cumulative distributions of the flux with respect to $\Delta V$ show a more compact view of the flux distribution across the spiral arms (Fig.~\ref{F:cumdistr}). Without considering the Galactic centre, a $\Delta V<10$\,\kms\ contains $\sim75\%$ of the flux within the full survey \lvmap{} and the \lvmap\ defined from the clouds. Approximately $95\%$ of the flux is observed for $\Delta V<22$\,\kms{}. %Those distributions do not include the flux towards the Galactic centre ($-2\leq l \leq 2$ deg).

Cloud association to a spiral arm is performed by matching the centroid position to the closest spiral arm in the $xyv$ space (\cxya\ method). Some clouds whose centroid is located within the spiral arms extend into the inter-arm region. This generates a slightly different cumulative  distribution (dotted line in Fig.~\ref{F:cumdistr}) compared to the one measured from the cloud $lv-$map (dashed-dotted line in Fig.~\ref{F:cumdistr}). In particular, with this representation, we observe that approximately $70\%$ of the cloud flux is within $\Delta V<10$\,\kms{}, but $95\%$ is reached around 50\,\kms{}.

%%%%%%%%%%%%%%%%%%%%%%%%%%%%%%%%%%%%%%%%%%%%%%%%%%%%%%%%%%%%%%%%%%%%%%
\subsection{Probability distribution functions}
\label{SS:pdf}
%%%%%%%%%%%%%%%%%%%%%%%%%%%%%%%%%%%%%%%%%%%%%%%%%%%%%%%%%%%%%%%%%%%%%%

\begin{figure*}
    \centering
        \includegraphics[width=\textwidth]{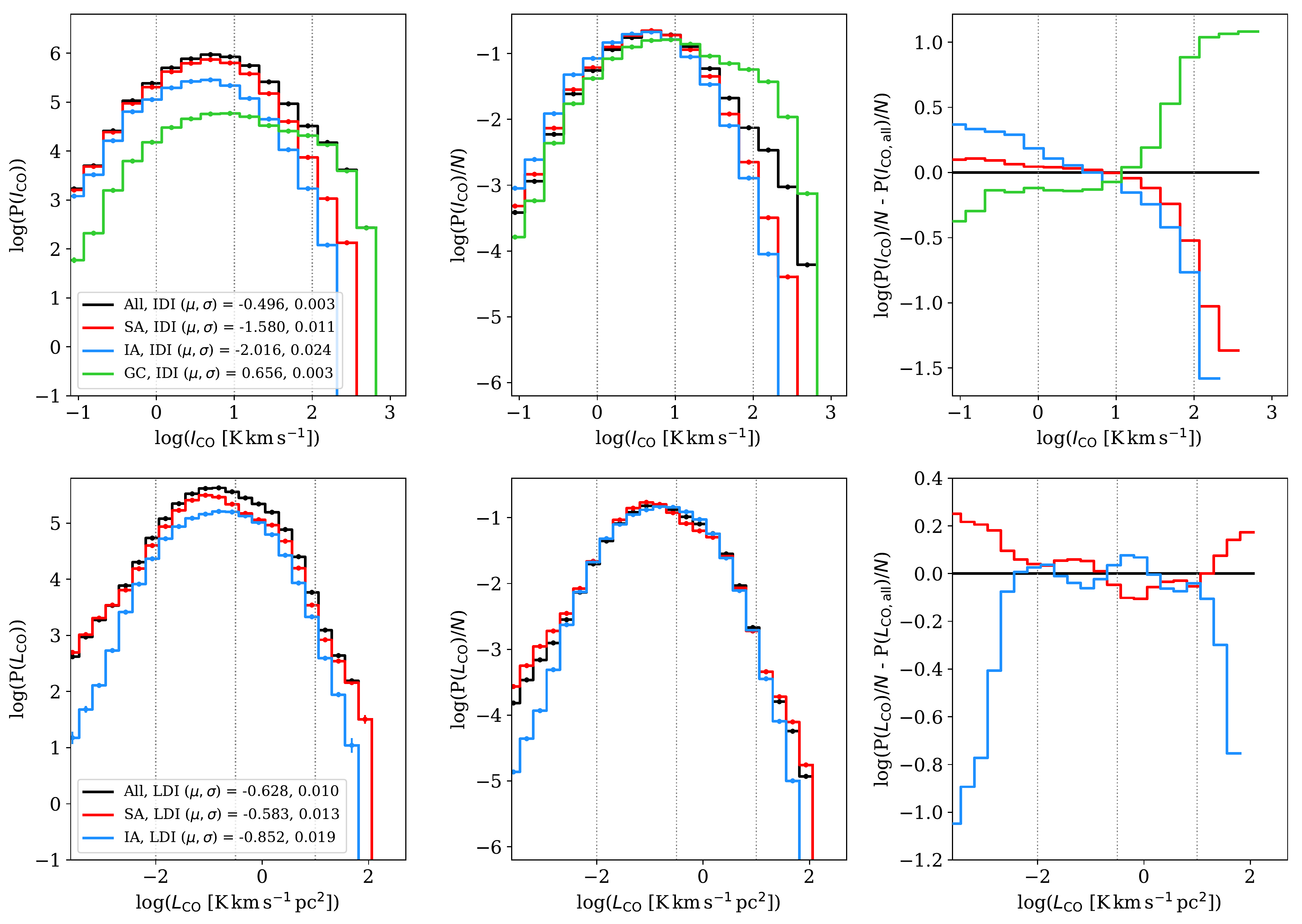}
   \caption{\emph{Top left:} Probability distribution functions (PDFs) of the integrated intensity emission for the full integrated intensity map (black), within the spiral arms (red), for inter-arm regions (blue), and towards the Galactic centre (green). \emph{Top middle:} PDFs normalised by the total counts ($N$) in the distribution. \emph{Top right:} Relative value of the normalised PDFs of each region with respect to the normalised total distribution (named `All' in the left panel). In the top-left panel legend, IDI median ($\mu$) and median absolute deviation ($\sigma$) are shown for each region (where SA, IA, and GC stand for spiral arms, inter-arm region, and Galactic centre, respectively). The IDI thresholds are indicated with grey vertical dotted lines. \emph{Bottom row:} Parallel distribution representations using CO luminosity from clouds. Statistics for the LDI from each considered region are indicated in the bottom-left panel legend.}
   \label{F:sparm_pdf}
\end{figure*}

To provide a more quantitative measure of the differences in the flux distributions between spiral arms, the inter-arm regions, and toward the Galactic centre, here we analyse the shape of the integrated flux PDFs (Fig.~\ref{F:sparm_pdf}, top row). To build the flux PDFs, we use logarithmic bins of 0.25 K\,km\,s$^{-1}$. Generally, the spiral arm PDF is relatively similar to the inter-arm PDF but is drastically different from the PDF built with the emission towards the Galactic centre. The flux PDF of the spiral arms has the highest peak among the different regions, and its extension to maximum intensity is very similar to that of the inter-arm region. The flux PDF from the ISM towards the Galactic centre (e.g. within $-2\leq l \leq 2$ deg) is the most discrepant, showing a significant lack of low-flux pixels, but a much larger amount of bright pixels compared to the other flux PDFs.  The differences between the flux PDFs appear more evident in the right panel of Fig.~\ref{F:sparm_pdf}, which illustrates the difference between the normalised PDF in the different regions with respect to the normalised total flux PDF (shown in the middle panel of the figure). The inter-arm region tentatively shows an excess of faint emission and a lack of bright emission with respect to the other regions, the Galactic centre flux PDF shows the opposite behaviour, and the spiral arm flux PDF sits in the middle of these extremes.

A technique developed in the context of Galactic survey data (\citealt{sawada12}) to provide a more quantitatively description and comparison of the PDFs is to use brightness distribution indices (BDIs). In \cite{sawada12}, the BDI was employed to discern between bright emission in the spiral arms from more diffuse emission in the inter-arm regions. \cite{hughes2013a} employed this index and the integrated intensity version of it (the integrated intensity distribution index, IDI) to compare the molecular gas distribution within the various galactic environments in the disc of M51. Here we apply the IDI to our flux PDFs. This index parameterises the ratio between bright and faint emission and is defined as:
\begin{equation}\label{E:idi}
    IDI = \log \left( \frac{\sum_{I_2<I_{\rm CO}<I_3} I_{\rm CO,i}}{\sum_{I_1<I_{\rm CO}<I_0} I_{\rm CO,i}} \right),
\end{equation}
where the thresholds are chosen arbitrarily in order to catch differences in the PDFs in certain ranges. In our case, we set ($I_0,\,I_1,\,I_2,\,I_3$) = (1,\,10,\,100,\,$\infty$)\,K\,km\,s$^{-1}$. We use a bootstrapping method in order to assess the uncertainty on the $IDI$. Briefly, we generated 1000 realisations of the integrated intensity distribution for each considered Galactic region. The values within a distribution are resampled allowing for repetitions and a new $IDI$ is calculated from the bootstrapped distribution. The uncertainty on the $IDI$ is given by the median absolute deviation of the $IDI$ distribution. This method is robust and widely used to calculate the uncertainties of emission-related structures (such as clouds, \citealt{rl06}). In Fig.~\ref{F:sparm_pdf},  we observe that the IDI from the spiral arm and the inter-arm flux distributions are quite similar (even if they are statistically distinct considering their relatively small uncertainties); the small difference in the IDIs indicates that the spiral arms contain a slightly higher amount of bright emission with respect to the inter-arm regions. The IDI from the emission towards the Galactic centre  is instead several times larger than that from the emission towards  spiral arms and inter-arm regions, indicating that the Galactic centre harbours a significantly higher amount of bright emission than the other two regions, which has been noticed before \citep[e.g. ][]{eden2020}.

This kind of analysis is performed assuming velocity constraints from the \lvmap{} (\flva{} method), which effectively mixes emission at different distances.  As the flux is a distance-independent parameter, in Fig.~\ref{F:sparm_pdf} (bottom row) we show PDFs calculated from the CO luminosity ($L_{\rm CO}$) in order to verify the finding from the flux PDFs. We inferred $L_{\rm CO}$ from each pixel of the integrated intensity maps as $L_{\rm CO} = \sum T_{\rm mb}\delta x \delta y \delta v$, where $\sum T_{\rm mb}$ is the velocity axis-integrated flux in K, $\delta x$ and $\delta y$ represent the size of the pixel in pc, and $\delta v$ is the original data channel width in km/s. As we know the distance of a given patch of gas, we can use the pixel sizes in pc units. As we do not possess a way to infer the distances in the inter-arm regions via the \flva{} method, we built the luminosity PDFs considering the cloud association from the \cxya{} method. Given the constraints of the \cxya{} method, here we considered only the clouds that have reliable distances and that are not associated with the 3~kpc arms. As such, we cannot infer the luminosity PDF from the Galactic centre. For the luminosity PDF, we assume a logarithmic bin size of 0.25 K\,km\,s$^{-1}$\,pc$^2$. The luminosity PDFs from spiral arms and inter-arm regions are qualitatively similar to the corresponding flux PDFs. However, the spiral arms appear to also contain  a larger fraction of low-luminosity pixels with respect to the inter-arm region. As in the case of the flux PDFs, we calculated an index (similar in scope to the IDI) that allows a more qualitative comparison of the luminosity PDFs. The luminosity distribution index is calculated as:
\begin{equation}\label{E:ldi}
    LDI = \log \left( \frac{\sum_{L_2<L_{\rm CO}<L_3} L_{\rm CO,i}}{\sum_{L_1<L_{\rm CO}<L_0} L_{\rm CO,i}} \right),
\end{equation}
where we chose the thresholds ($L_0,\,L_1,\,L_2,\,L_3$) = ($10^{-2}$,\,$10^{-0.5}$,\,$10^1$,\,$\infty$)\,K\,km\,s$^{-1}$\,pc$^{2}$. We took the bootstrapping method used to assess the $IDI$ uncertainty and used it in the same way to generate the $LDI$ uncertainties. Alternatively, we could assume the distance errors of the clouds (from DC21) as the dominant uncertainty in the calculation of the CO luminosity uncertainty and propagate them to measure the $LDI$ uncertainty. Both `bootstrapping' and `propagation' methods provided a comparable uncertainty measurement for the $LDIs$. We assumed the former for consistency with the calculation of the $IDI$ uncertainty.
As in the case of the flux PDFs, we observe that the spiral arm LDI is slightly higher than the inter-arm LDI, confirming the finding of the IDIs, and showing that there is not a significant difference between the luminosities of the spiral arms and the inter-arm regions\footnote{Varying the IDI (or LDI) threshold changes the value of the IDIs (LDIs) themselves. However, qualitatively, our conclusions on the similarity of spiral-arm and inter-arm region PDFs are robust.}.

\begin{table*}
\centering
\caption{Summary of the numbers and densities of the molecular ISM and clouds within given locations across the Galactic disc.}
\begin{tabular}{rc|ccccc|c|c}
\hline
Property & Units & 3~kpc & Nor-Out & Scu-Cen & Sag-Car & Perseus & Spiral arms & Inter-arm \\
(1) & (2) & (3) & (4) & (5) & (6) & (7) & (8) & (9) \\
\hline
\multicolumn{9}{c}{\flva\ method: non-overlapping spiral arm segments} \\
\hline
$\log$($M_{\rm full}$) & M$_{\odot}$ & 6.07 & 6.47 & 6.43 & 3.88 & 6.51 & 7.00/6.95 & - \\
length & kpc & 7.67 & 4.04 & 9.27 & 9.28 & 3.05 & 33.32/25.65 & - \\
$\log(M_{\rm full}$/length) & M$_{\odot}$\,kpc$^{-1}$ & 5.19 & 5.87 & 5.46 & 2.92 & 6.02 & 5.48/5.54 & - \\
\hline
\multicolumn{9}{c}{\cxya\ method: non-overlapping spiral arm segments} \\
\hline
$N_{\rm cloud}$ & - & - & 308 & 455 & 32 & 35 & 830 & - \\
$\log(M_{\rm cloud})$ & M$_{\odot}$ & - & 6.32 & 6.19 & 3.84 & 5.47 & 6.59 & - \\
length & kpc & - & 4.04 & 9.27 & 9.28 & 3.05 & 33.32 & - \\
$N_{\rm cloud}$/length & kpc$^{-1}$ & - & 75.85 & 48.98 & 3.46 & 11.48 & 25.12 & - \\
$\log(M_{\rm cloud}$/length) & M$_{\odot}$\,kpc$^{-1}$ & - & 5.72 & 5.22 & 2.87 & 4.99 & 5.07 & - \\
\hline
\multicolumn{9}{c}{\cxya\ method: full spiral arm extent} \\
\hline
$N_{\rm cloud}$ & - & - & 943 & 1742 & 543 & 263 & 3491 & 2973 \\
$\log(M_{\rm cloud})$ & M$_{\odot}$ & - & 6.69 & 6.83 & 6.38 & 6.13 & 7.19 & 7.01 \\
length & kpc & - & 16.92 & 27.23 & 22.13 & 4.01 & 70.29 & - \\
$N_{\rm cloud}$/length & kpc$^{-1}$ & - & 56.23 & 64.57 & 24.55 & 66.07 & 50.12 & - \\
$\log(M_{\rm cloud}$/length) & M$_{\odot}$\,kpc$^{-1}$ & - & 5.46 & 5.40 & 5.03 & 5.53 & 5.34 & - \\
\hline
length & kpc & - & 16.92 & 27.23 & 22.13 & 4.01 & 70.29 & - \\
width$_{\rm Q_{50}}$\,(width$_{\rm IQR}$) & pc & - & 463\,(666) & 531\,(810) & 951\,(952) & 682\,(885) & 579\,(832) & - \\
area & kpc$^2$ & - & 7.84 & 14.48 & 21.06 & 2.74 & 46.12 & 51.35 \\
$N_{\rm cloud}$ & - & - & 943 & 1742 & 543 & 263 & 3491 & 2973 \\
$N_{\rm cloud}$/area & kpc$^{-2}$ & - & 120.23 & 120.23 & 25.70 & 95.50 & 75.86 & 57.54 \\
$\log(M_{\rm cloud})$ & M$_{\odot}$ & - & 6.69 & 6.83 & 6.38 & 6.13 & 7.19 & 7.01 \\
$\log(M_{\rm cloud}$/area) & M$_{\odot}$\,kpc$^{-2}$ & - & 5.80 & 5.67 & 5.05 & 5.70 & 5.53 & 5.30 \\
\hline
\end{tabular}
\tablefoot{From left to right. Column (1): considered property: mass of the molecular ISM calculated from the \lvmap{} ($M_{\rm full}$), length of the spiral arm or spiral arm segment across the SEDIGISM field (\emph{length}), mass of the molecular ISM per unit length (of a spiral arm segment, $M_{\rm full}$/\emph{length}), number of clouds ($N_{\rm cloud}$), mass in clouds $M_{\rm cloud}$, number of clouds per unit length (of a spiral arm/spiral arm segment, $N_{\rm cloud}$/\emph{length}), mass in clouds per unit length (of a spiral arm/spiral arm segment, $M_{\rm cloud}$/\emph{length}), medians (width$_{\rm Q_{50}}$) and inter-quartile ranges (width$_{\rm IQR}$) of two times the absolute offset with respect to the spiral-arm ridge line, length of the spiral arms multiplied by the median width (\emph{area}), cloud number surface density ($N_{\rm cloud}$/\emph{area}), and cloud mass surface density ($M_{\rm cloud}$/\emph{area}). Column (2): considered property units. Columns (3)-(7): spiral arm or spiral arm segment name. Column (8): entire spiral-arm
region, the symbol `/' separates quantities calculated with the 3~kpc arms contribution included and excluded, respectively. Column (9): entire inter-arm region. Quantities are calculated following the \flva{} method (Section~\ref{SS:flva}) and \cxya{} method (Section~\ref{SS:cxya}) considering the non-overlapping segment on the \lvmap{} (see Figure~\ref{F:sparm_lvmaps}), and for the \cxya{} method also the  extent of the full arm.}
\label{T:densities}
\end{table*}

%%%%%%%%%%%%%%%%%%%%%%%%%%%%%%%%%%%%%%%%%%%%%%%%%%%%%%%%%%%
%\subsection{Line-masses and surface molecular mass densities}
\subsection{Global integrated quantities}
\label{SS:sparm_linmass}
%%%%%%%%%%%%%%%%%%%%%%%%%%%%%%%%%%%%%%%%%%%%%%%%%%%%%%%%%%%

Through the \flva\ method, by associating each pixel of the \lvmap\ to a given (interpolated) point across the adopted spiral arm model, we also defined a heliocentric distance map which we can now use to convert the latitude-integrated flux in the \lvmap\ to the CO luminosity of the molecular gas within the spiral arms. The CO luminosity in a given pixel of the \lvmap\ is given by $L_{\rm CO} = \sum T_{\rm mb}\delta x \delta y \delta v$, where $\sum T_{\rm mb}$ is the latitude-integrated flux in K, $\delta x$ and $\delta y$ represent the size of the pixel in pc, and $\delta v$ is the original data channel width in km/s. 
The molecular gas mass within the arms follows by assuming a certain CO-to-H$_2$ conversion factor, $\alpha_{\rm CO}$. We use $\alpha_{\rm ^{13}CO\,(2-1)} = 5\alpha_{\rm ^{12}CO\,(1-0)}$ (where $\alpha_{^{12}\rm CO\,(1-0)}=4.35$\,M$_{\odot}$\,(K\,km/s\,pc$^2$)$^{-1}$, \citealt{bolatto2013}), consistent with the value derived from the SEDIGISM science demonstration field (\citealt{schuller2017}) and the same value as that used to infer the molecular cloud masses in DC21. Assuming a constant $\alpha_{\rm CO}$ for clouds in the different Galactic regions could be an oversimplification as this value has been shown to have a large scatter and dependency with respect to metallicity, opacity, excitation conditions, and line width \citep[e.g. ][]{barnes2018}. We therefore implicitly assume that those quantities (together with the $^{13}$CO-to-$^{12}$CO ratio) do not vary significantly between the spiral arm and inter-arm regions. However, addressing these issues is outside the scope of this paper.

To convert these masses into a mass per unit length (line-masses), we require the length of the spiral-arm segments for which the masses were estimated. The spiral arm segment lengths are calculated by deriving the Galactocentric $x$ and $y$ coordinates:

\begin{flalign}\label{eq:sparm_xygal}
    x_{\rm Gal} &= d\sin(l), &\\
    y_{\rm Gal} &= R_0 - d\cos(l),
\end{flalign}

\noindent where $R_0 = 8.34$\,kpc is the distance from the Sun to the Galactic centre \citep{reid2014}\footnote{In this paper we adopt $R_0 = 8.34$\,kpc for consistency with DC21, who produced the SEDIGISM cloud catalogue. Nevertheless, this value was recently updated by Reid et al. (2019) to $R_0 = 8.15$\,kpc.}, while the longitudes ($l$) and the heliocentric distance ($d$) of each spiral arm segment are provided by the model and are interpolated on the SEDIGISM survey datacube grid. Those distances are totally dependent on the spiral arm model used (TC93 in our case) and are independent of the distances estimated for SEDIGISM clouds. The length of each segment is the sum of the Euclidean distance between each spiral arm point, given by: \emph{length} $= \sum_i \sqrt{(x_{\rm Gal}^i - x_{\rm Gal}^{i-1})^2 + (y_{\rm Gal}^{i} - y_{\rm Gal}^{i-1})^2}$. For compatibility with the derivations from cloud measurements, we excluded the regions where the clouds do not have reliable distances (i.e. between $-5^{\circ} \leq l \leq 10^{\circ}$) to calculate the arm length.

For this analysis, we only consider the emission that is unambiguously associated to a single spiral arm segment, i.e. for which the velocity gap between one arm and the adjacent one is larger than 20~\kms (double our definition of an arm velocity `width'). We also exclude from the analysis the section of the \lvmap\ in the range $|l|<2$ degrees, because that area includes gas towards the Galactic centre, which has highly non-circular orbits, and is not included in the TC93 model. 
Furthermore, the SEDIGISM latitude coverage does not account for complete coverage of the molecular gas distribution between $-60<l<-42$ degrees at Galactocentric distances larger than 8\,kpc where the Galactic disc is warped towards lower latitudes \citep[e.g.][]{chen2019,romero-gomez2019}; so this region will also be excluded from the analyses. Considering all these aspects, the usable spiral-arm segments for the analysis employing the \flva{} method are shown in Fig.~\ref{F:sparm_lvmaps} as shaded grey areas.

Spiral arm segment lengths, molecular gas masses, and linear mass densities inferred from the \flva{} methods are listed in Table~\ref{T:densities}. Those segments generally contain approximately the mass of one GMC per kiloparsec for each arm. Spiral arm linear masses calculated in this way vary between $10^{3}-10^{6}$\,M$_{\odot}$\,kpc$^{-1}$. In Section~\ref{S:disc} we put these numbers in the context of recent findings from nearby galaxy studies. 

Similar line masses can be calculated considering the clouds identified across those non-overlapping segments, as part of the \cxya\ method. For this test, we use only clouds with reliable distances that also possess a well-defined mass calculated from the CO luminosity, by assuming the same luminosity-to-mass conversion factor, $\alpha_{\rm ^{13}CO\,(2-1)}$,  as previously used. Table~\ref{T:densities} reports the results of the analysis. As before, we exclude the clouds associated with the 3~kpc arms. The biggest discrepancy concerns the segments related to the Perseus arm, which appear to have a linear mass one order of magnitude lower than that estimated considering the full \lvmap{}. On average, however, we observe that the line mass from the \flva\ method is a factor of  approximately three larger than the one inferred from clouds. This discrepancy might arise from various elements, as the two methods are not completely comparable. For instance, several clouds (or parts of clouds) associated with the inter-arm region by the \cxya{} method have a velocity offset $<10$\,\kms{} to the closest spiral arm.

The \cxya\ method allows the inference of line masses and cloud densities across the full extent of the spiral arms imaged by SEDIGISM. This is possible because clouds are attributed to a particular spiral arm based on their position in the $xy$ plane, where the arms do not show significant overlap. In this section we used the distance-reliable sample, excluding the clouds with uncertain location. Linear masses derived from clouds in this way are almost consistently found to be between $10^{5.0}$ and $10^{5.5}$\,M$_{\odot}\,$kpc$^{-1}$. A few differences between the line masses calculated from the \lvmap\ and the cloud methods can be noted. For example, the line mass from the \cxya\ method across the full extent of the Sagittarium-Carina arm is almost two orders of magnitude larger than the one calculated across the \lvmap\ (on the non-overlapping regions), which might indicate that the non-overlapping regions where this quantity has been calculated might not be representative of the linear arm across the full arm. In other cases (e.g. the Perseus arm), the calculated line masses are higher than the ones inferred from the molecular cloud distribution, which might suggest a similar effect, the existence of a slightly more diffuse medium not included in the cloud segmentation, or that several inter-arm region clouds are accounted for in that particular spiral arm section of the \lvmap{}. In terms of pure discrete objects, it appears that the number of clouds per unit length ($N_{\rm clouds}$/length in Table~\ref{T:densities}) is only a few tens of objects per kiloparsec  without significant discrepancies between the arms.

\begin{figure*}
        \includegraphics[width=\textwidth]{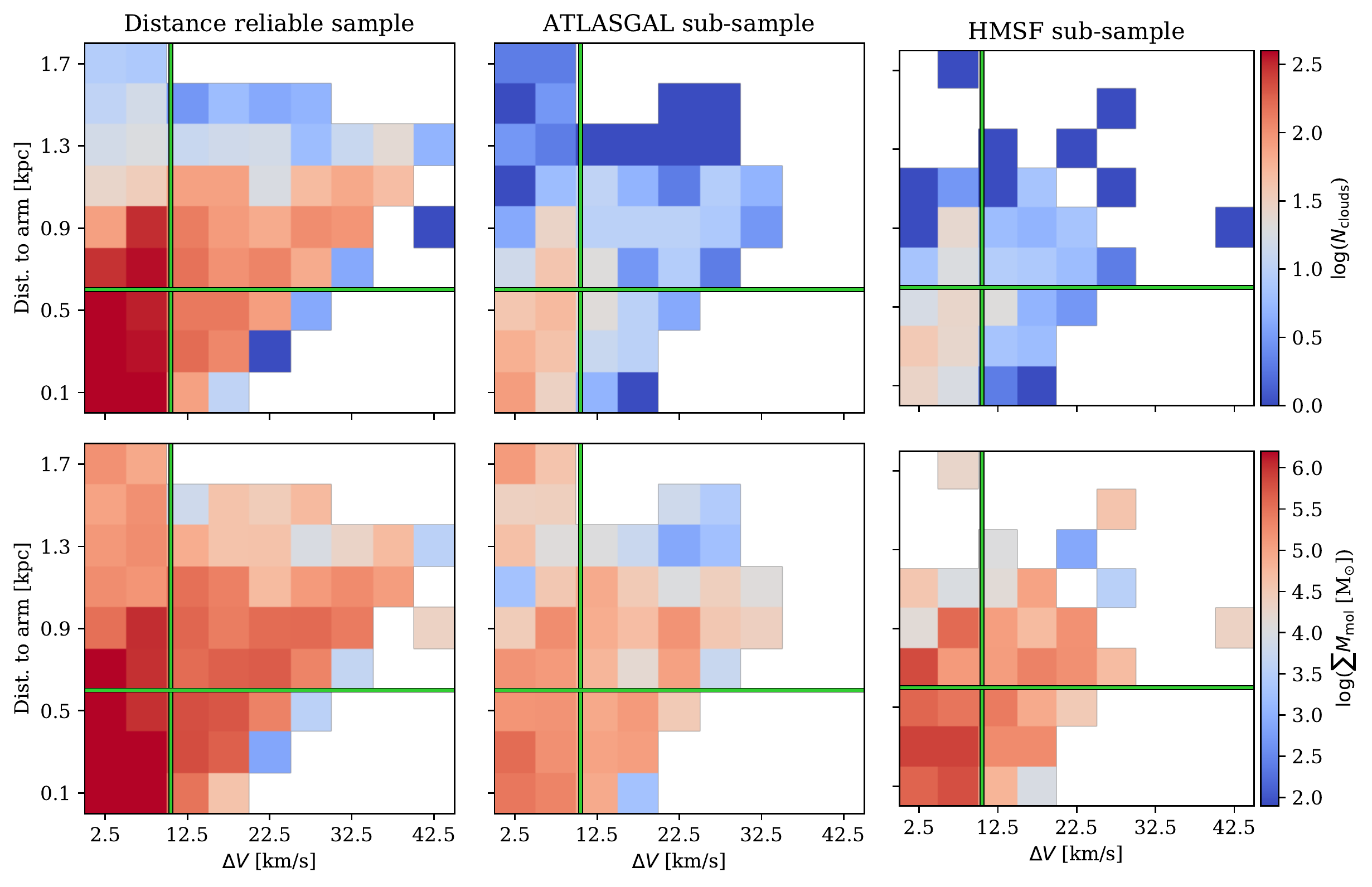}
   \caption{Bi-dimensional histograms of the total number of cloud counts (top row) and cumulative molecular gas mass in clouds (bottom row) within bins of 0.1\,kpc and 0.25\,\kms{}, considering the $xy-$plane distance and the velocity offset to the closest spiral arm. Different cloud subsamples are considered: the distance-reliable sample (left), the subset of this sample that contains an ATLASGAL source (middle), and the subset with a high-mass-star-formation (HMSF) signpost (right). The vertical green line indicates the $\Delta V=10$\,\kms\ that we use to define a spiral arm, and the horizontal green line indicates the median spiral width of 580\,pc.}
    \label{F:sparm_cloud_voff_xydist}
\end{figure*}

Having located clouds in particular positions in the Galactic disc, it is possible to measure additional quantities such as surface densities and to study the contrast between spiral arms and inter-arm regions. Figure~\ref{F:sparm_gmc_match} (top-left panel) gives some indications that certain regions of the Milky Way show enhanced densities of clouds. Those enhancements may correspond to the location of the arms, but not always. Over-densities are also observed between the arms defined by TC93. An additional analysis of the distribution of clouds around arms is provided in Fig.~\ref{F:sparm_cloud_voff_xydist} where the cumulative number of clouds and mass within clouds is binned according to the $xy-$plane distance of the clouds and their velocity offset with respect to the closest spiral arm. Generally, we observe that the number of clouds is indeed enhanced closer to the spiral arms, especially for clouds that host an ATLASGAL source or an HMSF signpost. This is not true for the mass contained within clouds, because a non-negligible amount of mass is observed in bins that cannot be attributed to spiral arms. The figure also shows that some clouds appear to be located away from the closest spiral arm both in terms of distance and velocity, for example up to maximum values of 1.8~kpc in distance on the $xy-$plane and with a velocity offset below 50\,\kms{}. Nevertheless, as noted above, several clouds associated with the inter-arm regions are observed within 10\,\kms{} of the closest arm. 

In the following, we attempt to better quantify these findings, first by calculating the areas of the different regions. The area spanned by each spiral arm is obtained by multiplying its length by its `width', with this latter being taken to be the median of two times the distribution of cloud offset to the closest (and associated) arm. 
In general, the width calculated in this way suggests that spiral arms are relatively wide, with a global median of $\sim580$\,pc, but with a broad inter-quartile range IQR$\sim830$\,pc. Those measurements are more than a factor of two larger than the spiral arm widths estimated by \cite{reid2019} (using maser locations), while being more similar to the measurements of \cite{vallee2017} who suggests arm widths of $\sim600$\,pc, considering a variety of tracers.
The global area of the inter-arm region is taken to be
\begin{equation}
A_{\rm inter-arm} = \left(\frac{\Delta l_{\rm survey} - \Delta l_{\rm unreliable}}{360^{\circ}}\right) \times \pi  d_{\rm max}^2 - \pi r_{\rm in}^2 - \sum A_{\rm spiral~arms}
,\end{equation}
where $\Delta l_{\rm survey}=78^{\circ}$ is the full longitudinal range spanned by the SEDIGISM survey, $\Delta l_{\rm unreliable}=15^{\circ}$ indicates the longitudinal region where cloud distances are not reliable, $d_{\rm max} = 15$\,kpc is the distance within which most clouds are located, $r_{\rm in} = 3$\,kpc is the radius where the clouds are associated to the 3~kpc arm, and $\sum A_{\rm spiral~arms}$ is the total Galactic disc area within the spiral arms. 

The number of clouds per unit area is slightly higher than the number of clouds per unit length, but it is still around several tens of clouds per kpc$^2$ within the spiral arms. At the same time, there is a difference between arms of less than an order of magnitude in the total amount of gas mass per unit area; $10^{5} - 10^{5.8}$\,M$_{\odot}$/kpc$^2$.
Interestingly, the contrast between spiral arms and the inter-arm regions in terms of mass is about 1.5, which is similar to the mass surface-density ratio. As a final note, these numbers do not change significantly if the ATLASGAL or HMSF samples are employed.

The molecular gas mass surface densities inferred from SEDIGISM clouds appear lower than the values measured Galaxy-wide. \cite{miville_deschenes2017} (their Fig.~9) show a collection of $\Sigma_{\rm mol}$ profiles measured using different techniques. In the inner Galaxy, all techniques obtained values consistent with $\Sigma_{\rm mol}\sim5$\,M$_{\odot}$\,pc$^{-2}$, including the analysis based on cloud segmentation. Considering the cloud distribution across the full field of SEDIGISM (excluding the inner 3~kpc region and the longitudes between -5$^{\circ}$ and 10$^{\circ}$ where clouds with uncertain locations are found), we calculated $\Sigma_{\rm mol}\sim0.3$\,M$_{\odot}$\,pc$^{-2}$, roughly an order of magnitude lower than the `canonical' value for the inner Galaxy. This discrepancy can arise from several factors. The cloud segmentation of \cite{miville_deschenes2017} is performed on \cite{dame01} data that include the whole Milky Way. Instead, our cloud catalogue is restricted to the IV quadrant and within $|b|<0.5^{\circ}$. By restricting the \cite{miville_deschenes2017} catalogue to the SEDIGISM coverage, we obtain $\Sigma_{\rm mol}\sim3.4$\,M$_{\odot}$\,pc$^{-2}$. In addition, the \cite{dame01} catalogue was built using a technique that attempts to allocate all CO flux into clouds, while the technique used to construct the SEDIGISM catalogue instead allows filtering of the emission, meaning that not all CO flux is necessarily assigned to discrete objects, but a fraction of emission is allowed to be in a diffuse form (as actually observed, \citealt{pety2013}). By considering Milky Way catalogues built using techniques more comparable to that used for SEDIGISM (from \citealt{rice2016} and \citealt{colombo2019}), restricted only to the clouds observed with the SEDIGISM field, we measure $\Sigma_{\rm mol}\sim0.8-0.9$\,M$_{\odot}$\,pc$^{-2}$. However, those cloud catalogues are built from $^{12}$CO data that trace a more diffuse medium with respect to the $^{13}$CO\,(2-1) emission used here and suffer from more severe optical depth effects. At the same time, $^{13}$CO suffers from more severe lower-beam-filling-factor effects than $^{12}$CO. We repeated the same test using data from the Galactic Ring Survey (GRS; \citealt{jackson2006}), where $^{13}$CO\,(1-0) was used to observe the Milky Way I quadrant. From the cloud measurements presented in \cite{roman_duval2009} and \cite{roman_duval2010}, we obtained $\Sigma_{\rm mol}\sim0.6$\,M$_{\odot}$\,pc$^{-2}$, which is fairly consistent with the $\Sigma_{\rm mol}$ measured across the SEDIGISM field. GRS and SEDIGISM both imaged  the inner Galaxy, but their derived cloud catalogues do not overlap in any Galactic region, and therefore a completely compatible comparison is not possible. As for the studies of \cite{rice2016} and \cite{colombo2019}, the technique used by the GRS collaboration to produce the cloud catalogues does not assign all CO emission to clouds, and is therefore similar in scope to the SEDIGISM cloud segmentation technique. From these experiments, it appears that the discrepancies noted in our $\Sigma_{\rm mol}$ measurements with respect to similar measurements in the Milky Way are attributable to the cloud segmentation and the tracer, with possibly a larger influence from the latter.

\begin{table*}
\centering
\caption{Cloud property distribution statistics across regions and subsamples}
\begin{tabular}{c|c|cc|cc|cc|cc}
\hline
& & \multicolumn{8}{|c}{Full sample} \\
& & \multicolumn{4}{|c|}{Science sample} & \multicolumn{4}{|c}{Dist. lim. sample} \\
& & \multicolumn{2}{|c|}{SA} & \multicolumn{2}{|c|}{IA} & \multicolumn{2}{|c|}{SA} & \multicolumn{2}{|c}{IA} \\
Property & Units & $\mu$ & $IQR$ & $\mu$ & $IQR$ & $\mu$ & $IQR$ & $\mu$ & $IQR$ \\
\hline
$\log(R_{\rm eff})$ & pc & 0.36 & 0.51 & 0.37 & 0.39 & 0.39 & 0.34 & 0.38 & 0.37 \\
$\log(\sigma_{\rm v})$ & km~s$^{-1}$ & -0.11 & 0.30 & -0.12 & 0.28 & 0.06 & 0.32 & 0.05 & 0.31 \\
$\log(M_{\rm mol})$ & M$_{\odot}$ & 3.11 & 1.08 & 3.11 & 0.87 & 3.23 & 0.82 & 3.28 & 0.88 \\
$\log(\Sigma_{\rm mol})$ & M$_{\odot}$~pc$^{-2}$ & 1.87 & 0.24 & 1.87 & 0.23 & 1.97 & 0.27 & 2.02 & 0.32 \\
$\log(\alpha_{\rm vir})$ &  & 0.13 & 0.44 & 0.08 & 0.42 & 0.26 & 0.41 & 0.24 & 0.48 \\
$\log(AR)$ & & 1.00 & 0.32 & 0.98 & 0.33 & 1.19 & 0.27 & 1.13 & 0.30 \\
\hline
\hline
& & \multicolumn{8}{|c}{ATLASGAL sample} \\
& & \multicolumn{4}{|c|}{Science sample} & \multicolumn{4}{|c}{Dist. lim. sample} \\
& & \multicolumn{2}{|c|}{SA} & \multicolumn{2}{|c|}{IA} & \multicolumn{2}{|c|}{SA} & \multicolumn{2}{|c}{IA} \\
Property & Units & $\mu$ & $IQR$ & $\mu$ & $IQR$ & $\mu$ & $IQR$ & $\mu$ & $IQR$ \\
\hline
$\log(R_{\rm eff})$ & pc & 0.38 & 0.50 & 0.47 & 0.35 & 0.36 & 0.33 & 0.40 & 0.32 \\
$\log(\sigma_{\rm v})$ & km~s$^{-1}$ & 0.04 & 0.20 & 0.03 & 0.22 & 0.08 & 0.21 & 0.06 & 0.25 \\
$\log(M_{\rm mol})$ & M$_{\odot}$ & 3.26 & 1.05 & 3.50 & 0.74 & 3.25 & 0.68 & 3.34 & 0.52 \\
$\log(\Sigma_{\rm mol})$ & M$_{\odot}$~pc$^{-2}$ & 2.04 & 0.22 & 2.09 & 0.21 & 2.03 & 0.22 & 2.11 & 0.20 \\
$\log(\alpha_{\rm vir})$ &  & 0.24 & 0.55 & 0.08 & 0.44 & 0.28 & 0.37 & 0.26 & 0.47 \\
$\log(AR)$ &  & 1.07 & 0.31 & 1.01 & 0.35 & 1.17 & 0.30 & 1.10 & 0.26 \\
\hline
\hline
& & \multicolumn{8}{|c}{HMSF sample} \\
& & \multicolumn{4}{|c|}{Science sample} & \multicolumn{4}{|c}{Dist. lim. sample} \\
& & \multicolumn{2}{|c|}{SA} & \multicolumn{2}{|c|}{IA} & \multicolumn{2}{|c|}{SA} & \multicolumn{2}{|c}{IA} \\
Property & Units & $\mu$ & $IQR$ & $\mu$ & $IQR$ & $\mu$ & $IQR$ & $\mu$ & $IQR$ \\
\hline
$\log(R_{\rm eff})$ & pc & 0.62 & 0.46 & 0.64 & 0.39 & 0.62 & 0.29 & 0.60 & 0.33 \\
$\log(\sigma_{\rm v})$ & km~s$^{-1}$ & 0.22 & 0.24 & 0.22 & 0.25 & 0.23 & 0.23 & 0.23 & 0.27 \\
$\log(M_{\rm mol})$ & M$_{\odot}$ & 3.97 & 1.03 & 4.07 & 0.89 & 3.98 & 0.63 & 4.05 & 0.92 \\
$\log(\Sigma_{\rm mol})$ & M$_{\odot}$~pc$^{-2}$ & 2.19 & 0.30 & 2.24 & 0.24 & 2.21 & 0.29 & 2.29 & 0.28 \\
$\log(\alpha_{\rm vir})$ &  & 0.19 & 0.63 & 0.08 & 0.54 & 0.22 & 0.42 & 0.15 & 0.50 \\
$\log(AR)$ &  & 1.17 & 0.29 & 1.22 & 0.35 & 1.21 & 0.25 & 1.28 & 0.30 \\
\hline
\hline
\end{tabular}
\tablefoot{Median ($\mu$) and inter-quartile range ($IQR$) of cloud property distributions (from top to bottom effective radius $R_{\rm eff}$, velocity dispersion $\sigma_{\rm v}$, molecular gas mass $M_{\rm mol}$, molecular gas mass surface density $\Sigma_{\rm mol}$, virial parameter $\alpha_{\rm vir}$, and aspect ratio $AR$) in the spiral arms (SA) with respect to the inter-arm regions (IA). The analysis is performed separately for the full sample, for the cloud sample that contains at least one ATLASGAL source (ATLASGAL sample), and for the sample with a high mass star formation (HMSF) signpost (HMSF sample); and their respective science and complete distance-limited subsamples.}
\label{T:med_iqr_taylor-cordes2003}
\end{table*}

\begin{table*}
\centering
\caption{KS test results on cloud properties for spiral arm and inter-arm distributions}
\begin{tabular}{c|cc|cc|cc}
\hline
Property & \multicolumn{2}{|c|}{Full} & \multicolumn{2}{|c|}{ATLASGAL} & \multicolumn{2}{|c}{HMSF} \\
& $p_{\rm val}$ & $D_{\rm stat}$ & $p_{\rm val}$ & $D_{\rm stat}$ & $p_{\rm val}$ & $D_{\rm stat}$ \\
\hline
\hline
\multicolumn{7}{c}{Science sample} \\
\hline
$R_{\rm eff}$ & <~0.0001 & 0.08 & 0.0008 & 0.17 & 0.2050 & 0.13 \\
$\sigma_{\rm v}$ & 0.3806 & 0.02 & 0.7908 & 0.06 & 0.7263 & 0.08 \\
$M_{\rm mol}$ & <~0.0001 & 0.07 & <~0.0001 & 0.22 & 0.1013 & 0.15 \\
$\Sigma_{\rm mol}$ & 0.9757 & 0.01 & 0.0027 & 0.16 & 0.2058 & 0.13 \\
$\alpha_{\rm vir}$ & <~0.0001 & 0.08 & <~0.0001 & 0.22 & 0.1525 & 0.13 \\
$AR$ & 0.0316 & 0.04 & 0.0846 & 0.11 & 0.3664 & 0.11 \\
\hline
\hline
\multicolumn{7}{c}{Complete distance-limited sample} \\
\hline
$R_{\rm eff}$ & 0.6720 & 0.05 & 0.8385 & 0.09 & 0.8380 & 0.10 \\
$\sigma_{\rm v}$ & 0.7971 & 0.04 & 0.9004 & 0.08 & 0.8246 & 0.11 \\
$M_{\rm mol}$ & 0.3791 & 0.06 & 0.0902 & 0.18 & 0.8088 & 0.11 \\
$\Sigma_{\rm mol}$ & 0.0088 & 0.11 & 0.0041 & 0.26 & 0.3077 & 0.16 \\
$\alpha_{\rm vir}$ & 0.3265 & 0.06 & 0.2856 & 0.15 & 0.4696 & 0.14 \\
$AR$ & <~0.0001 & 0.15 & 0.0968 & 0.18 & 0.5689 & 0.13 \\
\hline
\end{tabular}
\tablefoot{$p-$values ($p_{\rm val}$) and statistics ($D_{\rm stat}$) from the KS test comparing the distributions of cloud properties (from top to bottom: effective radius $R_{\rm eff}$, velocity dispersion $\sigma_{\rm v}$, molecular gas mass $M_{\rm mol}$, molecular gas mass surface density $\Sigma_{\rm mol}$, virial parameter $\alpha_{\rm vir}$, and aspect ratio $AR$) in the spiral arms (SA) with respect to the inter-arm regions (IA). The analysis is performed separately for the full sample, for the cloud sample that contains at least one ATLASGAL source, and for the cloud sample
that contains a HMSF signpost.}
\label{T:pvalues_taylor-cordes2003}
\end{table*}

%%%%%%%%%%%%%%%%%%%%%%%%%%%%%%%%%%%%%%%%%%%%%%%%%%%%%%%%%%%%%%%%%%%%%%%%%%%%%%
\subsection{Properties of the clouds in the spiral arms and in the inter-arm}
\label{SS:gmc_props}
%%%%%%%%%%%%%%%%%%%%%%%%%%%%%%%%%%%%%%%%%%%%%%%%%%%%%%%%%%%%%%%%%%%%%%%%%%%%%%

\begin{figure*}
        \includegraphics[width=0.95\textwidth]{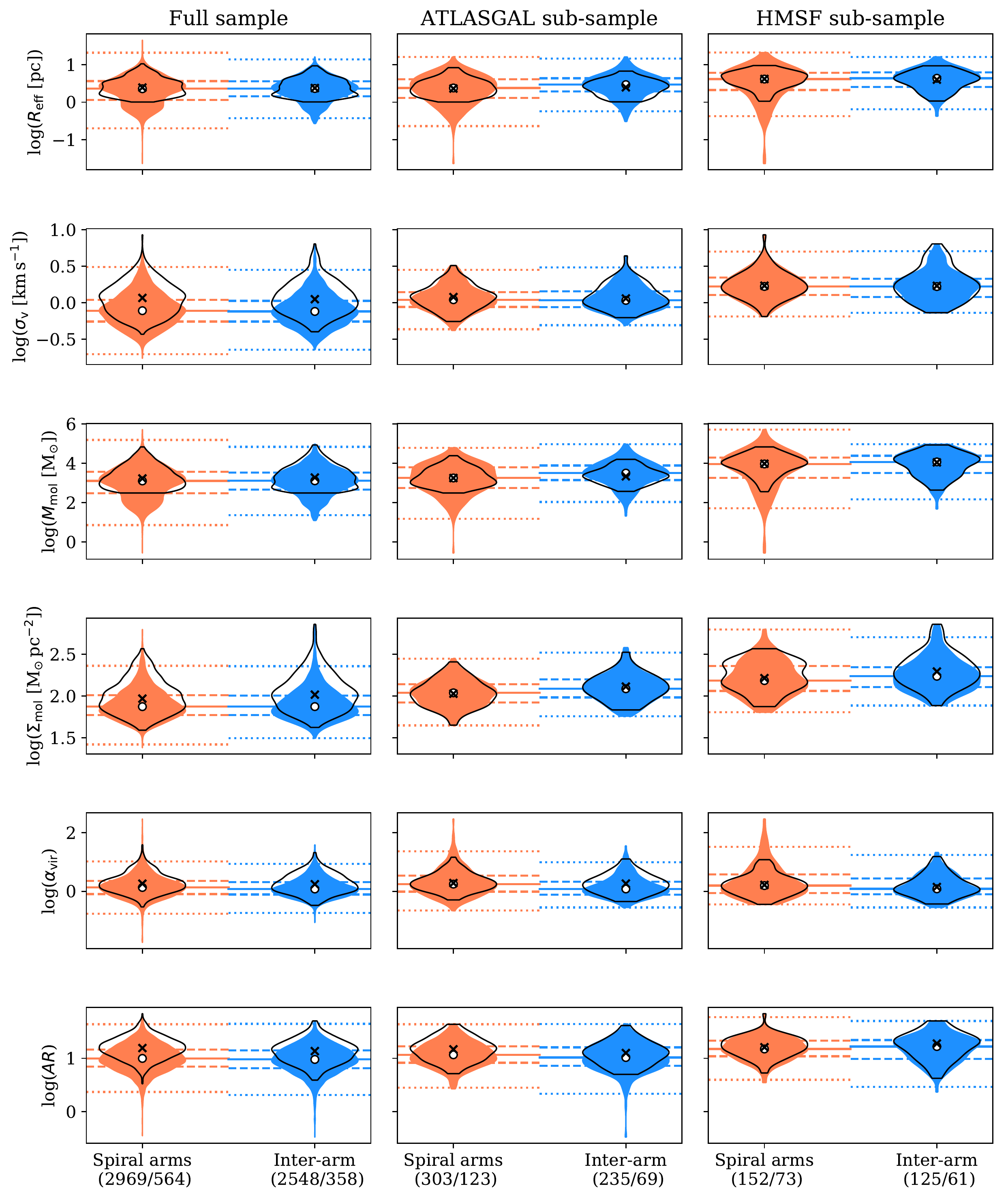}
   \caption{Distributions of the cloud properties in the spiral arms (red) and in the inter-arm regions (blue) drawn from the science sample and represented as violin plots (which show the relative shape of the distributions). The violins outlined in black show the distributions drawn from the complete distance-limited sample. From top to bottom rows: effective radius, velocity dispersion, molecular gas mass, molecular gas mass surface density, virial parameter, and aspect ratio, for the full (science or complete distance-limited) sample of clouds, clouds with an ATLASGAL source, and clouds with an HMSF signpost (from left to right). The full horizontal line shows the distribution medians of the science sample. The dashed lines display the position of the $Q_{25} = 25^{\rm th}$ and $Q_{75} = 75^{\rm th}$ percentiles. The dotted lines indicate the position of the lower and upper whiskers (defined as $Q_{25} - 1.5\,IQR$ and $Q_{75} + 1.5\,IQR$, respectively, where the interquartile range $IQR=Q_{75}-Q_{25}$). The horizontal lines have the same colours as the associated violin diagram. In the violins, the white circle indicates the median of the distribution from the science sample while the cross shows the median for the complete distance-limited sample. In the x-axis labels, for each environment (spiral-arms and inter-arm regions), the two numbers within the brackets indicate the number of clouds in the science sample and in the complete distance-limited sample, respectively.}
    \label{F:gmc_sparm_props}
\end{figure*}

The molecular cloud catalogue from the SEDIGISM survey reports a number of different cloud properties. It is interesting now to analyse whether the distributions of these properties vary between spiral arms and inter-arm regions. In this section, we consider the cloud effective radius ($R_{\rm eff}$), velocity dispersion ($\sigma_{\rm v}$), molecular gas mass ($M_{\rm mol}$), molecular gas mass surface density ($\Sigma_{\rm mol}$), virial parameter ($\alpha_{\rm vir}$), and aspect ratio ($AR$), as defined in DC21 (their Section 5.1; see also \citealt{colombo2019} for further details). We use the science sample and the complete distance-limited sample (that we refer to as `full' samples in this section), described in Section~\ref{SSS:gmc_cat}, from which we removed the clouds attributed to the 3~kpc arms and the ones with uncertain locations. We also differentiate between two subsamples of the full science sample and the complete distance-limited sample: a subsample that contains at least one ATLASGAL clump (ATLASGAL sample), and a  subsample with at least one HMSF signpost (HMSF sample; see Section~\ref{SSS:gmc_cat}). Figure~\ref{F:gmc_sparm_props} and Table~\ref{T:med_iqr_taylor-cordes2003} report the results of the analysis. 

Regarding the full science sample, we find slightly more clouds in the spiral arms ($\sim55$\% of the sample considered here) than in the inter-arm regions ($\sim45$\% of the sample). Similar percentages are observed for the subsets of clouds with one or more ATLASGAL sources or HMSF regions. Generally speaking, the distributions of properties differ slightly in shape from spiral arms to the inter-arm regions (especially at the distribution tails), but the median and the 25$^{\rm th}$ and 75$^{\rm th}$ percentiles are similar everywhere. Additionally, we observed that the median $\sigma_{\rm v}$ and $\Sigma_{\rm mol}$ appear to increase across the three subsamples (as noticed by DC21), independently of whether the clouds are in the spiral arms or in the inter-arm regions. Considering the complete distance-limited sample and its subsets that contain ATLASGAL or HMSF sources, the distributions drawn from both spiral-arm and inter-arm-region clouds appear narrower than for the corresponding science samples, especially regarding the spiral arm distributions. However, the median values of complete distance-limited and science samples appear the same everywhere, except for the velocity dispersion distribution (full sample). We still observe slightly more clouds in the spiral arms compared to the inter-arm region (in particular, 60\% of the total number of objects of the full sample are in the spiral arms, 65\% considering the ATLASGAL subsample, and 55\% considering the HMSF subsample).

To assess whether the property distributions for clouds in the spiral arms and inter-arm regions can be drawn from the same parental distribution, we use the two-sample Kolmogorov-Smirnov (KS) test as implemented within the {\sc SCIPY} package\footnote{\url{https://docs.scipy.org/doc/scipy/reference/generated/scipy.stats.ks\_2samp.html}}. This test provides two quantities, the KS statistic ($D_{\rm stat}$), which quantifies the absolute maximum distance between the cumulative distribution functions from the two samples, and the $p-$value ($p_{\rm val}$), which can be used to reject the null hypothesis that the two samples are drawn from the same parental distribution if the p-value is less than the significance level (generally considered to be equal to 0.05 as in Section~\ref{SSS:gmc_match}). The KS tests performed on both science and complete-distance limited samples indicate that the differences between the properties of the clouds in the two Galactic environments observed in the science sample can be largely attributed to a distance bias (see Table~\ref{T:pvalues_taylor-cordes2003}). In particular, the $p$-values for $R_{\rm eff}$, $M_{\rm mol}$, and $\alpha_{\rm vir}$ are very low (well below $10^{-4}$) for the full sample and for the ATLASGAL sample, but this is not observed in the complete distance-limited sample. Interestingly, the $p-$values for two of the properties least influenced by distance biases, e.g. $AR$ and $\Sigma_{\rm mol}$, are below the significance level ($p_{\rm val}=0.05$) in both the science and complete distance-limited samples. The $p$-values for the HMSF sample are significantly higher for every property and sample, implying that there is no significant difference between the distributions of clouds in the spiral arms and the inter-arm regions when considering clouds that contain at least one HMSF signpost. 

A similar analysis was carried out by \cite{rigby2019}, who investigated the clump distributions between spiral arms and inter-arm regions across the CO Heterodyne Inner Milky Way Plane Survey (CHIMPS). In contrast to our findings, these latter authors observed a low $p-$value for the spiral arm and inter-arm region distributions $\sigma_{\rm v}$, but, as in our case, they obtained a low $p-$value for the $\alpha_{\rm vir}$ distributions. However, given the difference in tracers, clump--cloud segmentation techniques, and property calculation methods, a fully robust comparison between the results is not possible.

%%%%%%%%%%%%%%%%%%%%%%%%%%%%%%%%%%%%%%%%%%%
\section{Discussion: the Milky Way, a grand-design or flocculent spiral galaxy?}
\label{S:disc}
%%%%%%%%%%%%%%%%%%%%%%%%%%%%%%%%%%%%%%%%%%%

\begin{figure*}
    \centering
        \includegraphics[width=0.95\textwidth]{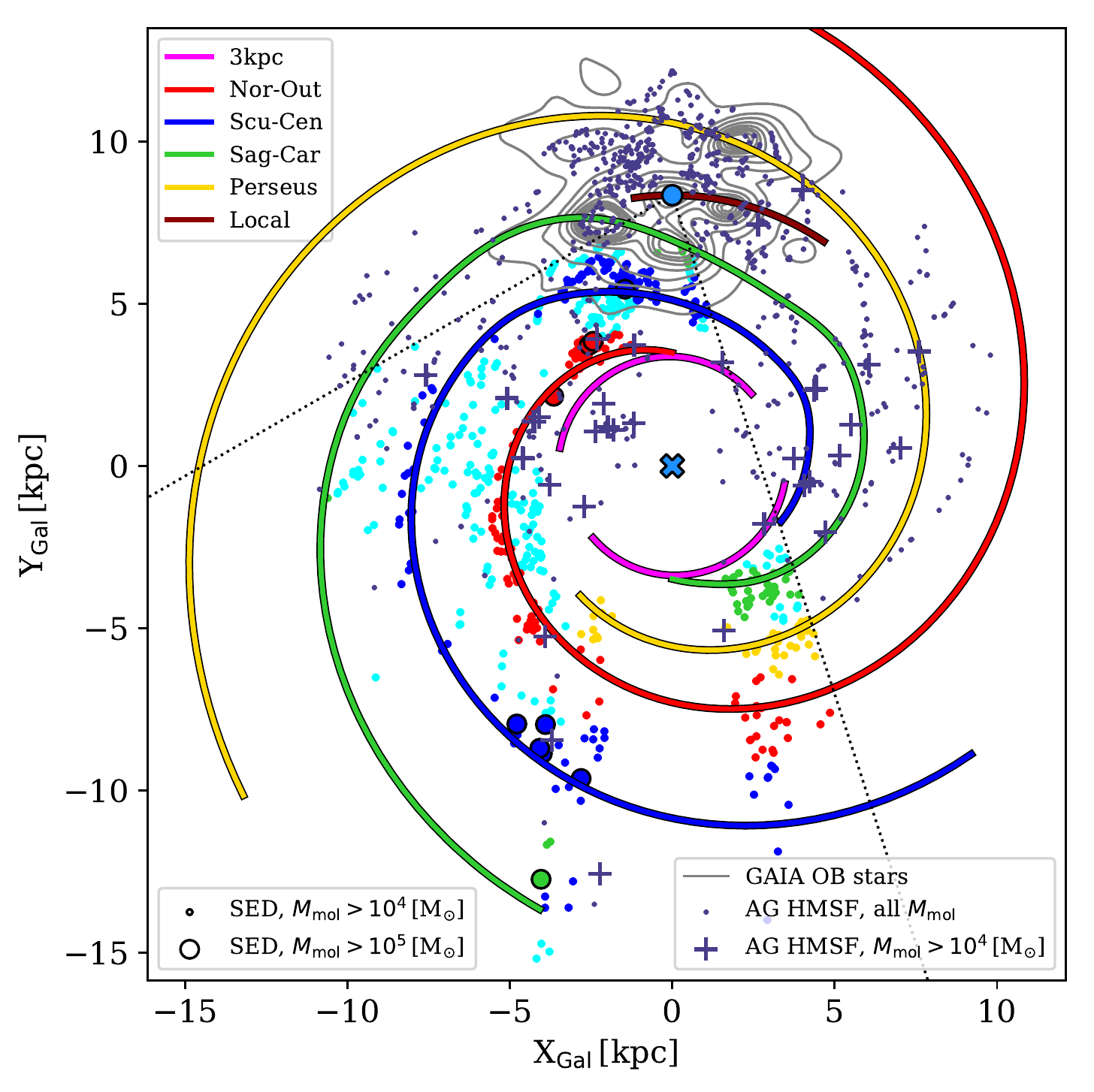}
   \caption{Distribution of the most massive clouds from SEDIGISM (SED, coloured circles) ATLASGAL HMSF (AG HMSF) clumps (purple dots and crosses, from \citealt{urquhart2018}) and GAIA distribution of OB stars (grey contours, from \citealt{xu2021}), overlaid with the spiral arm location from TC93 models. Local spur parameters are taken from \cite{reid2019}. Other symbol conventions follow Fig.~\ref{F:sparm_gmc_match} (top panels).}
    \label{F:gaia}
\end{figure*}

Despite many decades of observational and theoretical work, several aspects of the Milky Way remain shrouded in mystery. Because of the position of the Sun within the Galactic disc, the Milky Way appears to us as an edge-on galaxy, though observed at very high resolution. The determination of its large-scale structure, for example, is extremely challenging, due to a series of effects such as the kinematic distance ambiguity (for the inner Galaxy), velocity crowding, and spiral arm streaming motions. Nevertheless, in recent years, the increase in the coverage of spectroscopic large-scale Galactic plane surveys (see e.g., Table~1 in \citealt{schuller2021} and Fig.~1 in \citealt{stanke2019}), high-quality stellar data \citep[e.g.,][]{deng2012,kunder2017,majewski2017,gaia2021}, and maser measurements \citep{reid2014,xu2018,reid2019} has provided new insights into this complex issue. 

In particular, the mechanism that generated the Milky Way's spiral arms, as well as the number thereof, are still debated \citep{dobbs_baba2014}. The most common interpretation is that the Milky Way has four primary arm features: the Perseus, Sagittarius-Carina, Scutum-Centaurus, and Norma-Outer spiral arms \citep[e.g. ][]{georgelin_georgelin1976, urquhart2014, reid2019}, as we have adopted in this work. However, some studies have suggested that a two-armed structure (corresponding to the Perseus and Scutum-Centaurus arms) better fits their data \citep{drimmel2000,drimmel_spergel2001}. As highlighted by GLIMPSE data \citep{Benjamin2005,Churchwell2009}, it appears that the Milky Way spiral structure is tracer-dependent \citep{hou_han2014}: the old-stellar population (observed in the infrared bands) seems to be distributed across two spiral arms, while young stars, gas (\citealt{xu2018}, and references therein), and dust \citep{rezaei_kh2018} appear to follow a four-armed spiral structure, even if their positions across the disc cannot always be
unambiguously determined. One possible explanation is that the Perseus and Scutum-Centaurus arms were generated by a large-scale density wave, while the other ---slightly weaker--- arms are the result of resonances (e.g. \citealt{martos2004,pettitt2014}), visible only in the distribution of gas, dust, and young stars. As the Milky Way possesses a bar \citep[e.g. ][]{wegg2015}, this scenario might fit better with the observations of barred galaxies in the local Universe, which tend to show two arms originating from the two tips of the bar. Even this picture may be an oversimplification, with many secondary features observed (such as arm bridges and segments) in the distribution of the stars \citep{quillen2018} and of the cold gas \citep{amores_lepine2005}, especially towards the outer Galaxy \citep{koo2017}. An explicit example is the Local arm, which is often considered to be a spur rather than a large-scale arm that spans a significant fraction of the disc (e.g., \citealt{georgelin_georgelin1976}; though also see \citealt{xu2013}).
The question of the structure of the Milky Way spiral arms has been revisited on a number of recent occasions using \emph{Gaia} data, sometimes giving  divergent results. For example, \cite{castro-ginard2021} using the most updated catalogue of open clusters measured different pattern speeds for different spiral arms, favouring a flocculent structure with transient arms that co-rotate with the disc. Alternatively, \cite{martinez-medina2021}, studying the kinematics of selected stars in the Gaia EDR3 sample, found instead that stars do not always co-rotate with the disc: objects in the spiral arms tend to rotate more slowly than objects in the inter-arm region.  

Theoretical works have also attempted to shed light on the nature of the Galactic spiral pattern. Numerical methods in particular have been used in attempts to fit different structural models to a number of different galactic tracers. The studies of \citet{pettitt2014,Pettitt15} found that fewer grand-design discs tend to allow the more irregular emission features in the observed \lvmap\ to be matched to observations. Other authors have made similar findings, with many outer arm features ($>4$) needed to reproduce the observations \citep{2016ApJ...824...13L} with the time evolution of features such as the Perseus arm, pointing towards a dynamic kind of nature \citep{Tchernyshyov18,Baba18}. Stellar material has also become even more usable for studies of Galactic arms, with transient spiral features also providing a good match to observed stellar velocities \citep{Grand15,Sellwood19}.

Within the spiral arm sections studied here, we found surface densities of the order of $10^5-10^6$\,M$_{\odot}$\,kpc$^{-2}$. For example, in a region of 1\,kpc$^2$, fewer than ten GMCs of $10^5$\,M$_{\odot}$ are observed. Similar orders of magnitude are measured for linear masses. Such densities would indicate that large areas of the spiral arms are devoid of dense gas, or are filled with diffuse gas. Giant molecular clouds have sizes of up to $\sim100$\,pc, at least one order of magnitude lower than the unit length of the spiral arm tracks considered.

Although our results are self-consistent, the linear mass we obtain for Milky Way spiral arms, even after considering the possible variations introduced by the choice of tracer or segmentation technique (see Section~\ref{SS:sparm_linmass}), are still largely short of the $10^{8}$\,M$_{\odot}$\,kpc$^{-1}$ linear mass values obtained for M51 \citep[e.g.][]{colombo14a}, the most prototypical grand-design galaxy in the local Universe. Nevertheless, such values are not uncommon in the local Universe. For instance, \citet{sun2018} obtained molecular gas surface densities from high-resolution observations of a sample of 15 nearby galaxies. This sample contains grand-design spiral galaxies (such as M51 and NGC628), flocculent and multi-armed galaxies (e.g. NGC2835, NGC6744), as well as atomic-dominated galaxies (M31 and M33) and mergers (Antennae). Considering their beam sizes and the surface densities measured, and assuming that most of the molecular gas emission comes from spiral arms, we would get linear masses in the range of $10^5-10^7$\,M$_{\odot}$\,kpc$^{-1}$, more in line with our Milky Way values. Additionally, \cite{rosolowsky2021} performed molecular cloud segmentation for a sample of ten nearby galaxies, observing a spiral arm--inter-arm region molecular gas mass contrast of between $\sim1-2.3$ (E. Rosolowsky, private communication) considering the mass within clouds, which is similar to what we measure here (1.5, see Section~\ref{SS:sparm_linmass}). 

In Section~\ref{SS:pdf}, we show that the PDF drawn from the integrated intensity from the CO emission within the spiral arms is similar in shape to the inter-arm region PDF. This evidence has been quantified via IDIs (and LDIs) that show largely similar values for the two PDFs. Clearly, these results are qualitatively at variance with PDF studies of grand-design galaxies. In M51, the PDFs observed for the emission within the spiral arms are significantly wider than those of the inter-arm region emission \citep{hughes2013a}. This difference is also reflected in the values of the IDIs (see \citealt{hughes2013a}, their Table 2) drawn from the spiral arm and inter-arm environments. Nevertheless, nearby galaxy studies of this kind are not completely comparable with Milky Way studies of the same kind: differences in resolution, tracers, and our inability to perfectly separate the spiral arm emission from the bulk emission might play a role in homogenising the properties of the molecular ISM within and outside the spiral arms. In addition, the brightness of the CO emission, being a function of column density, is prone to projection effects, potentially influencing the shape of emission PDFs. Flux along a given line of sight in face-on nearby galaxies is difficult to separate in velocity because of the limited spectral resolution and the lower vertical velocity structure, which blends the emission from clouds along the same sight line. Therefore, multiple CO emission components can contribute to the global intensity observed, potentially enhancing the amount of brighter emission, particularly towards the spiral arms. This is starkly different for the Milky Way for which the velocity structure
can be clearly resolved and the CO emission splits across the various spiral arms, thus resulting in emission PDFs that are less skewed towards high values.

In Section~\ref{SS:gmc_props} we observe that, for clouds in the science sample, the distributions of $R_{\rm eff}$, $M_{\rm mol}$, $\Sigma_{\rm mol}$, and $\alpha_{\rm vir}$ for clouds in the spiral arms differ from those in the inter-arm regions. However, most of those differences tend to disappear when we consider the complete distance-limited sample, indicating that the statistics might be affected by a distance bias in the science sample. The only cloud property that shows significant distribution differences in both science and complete distance-limited samples is $AR$. This evidence is similar to that found with molecular cloud simulations from \cite{duarte-cabral2016} and \cite{duarte-cabral2017}, who showed that, based on aspect ratio, the cloud populations in their spiral arm and inter-arm regions do not descend from the same parental distribution. However, we did not observe a surplus of elongated structures in the inter-arm region compared to the spiral arms as in their simulations. 
When we look at the subsamples of clouds with an ATLASGAL counterpart or an HMSF signpost, we also find no significant difference between arms and the inter-arm medium. This suggests that star formation is taking place in clouds regardless of the environment. In other words, the conditions necessary to trigger HMSF can occur in both environments, although this does not exclude the possibility that spiral arms might be more conducive to the process. We do  see consistently slightly higher fractions of clouds in the spiral arms compared to the inter-arm region for all subsamples. However, the relative ratios of arm to inter-arm clouds are not significantly enhanced for the ATLASGAL or HMSF subsamples with respect to the full sample, as we would expect if spiral arms were to effectively drive a more efficient formation of dense gas and stars. Although our number statistics might be too limited to come to any robust conclusions in this regard (particularly as completeness limits could play a role in the ATLASGAL and HMSF associations), our findings could be an indication that the spiral arms in the Milky Way work as gas accumulators, as indicated by \cite{wang2020_thor}, who showed that the atomic-to-molecular gas ratio increases by a factor of six from spiral arms to the inter-arm region in the inner Galaxy. However, spiral arms might not be significantly determining the cloud properties or the conditions for star formation.  
This conclusion is consistent with other similar experiments in the Milky Way (see also \citealt{querejeta2021} for a nearby galaxy perspective). Indeed, although star-forming clumps seem to be closely associated with the spiral arm structure \citep{urquhart2018}, this might simply be due to source crowding \citep{moore2012}. Several studies involving independent datasets have shown that there is no enhancement of the star formation efficiency as a function of spiral arm locus \citep[e.g. ][]{moore2012,eden2013,eden2015,ragan2018,urquhart2021}. In particular, the star formation efficiency in the inner Galaxy does not show any particular trend compared to the velocity offset from the spiral arms or the Galactocentric radius \citep{urquhart2021}. Additionally, while the distribution of the HII regions tentatively follows a defined spiral pattern in the I and IV Milky Way quadrant, it becomes more sparse in the other quadrants \citep[see ][ and references therein]{xu2018}. More recently, \cite{xu2021} presented a comparison between the positions of OB stars
and the spiral arm loci traced by masers \citep[from ][]{reid2019}. While maser locations can be clearly fitted by logarithmic spirals, OB stars appear to be found everywhere across the surveyed area, and their distribution possibly follows the gas over-densities described by their natal clouds rather than a spiral structure. Some of these studies are summarised in Fig.~\ref{F:gaia}, which collects the most massive SEDIGISM clouds, the HMSF clumps from ATLASGAL \citep{urquhart2018}, and the OB stars from GAIA \citep{xu2021}. Interestingly, SEDIGISM clouds with masses above $10^5$\,M$_{\odot}$ are tightly associated to the spiral arms, while objects with $10^4<M_{\rm mol}<10^5$\,M$_{\odot}$ are distributed more homogeneously across the Galactic plane region surveyed by SEDIGISM. Similarly, the peaks in the distribution of the GAIA sources clearly correspond to the bottom of the spiral arm potential traced by TC93 models. At the same time, many OB stars are observed in the inter-arm region. Some high-mass star-forming regions from ATLASGAL clearly follow the spiral structure, however several of these clumps are detected everywhere across the Milky Way plane. These findings reinforce the idea that spiral arms in the Milky Way represent an environment that favours some physical processes (such as GMC and star formation), but they are not the exclusive place where those phenomena occur.
These pieces of evidence are in stark contrast with the HMSF picture present in M51, which displays a clear pattern and relationship with the spiral arm influence. In the inner part of this latter galaxy, the HMSF regions appear offset from the CO spiral arms \citep{schinnerer2013} and are distributed in discretely-spaced features that stretch in the downstream region of the inter-arm called spurs. As no age gradient of the stars is observed across the spurs, it appears that star formation happens in situ \citep{schinnerer2017}. This seems to occur in response to the strong streaming motion in M51's spiral arms, which provides an additional stabilisation element against the collapse of the clouds located at the bottom of the spiral arm potential \citep{meidt13}.  

In conclusion, the work we present here adds strong support to the mounting lines of evidence that the Milky Way is closer to a flocculent or multi-armed spiral galaxy (in line with earlier studies of \citealt{quillen2002}) rather than having a well-defined grand-design morphology. 

%%%%%%%%%%%%%%%%%%%%%%%
\section{Summary}
\label{S:summary}
%%%%%%%%%%%%%%%%%%%%%%%

We present an analysis of the molecular gas distribution in the inner Galaxy and in particular its relationship with the Milky Way spiral arms. We used $^{13}$CO\,(2-1) spectral line data from the high-resolution SEDIGISM survey and compared the properties of the global emission  from the spiral arms and those of the emission enclosed in molecular clouds with the emission from the inter-arm region. Our main results can be summarised as follows.

\begin{itemize}

    \item By uniquely associating clouds to given spiral arms, we are able to solve the KDA for 139 objects that had an unreliable distance attribution in the original SEDIGISM cloud catalogue.  
    
    \item The flux PDFs constructed from different Galactic regions show that globally the distribution of emission from the spiral arms does not largely differ from the distribution observed from the inter-arm regions. However, the flux PDFs from both regions are profoundly different from the PDF constructed from the flux towards the Galactic centre. Nevertheless, we do see that the inter-arm region contains more faint emission flux and less bright flux than the other regions, while the opposite behaviour is shown by the Galactic centre flux PDF. Similar conclusions can be drawn from luminosity PDFs built from cloud association.

    \item We calculate a spiral arm linear molecular gas mass that generally ranges between $10^5$ and $10^6$\,M$_{\odot}$\,kpc$^{-1}$ considering the global cloud emission from non-overlapping segments and cloud emission across the full  extent of the arms. Those values are similar to linear masses inferrable from spiral galaxies in the nearby Universe.
    
    \item Without the contribution of the clouds attributed to the 3~kpc arms (whose location with respect to the spiral structure is ambiguous), we find that $\sim10\%$ more clouds with reliable distances reside in the spiral arms than in the inter-arm regions. 
    The cloud number and gas mass per unit area is a factor of approximately 1.5 larger in the spiral arms than in the inter-arm regions, similar to other spiral-arm galaxies in the local Universe.  

    \item Tentatively, only the mass-surface-density and aspect-ratio distributions from the clouds in the spiral arms and inter-arm regions appear to show significant differences. For other properties (such as effective radius and mass), differences seems to be largely driven by a distance bias.
    
    \item We find that clouds with an HMSF signpost inside and outside the spiral arms  have identical properties, suggesting that the conditions needed to promote the formation of massive stars are achievable in both environments. The numbers of high-mass star-forming clouds in the arms are not enhanced beyond the expected enhancement due to higher numbers of clouds in the arms, suggesting that the spiral arms in the Milky Way are not increasing the efficiency of dense gas formation and high-mass star formation.
    
\end{itemize}

\noindent Taken together, these lines of evidences suggest that the Milky Way is a spiral galaxy with a flocculent or `multi-armed' morphology rather than a grand-design galaxy.

\begin{acknowledgements}
      The authors thank the anonymous referee, whose comments helped to improve the clearness of the paper. DC acknowledges support by the German
      \emph{Deut\-sche For\-schungs\-ge\-mein\-schaft, DFG\/} project
      number SFB956A. DC thanks Erik Rosolowsky for the useful discussion regarding the spiral arms in nearby galaxies. ADC acknowledges the support from the Royal Society University Research Fellowship (URF/R1/191609). ARP acknowledges the support of The Japanese Society for the Promotion of Science (JSPS) KAKENHI grant for Early Career Scientists (20K14456). HB acknowledges support from the European Research Council under the Horizon 2020 Framework Programme via the ERC Consolidator Grant CSF-648505. HB also acknowledges support from the Deutsche Forschungsgemeinschaft in the Collaborative Research Center (SFB 881) ``The Milky Way System'' (subproject B1). LB  acknowledges support from ANID project Basal AFB-170002. ASM acknowledges support by the Collaborative Research Centre 956, subproject A6, funded by the Deutsche Forschungsgemeinschaft (DFG), project ID 184018867. AG acknowledges support from NSF Grant 2008101. This research made use of Astropy,\footnote{http://www.astropy.org} a community-developed core Python package for Astronomy \citep{astropy2013, astropy2018}; matplotlib \citep{matplotlib2007}; numpy and scipy \citep{scipy2020}.

\end{acknowledgements}

\footnotesize{
\bibliographystyle{aa}
\bibliography{cold}

\begin{thebibliography}{137}
\expandafter\ifx\csname natexlab\endcsname\relax\def\natexlab#1{#1}\fi

\bibitem[{{Abreu-Vicente} {et~al.}(2016){Abreu-Vicente}, {Ragan},
  {Kainulainen}, {Henning}, {Beuther}, \& {Johnston}}]{abreu-vicente2016}
{Abreu-Vicente}, J., {Ragan}, S., {Kainulainen}, J., {et~al.} 2016, \aap, 590,
  A131

\bibitem[{{Am{\^o}res} \& {L{\'e}pine}(2005)}]{amores_lepine2005}
{Am{\^o}res}, E.~B. \& {L{\'e}pine}, J.~R.~D. 2005, \aj, 130, 659

\bibitem[{{Astropy Collaboration} {et~al.}(2018){Astropy Collaboration},
  {Price-Whelan}, {Sip{\H{o}}cz}, {G{\"u}nther}, {Lim}, {Crawford}, {Conseil},
  {Shupe}, {Craig}, {Dencheva}, {Ginsburg}, {Vand erPlas}, {Bradley},
  {P{\'e}rez-Su{\'a}rez}, {de Val-Borro}, {Aldcroft}, {Cruz}, {Robitaille},
  {Tollerud}, {Ardelean}, {Babej}, {Bach}, {Bachetti}, {Bakanov}, {Bamford},
  {Barentsen}, {Barmby}, {Baumbach}, {Berry}, {Biscani}, {Boquien}, {Bostroem},
  {Bouma}, {Brammer}, {Bray}, {Breytenbach}, {Buddelmeijer}, {Burke},
  {Calderone}, {Cano Rodr{\'\i}guez}, {Cara}, {Cardoso}, {Cheedella}, {Copin},
  {Corrales}, {Crichton}, {D'Avella}, {Deil}, {Depagne}, {Dietrich}, {Donath},
  {Droettboom}, {Earl}, {Erben}, {Fabbro}, {Ferreira}, {Finethy}, {Fox},
  {Garrison}, {Gibbons}, {Goldstein}, {Gommers}, {Greco}, {Greenfield},
  {Groener}, {Grollier}, {Hagen}, {Hirst}, {Homeier}, {Horton}, {Hosseinzadeh},
  {Hu}, {Hunkeler}, {Ivezi{\'c}}, {Jain}, {Jenness}, {Kanarek}, {Kendrew},
  {Kern}, {Kerzendorf}, {Khvalko}, {King}, {Kirkby}, {Kulkarni}, {Kumar},
  {Lee}, {Lenz}, {Littlefair}, {Ma}, {Macleod}, {Mastropietro}, {McCully},
  {Montagnac}, {Morris}, {Mueller}, {Mumford}, {Muna}, {Murphy}, {Nelson},
  {Nguyen}, {Ninan}, {N{\"o}the}, {Ogaz}, {Oh}, {Parejko}, {Parley}, {Pascual},
  {Patil}, {Patil}, {Plunkett}, {Prochaska}, {Rastogi}, {Reddy Janga},
  {Sabater}, {Sakurikar}, {Seifert}, {Sherbert}, {Sherwood-Taylor}, {Shih},
  {Sick}, {Silbiger}, {Singanamalla}, {Singer}, {Sladen}, {Sooley},
  {Sornarajah}, {Streicher}, {Teuben}, {Thomas}, {Tremblay}, {Turner},
  {Terr{\'o}n}, {van Kerkwijk}, {de la Vega}, {Watkins}, {Weaver}, {Whitmore},
  {Woillez}, {Zabalza}, \& {Astropy Contributors}}]{astropy2018}
{Astropy Collaboration}, {Price-Whelan}, A.~M., {Sip{\H{o}}cz}, B.~M., {et~al.}
  2018, \aj, 156, 123

\bibitem[{{Astropy Collaboration} {et~al.}(2013){Astropy Collaboration},
  {Robitaille}, {Tollerud}, {Greenfield}, {Droettboom}, {Bray}, {Aldcroft},
  {Davis}, {Ginsburg}, {Price-Whelan}, {Kerzendorf}, {Conley}, {Crighton},
  {Barbary}, {Muna}, {Ferguson}, {Grollier}, {Parikh}, {Nair}, {Unther},
  {Deil}, {Woillez}, {Conseil}, {Kramer}, {Turner}, {Singer}, {Fox}, {Weaver},
  {Zabalza}, {Edwards}, {Azalee Bostroem}, {Burke}, {Casey}, {Crawford},
  {Dencheva}, {Ely}, {Jenness}, {Labrie}, {Lim}, {Pierfederici}, {Pontzen},
  {Ptak}, {Refsdal}, {Servillat}, \& {Streicher}}]{astropy2013}
{Astropy Collaboration}, {Robitaille}, T.~P., {Tollerud}, E.~J., {et~al.} 2013,
  \aap, 558, A33

\bibitem[{{Baba} {et~al.}(2018){Baba}, {Kawata}, {Matsunaga}, {Grand }, \&
  {Hunt}}]{Baba18}
{Baba}, J., {Kawata}, D., {Matsunaga}, N., {Grand }, R. J.~J., \& {Hunt}, J.
  A.~S. 2018, \apjl, 853, L23

\bibitem[{{Barnes} {et~al.}(2018){Barnes}, {Hernandez}, {Muller}, \&
  {Pitts}}]{barnes2018}
{Barnes}, P.~J., {Hernandez}, A.~K., {Muller}, E., \& {Pitts}, R.~L. 2018,
  \apj, 866, 19

\bibitem[{{Benjamin} {et~al.}(2005){Benjamin}, {Churchwell}, {Babler},
  {Indebetouw}, {Meade}, {Whitney}, {Watson}, {Wolfire}, {Wolff}, {Ignace},
  {Bania}, {Bracker}, {Clemens}, {Chomiuk}, {Cohen}, {Dickey}, {Jackson},
  {Kobulnicky}, {Mercer}, {Mathis}, {Stolovy}, \& {Uzpen}}]{Benjamin2005}
{Benjamin}, R.~A., {Churchwell}, E., {Babler}, B.~L., {et~al.} 2005, \apjl,
  630, L149

\bibitem[{{Bolatto} {et~al.}(2013){Bolatto}, {Wolfire}, \&
  {Leroy}}]{bolatto2013}
{Bolatto}, A.~D., {Wolfire}, M., \& {Leroy}, A.~K. 2013, \araa, 51, 207

\bibitem[{{Bolatto} {et~al.}(2017){Bolatto}, {Wong}, {Utomo}, {Blitz}, {Vogel},
  {S{\'a}nchez}, {Barrera-Ballesteros}, {Cao}, {Colombo}, {Dannerbauer},
  {Garc{\'\i}a-Benito}, {Herrera-Camus}, {Husemann}, {Kalinova}, {Leroy},
  {Leung}, {Levy}, {Mast}, {Ostriker}, {Rosolowsky}, {Sandstrom}, {Teuben},
  {van de Ven}, \& {Walter}}]{bolatto2017}
{Bolatto}, A.~D., {Wong}, T., {Utomo}, D., {et~al.} 2017, \apj, 846, 159

\bibitem[{{Braine} {et~al.}(2020){Braine}, {Hughes}, {Rosolowsky}, {Gratier},
  {Colombo}, {Meidt}, \& {Schinnerer}}]{braine2020}
{Braine}, J., {Hughes}, A., {Rosolowsky}, E., {et~al.} 2020, \aap, 633, A17

\bibitem[{{Castro-Ginard} {et~al.}(2021){Castro-Ginard}, {McMillan}, {Luri},
  {Jordi}, {Romero-G{\'o}mez}, {Cantat-Gaudin}, {Casamiquela}, {Tarricq},
  {Soubiran}, \& {Anders}}]{castro-ginard2021}
{Castro-Ginard}, A., {McMillan}, P.~J., {Luri}, X., {et~al.} 2021, \aap, 652,
  A162

\bibitem[{{Chen} {et~al.}(2019){Chen}, {Wang}, {Deng}, {de Grijs}, {Liu}, \&
  {Tian}}]{chen2019}
{Chen}, X., {Wang}, S., {Deng}, L., {et~al.} 2019, Nature Astronomy, 3, 320

\bibitem[{{Churchwell} {et~al.}(2009){Churchwell}, {Babler}, {Meade},
  {Whitney}, {Benjamin}, {Indebetouw}, {Cyganowski}, {Robitaille}, {Povich},
  {Watson}, \& {Bracker}}]{Churchwell2009}
{Churchwell}, E., {Babler}, B.~L., {Meade}, M.~R., {et~al.} 2009, \pasp, 121,
  213

\bibitem[{{Clark} \& {Porter}(2004)}]{clark_porter2004}
{Clark}, J.~S. \& {Porter}, J.~M. 2004, \aap, 427, 839

\bibitem[{{Colombo} {et~al.}(2014{\natexlab{a}}){Colombo}, {Hughes},
  {Schinnerer}, {Meidt}, {Leroy}, {Pety}, {Dobbs}, {Garc{\'{\i}}a-Burillo},
  {Dumas}, {Thompson}, {Schuster}, \& {Kramer}}]{colombo14a}
{Colombo}, D., {Hughes}, A., {Schinnerer}, E., {et~al.} 2014{\natexlab{a}},
  \apj, 784, 3

\bibitem[{{Colombo} {et~al.}(2014{\natexlab{b}}){Colombo}, {Meidt},
  {Schinnerer}, {Garc{\'\i}a-Burillo}, {Hughes}, {Pety}, {Leroy}, {Dobbs},
  {Dumas}, {Thompson}, {Schuster}, \& {Kramer}}]{colombo2014b}
{Colombo}, D., {Meidt}, S.~E., {Schinnerer}, E., {et~al.} 2014{\natexlab{b}},
  \apj, 784, 4

\bibitem[{{Colombo} {et~al.}(2019){Colombo}, {Rosolowsky}, {Duarte-Cabral},
  {Ginsburg}, {Glenn}, {Zetterlund}, {Hernand ez}, {Dempsey}, \&
  {Currie}}]{colombo2019}
{Colombo}, D., {Rosolowsky}, E., {Duarte-Cabral}, A., {et~al.} 2019, \mnras,
  483, 4291

\bibitem[{{Colombo} {et~al.}(2015){Colombo}, {Rosolowsky}, {Ginsburg},
  {Duarte-Cabral}, \& {Hughes}}]{colombo15}
{Colombo}, D., {Rosolowsky}, E., {Ginsburg}, A., {Duarte-Cabral}, A., \&
  {Hughes}, A. 2015, \mnras, 454, 2067

\bibitem[{{Cordes}(2004)}]{cordes2004}
{Cordes}, J.~M. 2004, in Astronomical Society of the Pacific Conference Series,
  Vol. 317, Milky Way Surveys: The Structure and Evolution of our Galaxy, ed.
  D.~{Clemens}, R.~{Shah}, \& T.~{Brainerd}, 211

\bibitem[{{Currie} {et~al.}(2014){Currie}, {Berry}, {Jenness}, {Gibb}, {Bell},
  \& {Draper}}]{currie2014}
{Currie}, M.~J., {Berry}, D.~S., {Jenness}, T., {et~al.} 2014, in Astronomical
  Society of the Pacific Conference Series, Vol. 485, Astronomical Data
  Analysis Software and Systems XXIII, ed. N.~{Manset} \& P.~{Forshay}, 391

\bibitem[{{Dame} {et~al.}(1986){Dame}, {Elmegreen}, {Cohen}, \&
  {Thaddeus}}]{dame86}
{Dame}, T.~M., {Elmegreen}, B.~G., {Cohen}, R.~S., \& {Thaddeus}, P. 1986,
  \apj, 305, 892

\bibitem[{{Dame} {et~al.}(2001){Dame}, {Hartmann}, \& {Thaddeus}}]{dame01}
{Dame}, T.~M., {Hartmann}, D., \& {Thaddeus}, P. 2001, \apj, 547, 792

\bibitem[{{Dame} \& {Thaddeus}(2011)}]{dame-thaddeus2011}
{Dame}, T.~M. \& {Thaddeus}, P. 2011, \apjl, 734, L24

\bibitem[{{de Vaucouleurs}(1959)}]{de_vaucouleurs1959}
{de Vaucouleurs}, G. 1959, Handbuch der Physik, 53, 275

\bibitem[{{Deng} {et~al.}(2012){Deng}, {Newberg}, {Liu}, {Carlin}, {Beers},
  {Chen}, {Chen}, {Christlieb}, {Grillmair}, {Guhathakurta}, {Han}, {Hou},
  {Lee}, {L{\'e}pine}, {Li}, {Liu}, {Pan}, {Sellwood}, {Wang}, {Wang}, {Yang},
  {Yanny}, {Zhang}, {Zhang}, {Zheng}, \& {Zhu}}]{deng2012}
{Deng}, L.-C., {Newberg}, H.~J., {Liu}, C., {et~al.} 2012, Research in
  Astronomy and Astrophysics, 12, 735

\bibitem[{{Dobbs} \& {Baba}(2014)}]{dobbs_baba2014}
{Dobbs}, C. \& {Baba}, J. 2014, \pasa, 31, e035

\bibitem[{{Dobbs}(2008)}]{dobbs08}
{Dobbs}, C.~L. 2008, \mnras, 391, 844

\bibitem[{{Dobbs} {et~al.}(2006){Dobbs}, {Bonnell}, \&
  {Pringle}}]{dobbs_bonnell2006}
{Dobbs}, C.~L., {Bonnell}, I.~A., \& {Pringle}, J.~E. 2006, \mnras, 371, 1663

\bibitem[{{Dobbs} {et~al.}(2018){Dobbs}, {Pettitt}, {Corbelli}, \&
  {Pringle}}]{dobbs2018}
{Dobbs}, C.~L., {Pettitt}, A.~R., {Corbelli}, E., \& {Pringle}, J.~E. 2018,
  \mnras, 478, 3793

\bibitem[{{Dobbs} \& {Pringle}(2013)}]{dobbs_pringle2013}
{Dobbs}, C.~L. \& {Pringle}, J.~E. 2013, \mnras, 432, 653

\bibitem[{{Dobbs} {et~al.}(2015){Dobbs}, {Pringle}, \&
  {Duarte-Cabral}}]{dobbs_pdc2015}
{Dobbs}, C.~L., {Pringle}, J.~E., \& {Duarte-Cabral}, A. 2015, \mnras, 446,
  3608

\bibitem[{{Donovan Meyer} {et~al.}(2013){Donovan Meyer}, {Koda}, {Momose},
  {Mooney}, {Egusa}, {Carty}, {Kennicutt}, {Kuno}, {Rebolledo}, {Sawada},
  {Scoville}, \& {Wong}}]{donovan_meyer2013}
{Donovan Meyer}, J., {Koda}, J., {Momose}, R., {et~al.} 2013, \apj, 772, 107

\bibitem[{{Drimmel}(2000)}]{drimmel2000}
{Drimmel}, R. 2000, \aap, 358, L13

\bibitem[{{Drimmel} \& {Spergel}(2001)}]{drimmel_spergel2001}
{Drimmel}, R. \& {Spergel}, D.~N. 2001, \apj, 556, 181

\bibitem[{{Druard} {et~al.}(2014){Druard}, {Braine}, {Schuster}, {Schneider},
  {Gratier}, {Bontemps}, {Boquien}, {Combes}, {Corbelli}, {Henkel}, {Herpin},
  {Kramer}, {van der Tak}, \& {van der Werf}}]{druard2014}
{Druard}, C., {Braine}, J., {Schuster}, K.~F., {et~al.} 2014, \aap, 567, A118

\bibitem[{{Duarte-Cabral} {et~al.}(2021){Duarte-Cabral}, {Colombo}, {Urquhart},
  {Ginsburg}, {Russeil}, {Schuller}, {Anderson}, {Barnes}, {Beltr{\'a}n},
  {Beuther}, {Bontemps}, {Bronfman}, {Csengeri}, {Dobbs}, {Eden}, {Giannetti},
  {Kauffmann}, {Mattern}, {Medina}, {Menten}, {Lee}, {Pettitt}, {Riener},
  {Rigby}, {Traficante}, {Veena}, {Wienen}, {Wyrowski}, {Agurto}, {Azagra},
  {Cesaroni}, {Finger}, {Gonzalez}, {Henning}, {Hernandez}, {Kainulainen},
  {Leurini}, {Lopez}, {Mac-Auliffe}, {Mazumdar}, {Molinari}, {Motte}, {Muller},
  {Nguyen-Luong}, {Parra}, {Perez-Beaupuits}, {Montenegro-Montes}, {Moore},
  {Ragan}, {S{\'a}nchez-Monge}, {Sanna}, {Schilke}, {Schisano}, {Schneider},
  {Suri}, {Testi}, {Torstensson}, {Venegas}, {Wang}, \&
  {Zavagno}}]{duarte-cabral2021}
{Duarte-Cabral}, A., {Colombo}, D., {Urquhart}, J.~S., {et~al.} 2021, \mnras,
  500, 3027

\bibitem[{{Duarte-Cabral} \& {Dobbs}(2016)}]{duarte-cabral2016}
{Duarte-Cabral}, A. \& {Dobbs}, C.~L. 2016, \mnras, 458, 3667

\bibitem[{{Duarte-Cabral} \& {Dobbs}(2017)}]{duarte-cabral2017}
{Duarte-Cabral}, A. \& {Dobbs}, C.~L. 2017, \mnras, 470, 4261

\bibitem[{{Eden} {et~al.}(2020){Eden}, {Moore}, {Currie}, {Rigby},
  {Rosolowsky}, {Su}, {Kim}, {Parsons}, {Morata}, {Chen}, {Minamidani}, {Park},
  {Ragan}, {Urquhart}, {Rani}, {Tahani}, {Billington}, {Deb}, {Figura},
  {Fujiyoshi}, {Joncas}, {Liao}, {Liu}, {Ma}, {Tuan-Anh}, {Yun}, {Zhang},
  {Zhu}, {Henshaw}, {Longmore}, {Kobayashi}, {Thompson}, {Ao},
  {Campbell-White}, {Ching}, {Chung}, {Duarte-Cabral}, {Fich}, {Gao}, {Graves},
  {Jiang}, {Kemper}, {Kuan}, {Kwon}, {Lee}, {Lee}, {Liu}, {Pe{\~n}aloza},
  {Peretto}, {Phuong}, {Pineda}, {Plume}, {Puspitaningrum}, {Samal}, {Soam},
  {Sun}, {Tang}, {Traficante}, {White}, {Yan}, {Yang}, {Yuan}, {Yue}, {Bemis},
  {Brunt}, {Chen}, {Cho}, {Clark}, {Cyganowski}, {Friberg}, {Fuller}, {Han},
  {Hoare}, {Izumi}, {Kim}, {Kim}, {Kim}, {Koch}, {Kuno}, {Lacialle}, {Lai},
  {Lee}, {Lee}, {Li}, {Liu}, {Mairs}, {Pan}, {Qian}, {Scicluna}, {Shi}, {Shi},
  {Srinivasan}, {Tan}, {Thomas}, {Torii}, {Trejo}, {Umemoto}, {Violino},
  {Wallstr{\"o}m}, {Wang}, {Wu}, {Yuan}, {Zhang}, {Zhang}, {Zhou}, \&
  {Zhou}}]{eden2020}
{Eden}, D.~J., {Moore}, T.~J.~T., {Currie}, M.~J., {et~al.} 2020, \mnras, 498,
  5936

\bibitem[{{Eden} {et~al.}(2013){Eden}, {Moore}, {Morgan}, {Thompson}, \&
  {Urquhart}}]{eden2013}
{Eden}, D.~J., {Moore}, T.~J.~T., {Morgan}, L.~K., {Thompson}, M.~A., \&
  {Urquhart}, J.~S. 2013, \mnras, 431, 1587

\bibitem[{{Eden} {et~al.}(2015){Eden}, {Moore}, {Urquhart}, {Elia}, {Plume},
  {Rigby}, \& {Thompson}}]{eden2015}
{Eden}, D.~J., {Moore}, T.~J.~T., {Urquhart}, J.~S., {et~al.} 2015, \mnras,
  452, 289

\bibitem[{{Egusa} {et~al.}(2018){Egusa}, {Hirota}, {Baba}, \&
  {Muraoka}}]{egusa2018}
{Egusa}, F., {Hirota}, A., {Baba}, J., \& {Muraoka}, K. 2018, \apj, 854, 90

\bibitem[{{Elmegreen}(1990)}]{elmegreen1990}
{Elmegreen}, B.~G. 1990, Annals of the New York Academy of Sciences, 596, 40

\bibitem[{{Elmegreen} {et~al.}(2017){Elmegreen}, {Elmegreen}, {Kaufman},
  {Brinks}, {Struck}, {Bournaud}, {Sheth}, \& {Juneau}}]{elmegreen2017}
{Elmegreen}, D.~M., {Elmegreen}, B.~G., {Kaufman}, M., {et~al.} 2017, \apj,
  841, 43

\bibitem[{{Elmegreen} {et~al.}(2011){Elmegreen}, {Elmegreen}, {Yau},
  {Athanassoula}, {Bosma}, {Buta}, {Helou}, {Ho}, {Gadotti}, {Knapen},
  {Laurikainen}, {Madore}, {Masters}, {Meidt}, {Men{\'e}ndez-Delmestre},
  {Regan}, {Salo}, {Sheth}, {Zaritsky}, {Aravena}, {Skibba}, {Hinz}, {Laine},
  {Gil de Paz}, {Mu{\~n}oz-Mateos}, {Seibert}, {Mizusawa}, {Kim}, \& {Erroz
  Ferrer}}]{elmegreen2011}
{Elmegreen}, D.~M., {Elmegreen}, B.~G., {Yau}, A., {et~al.} 2011, \apj, 737, 32

\bibitem[{{Fujimoto} {et~al.}(2014){Fujimoto}, {Tasker}, \&
  {Habe}}]{fujimoto2014}
{Fujimoto}, Y., {Tasker}, E.~J., \& {Habe}, A. 2014, \mnras, 445, L65

\bibitem[{{Gaia Collaboration} {et~al.}(2021){Gaia Collaboration}, {Brown},
  {Vallenari}, {Prusti}, {de Bruijne}, {Babusiaux}, {Biermann}, {Creevey},
  {Evans}, {Eyer}, {Hutton}, {Jansen}, {Jordi}, {Klioner}, {Lammers},
  {Lindegren}, {Luri}, {Mignard}, {Panem}, {Pourbaix}, {Randich}, {Sartoretti},
  {Soubiran}, {Walton}, {Arenou}, {Bailer-Jones}, {Bastian}, {Cropper},
  {Drimmel}, {Katz}, {Lattanzi}, {van Leeuwen}, {Bakker}, {Cacciari},
  {Casta{\~n}eda}, {De Angeli}, {Ducourant}, {Fabricius}, {Fouesneau},
  {Fr{\'e}mat}, {Guerra}, {Guerrier}, {Guiraud}, {Jean-Antoine Piccolo},
  {Masana}, {Messineo}, {Mowlavi}, {Nicolas}, {Nienartowicz}, {Pailler},
  {Panuzzo}, {Riclet}, {Roux}, {Seabroke}, {Sordo}, {Tanga}, {Th{\'e}venin},
  {Gracia-Abril}, {Portell}, {Teyssier}, {Altmann}, {Andrae}, {Bellas-Velidis},
  {Benson}, {Berthier}, {Blomme}, {Brugaletta}, {Burgess}, {Busso}, {Carry},
  {Cellino}, {Cheek}, {Clementini}, {Damerdji}, {Davidson}, {Delchambre},
  {Dell'Oro}, {Fern{\'a}ndez-Hern{\'a}ndez}, {Galluccio}, {Garc{\'\i}a-Lario},
  {Garcia-Reinaldos}, {Gonz{\'a}lez-N{\'u}{\~n}ez}, {Gosset}, {Haigron},
  {Halbwachs}, {Hambly}, {Harrison}, {Hatzidimitriou}, {Heiter},
  {Hern{\'a}ndez}, {Hestroffer}, {Hodgkin}, {Holl}, {Jan{\ss}en}, {Jevardat de
  Fombelle}, {Jordan}, {Krone-Martins}, {Lanzafame}, {L{\"o}ffler}, {Lorca},
  {Manteiga}, {Marchal}, {Marrese}, {Moitinho}, {Mora}, {Muinonen}, {Osborne},
  {Pancino}, {Pauwels}, {Petit}, {Recio-Blanco}, {Richards}, {Riello},
  {Rimoldini}, {Robin}, {Roegiers}, {Rybizki}, {Sarro}, {Siopis}, {Smith},
  {Sozzetti}, {Ulla}, {Utrilla}, {van Leeuwen}, {van Reeven}, {Abbas}, {Abreu
  Aramburu}, {Accart}, {Aerts}, {Aguado}, {Ajaj}, {Altavilla}, {{\'A}lvarez},
  {{\'A}lvarez Cid-Fuentes}, {Alves}, {Anderson}, {Anglada Varela}, {Antoja},
  {Audard}, {Baines}, {Baker}, {Balaguer-N{\'u}{\~n}ez}, {Balbinot}, {Balog},
  {Barache}, {Barbato}, {Barros}, {Barstow}, {Bartolom{\'e}}, {Bassilana},
  {Bauchet}, {Baudesson-Stella}, {Becciani}, {Bellazzini}, {Bernet}, {Bertone},
  {Bianchi}, {Blanco-Cuaresma}, {Boch}, {Bombrun}, {Bossini}, {Bouquillon},
  {Bragaglia}, {Bramante}, {Breedt}, {Bressan}, {Brouillet}, {Bucciarelli},
  {Burlacu}, {Busonero}, {Butkevich}, {Buzzi}, {Caffau}, {Cancelliere},
  {C{\'a}novas}, {Cantat-Gaudin}, {Carballo}, {Carlucci}, {Carnerero},
  {Carrasco}, {Casamiquela}, {Castellani}, {Castro-Ginard}, {Castro Sampol},
  {Chaoul}, {Charlot}, {Chemin}, {Chiavassa}, {Cioni}, {Comoretto}, {Cooper},
  {Cornez}, {Cowell}, {Crifo}, {Crosta}, {Crowley}, {Dafonte}, {Dapergolas},
  {David}, {David}, {de Laverny}, {De Luise}, {De March}, {De Ridder}, {de
  Souza}, {de Teodoro}, {de Torres}, {del Peloso}, {del Pozo}, {Delbo},
  {Delgado}, {Delgado}, {Delisle}, {Di Matteo}, {Diakite}, {Diener},
  {Distefano}, {Dolding}, {Eappachen}, {Edvardsson}, {Enke}, {Esquej}, {Fabre},
  {Fabrizio}, {Faigler}, {Fedorets}, {Fernique}, {Fienga}, {Figueras},
  {Fouron}, {Fragkoudi}, {Fraile}, {Franke}, {Gai}, {Garabato},
  {Garcia-Gutierrez}, {Garc{\'\i}a-Torres}, {Garofalo}, {Gavras}, {Gerlach},
  {Geyer}, {Giacobbe}, {Gilmore}, {Girona}, {Giuffrida}, {Gomel}, {Gomez},
  {Gonzalez-Santamaria}, {Gonz{\'a}lez-Vidal}, {Granvik},
  {Guti{\'e}rrez-S{\'a}nchez}, {Guy}, {Hauser}, {Haywood}, {Helmi}, {Hidalgo},
  {Hilger}, {H{\l}adczuk}, {Hobbs}, {Holland}, {Huckle}, {Jasniewicz},
  {Jonker}, {Juaristi Campillo}, {Julbe}, {Karbevska}, {Kervella}, {Khanna},
  {Kochoska}, {Kontizas}, {Kordopatis}, {Korn}, {Kostrzewa-Rutkowska},
  {Kruszy{\'n}ska}, {Lambert}, {Lanza}, {Lasne}, {Le Campion}, {Le Fustec},
  {Lebreton}, {Lebzelter}, {Leccia}, {Leclerc}, {Lecoeur-Taibi}, {Liao},
  {Licata}, {Lindstr{\o}m}, {Lister}, {Livanou}, {Lobel}, {Madrero Pardo},
  {Managau}, {Mann}, {Marchant}, {Marconi}, {Marcos Santos}, {Marinoni},
  {Marocco}, {Marshall}, {Martin Polo}, {Mart{\'\i}n-Fleitas}, {Masip},
  {Massari}, {Mastrobuono-Battisti}, {Mazeh}, {McMillan}, {Messina},
  {Michalik}, {Millar}, {Mints}, {Molina}, {Molinaro}, {Moln{\'a}r},
  {Montegriffo}, {Mor}, {Morbidelli}, {Morel}, {Morris}, {Mulone}, {Munoz},
  {Muraveva}, {Murphy}, {Musella}, {Noval}, {Ord{\'e}novic}, {Orr{\`u}},
  {Osinde}, {Pagani}, {Pagano}, {Palaversa}, {Palicio}, {Panahi}, {Pawlak},
  {Pe{\~n}alosa Esteller}, {Penttil{\"a}}, {Piersimoni}, {Pineau}, {Plachy},
  {Plum}, {Poggio}, {Poretti}, {Poujoulet}, {Pr{\v{s}}a}, {Pulone}, {Racero},
  {Ragaini}, {Rainer}, {Raiteri}, {Rambaux}, {Ramos}, {Ramos-Lerate}, {Re
  Fiorentin}, {Regibo}, {Reyl{\'e}}, {Ripepi}, {Riva}, {Rixon}, {Robichon},
  {Robin}, {Roelens}, {Rohrbasser}, {Romero-G{\'o}mez}, {Rowell}, {Royer},
  {Rybicki}, {Sadowski}, {Sagrist{\`a} Sell{\'e}s}, {Sahlmann}, {Salgado},
  {Salguero}, {Samaras}, {Sanchez Gimenez}, {Sanna}, {Santove{\~n}a},
  {Sarasso}, {Schultheis}, {Sciacca}, {Segol}, {Segovia}, {S{\'e}gransan},
  {Semeux}, {Shahaf}, {Siddiqui}, {Siebert}, {Siltala}, {Slezak}, {Smart},
  {Solano}, {Solitro}, {Souami}, {Souchay}, {Spagna}, {Spoto}, {Steele},
  {Steidelm{\"u}ller}, {Stephenson}, {S{\"u}veges}, {Szabados}, {Szegedi-Elek},
  {Taris}, {Tauran}, {Taylor}, {Teixeira}, {Thuillot}, {Tonello}, {Torra},
  {Torra}, {Turon}, {Unger}, {Vaillant}, {van Dillen}, {Vanel}, {Vecchiato},
  {Viala}, {Vicente}, {Voutsinas}, {Weiler}, {Wevers}, {Wyrzykowski}, {Yoldas},
  {Yvard}, {Zhao}, {Zorec}, {Zucker}, {Zurbach}, \& {Zwitter}}]{gaia2021}
{Gaia Collaboration}, {Brown}, A.~G.~A., {Vallenari}, A., {et~al.} 2021, \aap,
  649, A1

\bibitem[{{Gallagher} {et~al.}(2018){Gallagher}, {Leroy}, {Bigiel}, {Cormier},
  {Jim{\'e}nez-Donaire}, {Ostriker}, {Usero}, {Bolatto}, {Garc{\'\i}a-Burillo},
  {Hughes}, {Kepley}, {Krumholz}, {Meidt}, {Meier}, {Murphy}, {Pety},
  {Rosolowsky}, {Schinnerer}, {Schruba}, \& {Walter}}]{gallagher2018}
{Gallagher}, M.~J., {Leroy}, A.~K., {Bigiel}, F., {et~al.} 2018, \apj, 858, 90

\bibitem[{{Georgelin} \& {Georgelin}(1976)}]{georgelin_georgelin1976}
{Georgelin}, Y.~M. \& {Georgelin}, Y.~P. 1976, \aap, 49, 57

\bibitem[{{Goodman} {et~al.}(2014){Goodman}, {Alves}, {Beaumont}, {Benjamin},
  {Borkin}, {Burkert}, {Dame}, {Jackson}, {Kauffmann}, {Robitaille}, \&
  {Smith}}]{goodman2014}
{Goodman}, A.~A., {Alves}, J., {Beaumont}, C.~N., {et~al.} 2014, \apj, 797, 53

\bibitem[{{Grand} {et~al.}(2015){Grand}, {Bovy}, {Kawata}, {Hunt}, {Famaey},
  {Siebert}, {Monari}, \& {Cropper}}]{Grand15}
{Grand}, R.~J.~J., {Bovy}, J., {Kawata}, D., {et~al.} 2015, \mnras, 453, 1867

\bibitem[{{Gratier} {et~al.}(2012){Gratier}, {Braine}, {Rodriguez-Fernandez},
  {Schuster}, {Kramer}, {Corbelli}, {Combes}, {Brouillet}, {van der Werf}, \&
  {R{\"o}llig}}]{gratier12}
{Gratier}, P., {Braine}, J., {Rodriguez-Fernandez}, N.~J., {et~al.} 2012, \aap,
  542, A108

\bibitem[{{Grosb{\o}l} \& {Carraro}(2018)}]{grosbol_carraro2018}
{Grosb{\o}l}, P. \& {Carraro}, G. 2018, \aap, 619, A50

\bibitem[{{G{\"u}sten} {et~al.}(2006){G{\"u}sten}, {Nyman}, {Schilke},
  {Menten}, {Cesarsky}, \& {Booth}}]{guesten2006}
{G{\"u}sten}, R., {Nyman}, L.~{\r{A}}., {Schilke}, P., {et~al.} 2006, \aap,
  454, L13

\bibitem[{{Helfer} {et~al.}(2003){Helfer}, {Thornley}, {Regan}, {Wong},
  {Sheth}, {Vogel}, {Blitz}, \& {Bock}}]{helfer03}
{Helfer}, T.~T., {Thornley}, M.~D., {Regan}, M.~W., {et~al.} 2003, \apjs, 145,
  259

\bibitem[{{Hou} \& {Han}(2014{\natexlab{a}})}]{hou-han2014}
{Hou}, L.~G. \& {Han}, J.~L. 2014{\natexlab{a}}, \aap, 569, A125

\bibitem[{{Hou} \& {Han}(2014{\natexlab{b}})}]{hou_han2014}
{Hou}, L.~G. \& {Han}, J.~L. 2014{\natexlab{b}}, \aap, 569, A125

\bibitem[{{Hubble}(1926)}]{hubble1926}
{Hubble}, E.~P. 1926, \apj, 64, 321

\bibitem[{{Hughes} {et~al.}(2013){Hughes}, {Meidt}, {Schinnerer}, {Colombo},
  {Pety}, {Leroy}, {Dobbs}, {Garc{\'\i}a-Burillo}, {Thompson}, {Dumas},
  {Schuster}, \& {Kramer}}]{hughes2013a}
{Hughes}, A., {Meidt}, S.~E., {Schinnerer}, E., {et~al.} 2013, \apj, 779, 44

\bibitem[{Hunter(2007)}]{matplotlib2007}
Hunter, J.~D. 2007, Computing in Science \& Engineering, 9, 90

\bibitem[{{Inutsuka} {et~al.}(2015){Inutsuka}, {Inoue}, {Iwasaki}, \&
  {Hosokawa}}]{inutsuka2015}
{Inutsuka}, S.-i., {Inoue}, T., {Iwasaki}, K., \& {Hosokawa}, T. 2015, \aap,
  580, A49

\bibitem[{{Jackson} {et~al.}(2010){Jackson}, {Finn}, {Chambers}, {Rathborne},
  \& {Simon}}]{jackson2010}
{Jackson}, J.~M., {Finn}, S.~C., {Chambers}, E.~T., {Rathborne}, J.~M., \&
  {Simon}, R. 2010, \apjl, 719, L185

\bibitem[{{Jackson} {et~al.}(2006){Jackson}, {Rathborne}, {Shah}, {Simon},
  {Bania}, {Clemens}, {Chambers}, {Johnson}, {Dormody}, {Lavoie}, \&
  {Heyer}}]{jackson2006}
{Jackson}, J.~M., {Rathborne}, J.~M., {Shah}, R.~Y., {et~al.} 2006, \apjs, 163,
  145

\bibitem[{{Koda} {et~al.}(2009){Koda}, {Scoville}, {Sawada}, {La Vigne},
  {Vogel}, {Potts}, {Carpenter}, {Corder}, {Wright}, {White}, {Zauderer},
  {Patience}, {Sargent}, {Bock}, {Hawkins}, {Hodges}, {Kemball}, {Lamb},
  {Plambeck}, {Pound}, {Scott}, {Teuben}, \& {Woody}}]{koda2009}
{Koda}, J., {Scoville}, N., {Sawada}, T., {et~al.} 2009, \apjl, 700, L132

\bibitem[{{Koo} {et~al.}(2017){Koo}, {Park}, {Kim}, {Lee}, {Balser}, \&
  {Wenger}}]{koo2017}
{Koo}, B.-C., {Park}, G., {Kim}, W.-T., {et~al.} 2017, \pasp, 129, 094102

\bibitem[{{Kunder} {et~al.}(2017){Kunder}, {Kordopatis}, {Steinmetz},
  {Zwitter}, {McMillan}, {Casagrande}, {Enke}, {Wojno}, {Valentini},
  {Chiappini}, {Matijevi{\v{c}}}, {Siviero}, {de Laverny}, {Recio-Blanco},
  {Bijaoui}, {Wyse}, {Binney}, {Grebel}, {Helmi}, {Jofre}, {Antoja}, {Gilmore},
  {Siebert}, {Famaey}, {Bienaym{\'e}}, {Gibson}, {Freeman}, {Navarro},
  {Munari}, {Seabroke}, {Anguiano}, {{\v{Z}}erjal}, {Minchev}, {Reid},
  {Bland-Hawthorn}, {Kos}, {Sharma}, {Watson}, {Parker}, {Scholz}, {Burton},
  {Cass}, {Hartley}, {Fiegert}, {Stupar}, {Ritter}, {Hawkins}, {Gerhard},
  {Chaplin}, {Davies}, {Elsworth}, {Lund}, {Miglio}, \& {Mosser}}]{kunder2017}
{Kunder}, A., {Kordopatis}, G., {Steinmetz}, M., {et~al.} 2017, \aj, 153, 75

\bibitem[{{Leroy} {et~al.}(2017){Leroy}, {Schinnerer}, {Hughes}, {Kruijssen},
  {Meidt}, {Schruba}, {Sun}, {Bigiel}, {Aniano}, {Blanc}, {Bolatto},
  {Chevance}, {Colombo}, {Gallagher}, {Garcia-Burillo}, {Kramer}, {Querejeta},
  {Pety}, {Thompson}, \& {Usero}}]{leroy2017}
{Leroy}, A.~K., {Schinnerer}, E., {Hughes}, A., {et~al.} 2017, \apj, 846, 71

\bibitem[{{Leroy} {et~al.}(2008){Leroy}, {Walter}, {Brinks}, {Bigiel}, {de
  Blok}, {Madore}, \& {Thornley}}]{leroy2008}
{Leroy}, A.~K., {Walter}, F., {Brinks}, E., {et~al.} 2008, \aj, 136, 2782

\bibitem[{{Levine} {et~al.}(2006){Levine}, {Blitz}, \& {Heiles}}]{levine2006}
{Levine}, E.~S., {Blitz}, L., \& {Heiles}, C. 2006, Science, 312, 1773

\bibitem[{{Li} {et~al.}(2016){Li}, {Gerhard}, {Shen}, {Portail}, \&
  {Wegg}}]{2016ApJ...824...13L}
{Li}, Z., {Gerhard}, O., {Shen}, J., {Portail}, M., \& {Wegg}, C. 2016, \apj,
  824, 13

\bibitem[{{Majewski} {et~al.}(2017){Majewski}, {Schiavon}, {Frinchaboy},
  {Allende Prieto}, {Barkhouser}, {Bizyaev}, {Blank}, {Brunner}, {Burton},
  {Carrera}, {Chojnowski}, {Cunha}, {Epstein}, {Fitzgerald}, {Garc{\'\i}a
  P{\'e}rez}, {Hearty}, {Henderson}, {Holtzman}, {Johnson}, {Lam}, {Lawler},
  {Maseman}, {M{\'e}sz{\'a}ros}, {Nelson}, {Nguyen}, {Nidever}, {Pinsonneault},
  {Shetrone}, {Smee}, {Smith}, {Stolberg}, {Skrutskie}, {Walker}, {Wilson},
  {Zasowski}, {Anders}, {Basu}, {Beland}, {Blanton}, {Bovy}, {Brownstein},
  {Carlberg}, {Chaplin}, {Chiappini}, {Eisenstein}, {Elsworth}, {Feuillet},
  {Fleming}, {Galbraith-Frew}, {Garc{\'\i}a}, {Garc{\'\i}a-Hern{\'a}ndez},
  {Gillespie}, {Girardi}, {Gunn}, {Hasselquist}, {Hayden}, {Hekker}, {Ivans},
  {Kinemuchi}, {Klaene}, {Mahadevan}, {Mathur}, {Mosser}, {Muna}, {Munn},
  {Nichol}, {O'Connell}, {Parejko}, {Robin}, {Rocha-Pinto}, {Schultheis},
  {Serenelli}, {Shane}, {Silva Aguirre}, {Sobeck}, {Thompson}, {Troup},
  {Weinberg}, \& {Zamora}}]{majewski2017}
{Majewski}, S.~R., {Schiavon}, R.~P., {Frinchaboy}, P.~M., {et~al.} 2017, \aj,
  154, 94

\bibitem[{{Martinez-Medina} {et~al.}(2021){Martinez-Medina},
  {P{\'e}rez-Villegas}, \& {Peimbert}}]{martinez-medina2021}
{Martinez-Medina}, L., {P{\'e}rez-Villegas}, A., \& {Peimbert}, A. 2021, arXiv
  e-prints, arXiv:2109.04696

\bibitem[{{Martos} {et~al.}(2004){Martos}, {Hernandez}, {Y{\'a}{\~n}ez},
  {Moreno}, \& {Pichardo}}]{martos2004}
{Martos}, M., {Hernandez}, X., {Y{\'a}{\~n}ez}, M., {Moreno}, E., \&
  {Pichardo}, B. 2004, \mnras, 350, L47

\bibitem[{{Mattern} {et~al.}(2018){Mattern}, {Kainulainen}, {Zhang}, \&
  {Beuther}}]{mattern2018a}
{Mattern}, M., {Kainulainen}, J., {Zhang}, M., \& {Beuther}, H. 2018, \aap,
  616, A78

\bibitem[{{Meidt} {et~al.}(2013){Meidt}, {Schinnerer}, {Garc{\'{\i}}a-Burillo},
  {Hughes}, {Colombo}, {Pety}, {Dobbs}, {Schuster}, {Kramer}, {Leroy}, {Dumas},
  \& {Thompson}}]{meidt13}
{Meidt}, S.~E., {Schinnerer}, E., {Garc{\'{\i}}a-Burillo}, S., {et~al.} 2013,
  \apj, 779, 45

\bibitem[{{Mertsch} \& {Vittino}(2020)}]{mertsch-vittino2020}
{Mertsch}, P. \& {Vittino}, A. 2020, arXiv e-prints, arXiv:2012.15770

\bibitem[{{Miville-Desch{\^e}nes} {et~al.}(2017){Miville-Desch{\^e}nes},
  {Murray}, \& {Lee}}]{miville_deschenes2017}
{Miville-Desch{\^e}nes}, M.-A., {Murray}, N., \& {Lee}, E.~J. 2017, \apj, 834,
  57

\bibitem[{{Moore} {et~al.}(2012){Moore}, {Urquhart}, {Morgan}, \&
  {Thompson}}]{moore2012}
{Moore}, T.~J.~T., {Urquhart}, J.~S., {Morgan}, L.~K., \& {Thompson}, M.~A.
  2012, \mnras, 426, 701

\bibitem[{{Nguyen} {et~al.}(2018){Nguyen}, {Pettitt}, {Tasker}, \&
  {Okamoto}}]{nguyen2018}
{Nguyen}, N.~K., {Pettitt}, A.~R., {Tasker}, E.~J., \& {Okamoto}, T. 2018,
  \mnras, 475, 27

\bibitem[{{Pan} \& {Kuno}(2017)}]{pan2017}
{Pan}, H.-A. \& {Kuno}, N. 2017, \apj, 839, 133

\bibitem[{{Pettitt} {et~al.}(2015){Pettitt}, {Dobbs}, {Acreman}, \&
  {Bate}}]{Pettitt15}
{Pettitt}, A.~R., {Dobbs}, C.~L., {Acreman}, D.~M., \& {Bate}, M.~R. 2015,
  \mnras, 449, 3911

\bibitem[{{Pettitt} {et~al.}(2014){Pettitt}, {Dobbs}, {Acreman}, \&
  {Price}}]{pettitt2014}
{Pettitt}, A.~R., {Dobbs}, C.~L., {Acreman}, D.~M., \& {Price}, D.~J. 2014,
  \mnras, 444, 919

\bibitem[{{Pettitt} {et~al.}(2020){Pettitt}, {Dobbs}, {Baba}, {Colombo},
  {Duarte-Cabral}, {Egusa}, \& {Habe}}]{pettitt2020}
{Pettitt}, A.~R., {Dobbs}, C.~L., {Baba}, J., {et~al.} 2020, \mnras, 498, 1159

\bibitem[{{Pety} {et~al.}(2013){Pety}, {Schinnerer}, {Leroy}, {Hughes},
  {Meidt}, {Colombo}, {Dumas}, {Garc{\'{\i}}a-Burillo}, {Schuster}, {Kramer},
  {Dobbs}, \& {Thompson}}]{pety2013}
{Pety}, J., {Schinnerer}, E., {Leroy}, A.~K., {et~al.} 2013, \apj, 779, 43

\bibitem[{{Querejeta} {et~al.}(2021){Querejeta}, {Schinnerer}, {Meidt}, {Sun},
  {Leroy}, {Emsellem}, {Klessen}, {Munoz-Mateos}, {Salo}, {Laurikainen},
  {Beslic}, {Blanc}, {Chevance}, {Dale}, {Eibensteiner}, {Faesi},
  {Garcia-Rodriguez}, {Glover}, {Grasha}, {Henshaw}, {Herrera}, {Hughes},
  {Kreckel}, {Kruijssen}, {Liu}, {Murphy}, {Pan}, {Pety}, {Razza},
  {Rosolowsky}, {Saito}, {Schruba}, {Usero}, {Watkins}, \&
  {Williams}}]{querejeta2021}
{Querejeta}, M., {Schinnerer}, E., {Meidt}, S., {et~al.} 2021, arXiv e-prints,
  arXiv:2109.04491

\bibitem[{{Quillen}(2002)}]{quillen2002}
{Quillen}, A.~C. 2002, \aj, 124, 924

\bibitem[{{Quillen} {et~al.}(2018){Quillen}, {Carrillo}, {Anders}, {McMillan},
  {Hilmi}, {Monari}, {Minchev}, {Chiappini}, {Khalatyan}, \&
  {Steinmetz}}]{quillen2018}
{Quillen}, A.~C., {Carrillo}, I., {Anders}, F., {et~al.} 2018, \mnras, 480,
  3132

\bibitem[{{Ragan} {et~al.}(2014){Ragan}, {Henning}, {Tackenberg}, {Beuther},
  {Johnston}, {Kainulainen}, \& {Linz}}]{ragan2014}
{Ragan}, S.~E., {Henning}, T., {Tackenberg}, J., {et~al.} 2014, \aap, 568, A73

\bibitem[{{Ragan} {et~al.}(2018){Ragan}, {Moore}, {Eden}, {Hoare}, {Urquhart},
  {Elia}, \& {Molinari}}]{ragan2018}
{Ragan}, S.~E., {Moore}, T.~J.~T., {Eden}, D.~J., {et~al.} 2018, \mnras, 479,
  2361

\bibitem[{{Rahman} {et~al.}(2012){Rahman}, {Bolatto}, {Xue}, {Wong}, {Leroy},
  {Walter}, {Bigiel}, {Rosolowsky}, {Fisher}, {Vogel}, {Blitz}, {West}, \&
  {Ott}}]{rahman2012}
{Rahman}, N., {Bolatto}, A.~D., {Xue}, R., {et~al.} 2012, \apj, 745, 183

\bibitem[{{Ram{\'o}n-Fox} \& {Bonnell}(2018)}]{ramon-fox_bonnell2018}
{Ram{\'o}n-Fox}, F.~G. \& {Bonnell}, I.~A. 2018, \mnras, 474, 2028

\bibitem[{{Reid} {et~al.}(2019){Reid}, {Menten}, {Brunthaler}, {Zheng}, {Dame},
  {Xu}, {Li}, {Sakai}, {Wu}, {Immer}, {Zhang}, {Sanna}, {Moscadelli}, {Rygl},
  {Bartkiewicz}, {Hu}, {Quiroga-Nu{\~n}ez}, \& {van Langevelde}}]{reid2019}
{Reid}, M.~J., {Menten}, K.~M., {Brunthaler}, A., {et~al.} 2019, \apj, 885, 131

\bibitem[{{Reid} {et~al.}(2014){Reid}, {Menten}, {Brunthaler}, {Zheng}, {Dame},
  {Xu}, {Wu}, {Zhang}, {Sanna}, {Sato}, {Hachisuka}, {Choi}, {Immer},
  {Moscadelli}, {Rygl}, \& {Bartkiewicz}}]{reid2014}
{Reid}, M.~J., {Menten}, K.~M., {Brunthaler}, A., {et~al.} 2014, \apj, 783, 130

\bibitem[{{Rezaei Kh.} {et~al.}(2018){Rezaei Kh.}, {Bailer-Jones}, {Hogg}, \&
  {Schultheis}}]{rezaei_kh2018}
{Rezaei Kh.}, S., {Bailer-Jones}, C. A.~L., {Hogg}, D.~W., \& {Schultheis}, M.
  2018, \aap, 618, A168

\bibitem[{{Rice} {et~al.}(2016){Rice}, {Goodman}, {Bergin}, {Beaumont}, \&
  {Dame}}]{rice2016}
{Rice}, T.~S., {Goodman}, A.~A., {Bergin}, E.~A., {Beaumont}, C., \& {Dame},
  T.~M. 2016, \apj, 822, 52

\bibitem[{{Rigby} {et~al.}(2019){Rigby}, {Moore}, {Eden}, {Urquhart}, {Ragan},
  {Peretto}, {Plume}, {Thompson}, {Currie}, \& {Park}}]{rigby2019}
{Rigby}, A.~J., {Moore}, T.~J.~T., {Eden}, D.~J., {et~al.} 2019, \aap, 632, A58

\bibitem[{{Rigby} {et~al.}(2016){Rigby}, {Moore}, {Plume}, {Eden}, {Urquhart},
  {Thompson}, {Mottram}, {Brunt}, {Butner}, {Dempsey}, {Gibson}, {Hatchell},
  {Jenness}, {Kuno}, {Longmore}, {Morgan}, {Polychroni}, {Thomas}, {White}, \&
  {Zhu}}]{rigby2016}
{Rigby}, A.~J., {Moore}, T.~J.~T., {Plume}, R., {et~al.} 2016, \mnras, 456,
  2885

\bibitem[{{Roman-Duval} {et~al.}(2009){Roman-Duval}, {Jackson}, {Heyer},
  {Johnson}, {Rathborne}, {Shah}, \& {Simon}}]{roman_duval2009}
{Roman-Duval}, J., {Jackson}, J.~M., {Heyer}, M., {et~al.} 2009, \apj, 699,
  1153

\bibitem[{{Roman-Duval} {et~al.}(2010){Roman-Duval}, {Jackson}, {Heyer},
  {Rathborne}, \& {Simon}}]{roman_duval2010}
{Roman-Duval}, J., {Jackson}, J.~M., {Heyer}, M., {Rathborne}, J., \& {Simon},
  R. 2010, \apj, 723, 492

\bibitem[{{Romero-G{\'o}mez} {et~al.}(2019){Romero-G{\'o}mez}, {Mateu},
  {Aguilar}, {Figueras}, \& {Castro-Ginard}}]{romero-gomez2019}
{Romero-G{\'o}mez}, M., {Mateu}, C., {Aguilar}, L., {Figueras}, F., \&
  {Castro-Ginard}, A. 2019, \aap, 627, A150

\bibitem[{{Rosolowsky} {et~al.}(2021){Rosolowsky}, {Hughes}, {Leroy}, {Sun},
  {Querejeta}, {Schruba}, {Usero}, {Herrera}, {Liu}, {Pety}, {Saito},
  {Be{\v{s}}li{\'c}}, {Bigiel}, {Blanc}, {Chevance}, {Dale}, {Deger}, {Faesi},
  {Glover}, {Henshaw}, {Klessen}, {Kruijssen}, {Larson}, {Lee}, {Meidt}, {Mok},
  {Schinnerer}, {Thilker}, \& {Williams}}]{rosolowsky2021}
{Rosolowsky}, E., {Hughes}, A., {Leroy}, A.~K., {et~al.} 2021, \mnras, 502,
  1218

\bibitem[{{Rosolowsky} \& {Leroy}(2006)}]{rl06}
{Rosolowsky}, E. \& {Leroy}, A. 2006, \pasp, 118, 590

\bibitem[{{Rosolowsky} {et~al.}(2008){Rosolowsky}, {Pineda}, {Kauffmann}, \&
  {Goodman}}]{rosolowsky08}
{Rosolowsky}, E.~W., {Pineda}, J.~E., {Kauffmann}, J., \& {Goodman}, A.~A.
  2008, \apj, 679, 1338

\bibitem[{{Russeil}(2003)}]{russeil2003}
{Russeil}, D. 2003, \aap, 397, 133

\bibitem[{{Sawada} {et~al.}(2012){Sawada}, {Hasegawa}, \& {Koda}}]{sawada12}
{Sawada}, T., {Hasegawa}, T., \& {Koda}, J. 2012, \apjl, 759, L26

\bibitem[{{Schinnerer} {et~al.}(2017){Schinnerer}, {Meidt}, {Colombo},
  {Chandar}, {Dobbs}, {Garc{\'{\i}}a-Burillo}, {Hughes}, {Leroy}, {Pety},
  {Querejeta}, {Kramer}, \& {Schuster}}]{schinnerer2017}
{Schinnerer}, E., {Meidt}, S.~E., {Colombo}, D., {et~al.} 2017, \apj, 836, 62

\bibitem[{{Schinnerer} {et~al.}(2013){Schinnerer}, {Meidt}, {Pety}, {Hughes},
  {Colombo}, {Garc{\'\i}a-Burillo}, {Schuster}, {Dumas}, {Dobbs}, {Leroy},
  {Kramer}, {Thompson}, \& {Regan}}]{schinnerer2013}
{Schinnerer}, E., {Meidt}, S.~E., {Pety}, J., {et~al.} 2013, \apj, 779, 42

\bibitem[{{Schuller} {et~al.}(2017){Schuller}, {Csengeri}, {Urquhart},
  {Duarte-Cabral}, {Barnes}, {Giannetti}, {Hernandez}, {Leurini}, {Mattern},
  {Medina}, {Agurto}, {Azagra}, {Anderson}, {Beltr{\'a}n}, {Beuther},
  {Bontemps}, {Bronfman}, {Dobbs}, {Dumke}, {Finger}, {Ginsburg}, {Gonzalez},
  {Henning}, {Kauffmann}, {Mac-Auliffe}, {Menten}, {Montenegro-Montes},
  {Moore}, {Muller}, {Parra}, {Perez-Beaupuits}, {Pettitt}, {Russeil},
  {S{\'a}nchez-Monge}, {Schilke}, {Schisano}, {Suri}, {Testi}, {Torstensson},
  {Venegas}, {Wang}, {Wienen}, {Wyrowski}, \& {Zavagno}}]{schuller2017}
{Schuller}, F., {Csengeri}, T., {Urquhart}, J.~S., {et~al.} 2017, \aap, 601,
  A124

\bibitem[{{Schuller} {et~al.}(2009){Schuller}, {Menten}, {Contreras},
  {Wyrowski}, {Schilke}, {Bronfman}, {Henning}, {Walmsley}, {Beuther},
  {Bontemps}, {Cesaroni}, {Deharveng}, {Garay}, {Herpin}, {Lefloch}, {Linz},
  {Mardones}, {Minier}, {Molinari}, {Motte}, {Nyman}, {Reveret}, {Risacher},
  {Russeil}, {Schneider}, {Testi}, {Troost}, {Vasyunina}, {Wienen}, {Zavagno},
  {Kovacs}, {Kreysa}, {Siringo}, \& {Wei{\ss}}}]{schuller2009}
{Schuller}, F., {Menten}, K.~M., {Contreras}, Y., {et~al.} 2009, \aap, 504, 415

\bibitem[{{Schuller} {et~al.}(2021){Schuller}, {Urquhart}, {Csengeri},
  {Colombo}, {Duarte-Cabral}, {Mattern}, {Ginsburg}, {Pettitt}, {Wyrowski},
  {Anderson}, {Azagra}, {Barnes}, {Beltran}, {Beuther}, {Billington},
  {Bronfman}, {Cesaroni}, {Dobbs}, {Eden}, {Lee}, {Medina}, {Menten}, {Moore},
  {Montenegro-Montes}, {Ragan}, {Rigby}, {Riener}, {Russeil}, {Schisano},
  {Sanchez-Monge}, {Traficante}, {Zavagno}, {Agurto}, {Bontemps}, {Finger},
  {Giannetti}, {Gonzalez}, {Hernandez}, {Henning}, {Kainulainen}, {Kauffmann},
  {Leurini}, {Lopez}, {Mac-Auliffe}, {Mazumdar}, {Molinari}, {Motte}, {Muller},
  {Nguyen-Luong}, {Parra}, {Perez-Beaupuits}, {Schilke}, {Schneider}, {Suri},
  {Testi}, {Torstensson}, {Veena}, {Venegas}, {Wang}, \&
  {Wienen}}]{schuller2021}
{Schuller}, F., {Urquhart}, J.~S., {Csengeri}, T., {et~al.} 2021, \mnras, 500,
  3064

\bibitem[{{Sellwood} {et~al.}(2019){Sellwood}, {Trick}, {Carlberg}, {Coronado},
  \& {Rix}}]{Sellwood19}
{Sellwood}, J.~A., {Trick}, W.~H., {Carlberg}, R.~G., {Coronado}, J., \& {Rix},
  H.-W. 2019, \mnras, 484, 3154

\bibitem[{{Shane}(1972)}]{shane1972}
{Shane}, W.~W. 1972, \aap, 16, 118

\bibitem[{{Smith} {et~al.}(2014){Smith}, {Glover}, \& {Klessen}}]{smith2014}
{Smith}, R.~J., {Glover}, S. C.~O., \& {Klessen}, R.~S. 2014, \mnras, 445, 2900

\bibitem[{{Smith} {et~al.}(2020){Smith}, {Tre{\ss}}, {Sormani}, {Glover},
  {Klessen}, {Clark}, {Izquierdo}, {Duarte-Cabral}, \& {Zucker}}]{smith2020}
{Smith}, R.~J., {Tre{\ss}}, R.~G., {Sormani}, M.~C., {et~al.} 2020, \mnras,
  492, 1594

\bibitem[{{Sorai} {et~al.}(2019){Sorai}, {Kuno}, {Muraoka}, {Miyamoto},
  {Kaneko}, {Nakanishi}, {Nakai}, {Yanagitani}, {Tanaka}, {Sato}, {Salak},
  {Umei}, {Morokuma-Matsui}, {Matsumoto}, {Ueno}, {Pan}, {Noma}, {Takeuchi},
  {Yoda}, {Kuroda}, {Yasuda}, {Yajima}, {Oi}, {Shibata}, {Seta}, {Watanabe},
  {Kita}, {Komatsuzaki}, {Kajikawa}, {Yashima}, {Cooray}, {Baji}, {Segawa},
  {Tashiro}, {Takeda}, {Kishida}, {Hatakeyama}, {Tomiyasu}, \&
  {Saita}}]{sorai2019}
{Sorai}, K., {Kuno}, N., {Muraoka}, K., {et~al.} 2019, \pasj, 71, S14

\bibitem[{{Stanke} {et~al.}(2019){Stanke}, {Beuther}, {Kauffmann}, {Klaassen},
  {Perez-Beaupuits}, {Johnstone}, {Colombo}, {Schuller}, {Sadavoy}, {Soler},
  {Hatchell}, {Lumsden}, \& {Kulesa}}]{stanke2019}
{Stanke}, T., {Beuther}, H., {Kauffmann}, J., {et~al.} 2019, \baas, 51, 542

\bibitem[{{Stark} \& {Lee}(2006)}]{stark06}
{Stark}, A.~A. \& {Lee}, Y. 2006, \apjl, 641, L113

\bibitem[{{Sun} {et~al.}(2018){Sun}, {Leroy}, {Schruba}, {Rosolowsky},
  {Hughes}, {Kruijssen}, {Meidt}, {Schinnerer}, {Blanc}, {Bigiel}, {Bolatto},
  {Chevance}, {Groves}, {Herrera}, {Hygate}, {Pety}, {Querejeta}, {Usero}, \&
  {Utomo}}]{sun2018}
{Sun}, J., {Leroy}, A.~K., {Schruba}, A., {et~al.} 2018, \apj, 860, 172

\bibitem[{{Tasker} \& {Tan}(2009)}]{tasker09}
{Tasker}, E.~J. \& {Tan}, J.~C. 2009, \apj, 700, 358

\bibitem[{{Taylor} \& {Cordes}(1993)}]{taylor_cordes1993}
{Taylor}, J.~H. \& {Cordes}, J.~M. 1993, \apj, 411, 674

\bibitem[{{Tchernyshyov} {et~al.}(2018){Tchernyshyov}, {Peek}, \&
  {Zasowski}}]{Tchernyshyov18}
{Tchernyshyov}, K., {Peek}, J.~E.~G., \& {Zasowski}, G. 2018, \aj, 156, 248

\bibitem[{{Urquhart} {et~al.}(2021){Urquhart}, {Figura}, {Cross}, {Wells},
  {Moore}, {Eden}, {Ragan}, {Pettitt}, {Duarte-Cabral}, {Colombo}, {Schuller},
  {Csengeri}, {Mattern}, {Beuther}, {Menten}, {Wyrowski}, {Anderson}, {Barnes},
  {Beltr{\'a}n}, {Billington}, {Bronfman}, {Giannetti}, {Kainulainen},
  {Kauffmann}, {Lee}, {Leurini}, {Medina}, {Montenegro-Montes}, {Riener},
  {Rigby}, {S{\'a}nchez-Monge}, {Schilke}, {Schisano}, {Traficante}, \&
  {Wienen}}]{urquhart2021}
{Urquhart}, J.~S., {Figura}, C., {Cross}, J.~R., {et~al.} 2021, \mnras, 500,
  3050

\bibitem[{{Urquhart} {et~al.}(2018){Urquhart}, {K{\"o}nig}, {Giannetti},
  {Leurini}, {Moore}, {Eden}, {Pillai}, {Thompson}, {Braiding}, {Burton},
  {Csengeri}, {Dempsey}, {Figura}, {Froebrich}, {Menten}, {Schuller}, {Smith},
  \& {Wyrowski}}]{urquhart2018}
{Urquhart}, J.~S., {K{\"o}nig}, C., {Giannetti}, A., {et~al.} 2018, \mnras,
  473, 1059

\bibitem[{{Urquhart} {et~al.}(2014){Urquhart}, {Moore}, {Csengeri}, {Wyrowski},
  {Schuller}, {Hoare}, {Lumsden}, {Mottram}, {Thompson}, {Menten}, {Walmsley},
  {Bronfman}, {Pfalzner}, {K{\"o}nig}, \& {Wienen}}]{urquhart2014}
{Urquhart}, J.~S., {Moore}, T.~J.~T., {Csengeri}, T., {et~al.} 2014, \mnras,
  443, 1555

\bibitem[{{Vall{\'e}e}(2008)}]{vallee2008}
{Vall{\'e}e}, J.~P. 2008, \aj, 135, 1301

\bibitem[{{Vall{\'e}e}(2017)}]{vallee2017}
{Vall{\'e}e}, J.~P. 2017, The Astronomical Review, 13, 113

\bibitem[{{Virtanen} {et~al.}(2020){Virtanen}, {Gommers}, {Oliphant},
  {Haberland}, {Reddy}, {Cournapeau}, {Burovski}, {Peterson}, {Weckesser},
  {Bright}, {van der Walt}, {Brett}, {Wilson}, {Jarrod Millman}, {Mayorov},
  {Nelson}, {Jones}, {Kern}, {Larson}, {Carey}, {Polat}, {Feng}, {Moore}, {Vand
  erPlas}, {Laxalde}, {Perktold}, {Cimrman}, {Henriksen}, {Quintero}, {Harris},
  {Archibald}, {Ribeiro}, {Pedregosa}, {van Mulbregt}, \&
  {Contributors}}]{scipy2020}
{Virtanen}, P., {Gommers}, R., {Oliphant}, T.~E., {et~al.} 2020, Nature
  Methods, 17, 261

\bibitem[{{Wang} {et~al.}(2020{\natexlab{a}}){Wang}, {L{\'o}pez-Corredoira},
  {Huang}, {Carlin}, {Chen}, {Wang}, {Chang}, {Zhang}, {Xiang}, {Yuan}, {Sun},
  {Li}, {Yang}, \& {Deng}}]{wang2020_gaia}
{Wang}, H.~F., {L{\'o}pez-Corredoira}, M., {Huang}, Y., {et~al.}
  2020{\natexlab{a}}, \mnras, 491, 2104

\bibitem[{{Wang} {et~al.}(2015){Wang}, {Testi}, {Ginsburg}, {Walmsley},
  {Molinari}, \& {Schisano}}]{wang2015}
{Wang}, K., {Testi}, L., {Ginsburg}, A., {et~al.} 2015, \mnras, 450, 4043

\bibitem[{{Wang} {et~al.}(2020{\natexlab{b}}){Wang}, {Beuther}, {Rugel},
  {Soler}, {Stil}, {Ott}, {Bihr}, {McClure-Griffiths}, {Anderson}, {Klessen},
  {Goldsmith}, {Roy}, {Glover}, {Urquhart}, {Heyer}, {Linz}, {Smith}, {Bigiel},
  {Dempsey}, \& {Henning}}]{wang2020_thor}
{Wang}, Y., {Beuther}, H., {Rugel}, M.~R., {et~al.} 2020{\natexlab{b}}, \aap,
  634, A83

\bibitem[{{Wegg} {et~al.}(2015){Wegg}, {Gerhard}, \& {Portail}}]{wegg2015}
{Wegg}, C., {Gerhard}, O., \& {Portail}, M. 2015, \mnras, 450, 4050

\bibitem[{{Wilson} {et~al.}(2009){Wilson}, {Warren}, {Israel}, {Serjeant},
  {Bendo}, {Brinks}, {Clements}, {Courteau}, {Irwin}, {Knapen}, {Leech},
  {Matthews}, {M{\"u}hle}, {Mortier}, {Petitpas}, {Sinukoff}, {Spekkens},
  {Tan}, {Tilanus}, {Usero}, {van der Werf}, {Wiegert}, \& {Zhu}}]{wilson2009}
{Wilson}, C.~D., {Warren}, B.~E., {Israel}, F.~P., {et~al.} 2009, \apj, 693,
  1736

\bibitem[{{Xu} {et~al.}(2021){Xu}, {Hou}, {Bian}, {Hao}, {Liu}, {Li}, \&
  {Li}}]{xu2021}
{Xu}, Y., {Hou}, L.~G., {Bian}, S.~B., {et~al.} 2021, \aap, 645, L8

\bibitem[{{Xu} {et~al.}(2018){Xu}, {Hou}, \& {Wu}}]{xu2018}
{Xu}, Y., {Hou}, L.-G., \& {Wu}, Y.-W. 2018, Research in Astronomy and
  Astrophysics, 18, 146

\bibitem[{{Xu} {et~al.}(2013){Xu}, {Li}, {Reid}, {Menten}, {Zheng},
  {Brunthaler}, {Moscadelli}, {Dame}, \& {Zhang}}]{xu2013}
{Xu}, Y., {Li}, J.~J., {Reid}, M.~J., {et~al.} 2013, \apj, 769, 15

\bibitem[{{Zucker} {et~al.}(2015){Zucker}, {Battersby}, \&
  {Goodman}}]{zucker2015}
{Zucker}, C., {Battersby}, C., \& {Goodman}, A. 2015, \apj, 815, 23

\bibitem[{{Zucker} {et~al.}(2018){Zucker}, {Battersby}, \&
  {Goodman}}]{zucker2018}
{Zucker}, C., {Battersby}, C., \& {Goodman}, A. 2018, \apj, 864, 153

\end{thebibliography}
}

\appendix

\section{Assessing the robustness of the results with an alternative spiral arm model}
\label{A:reid19}

\begin{figure*}
    \centering
        \includegraphics[width=1\textwidth]{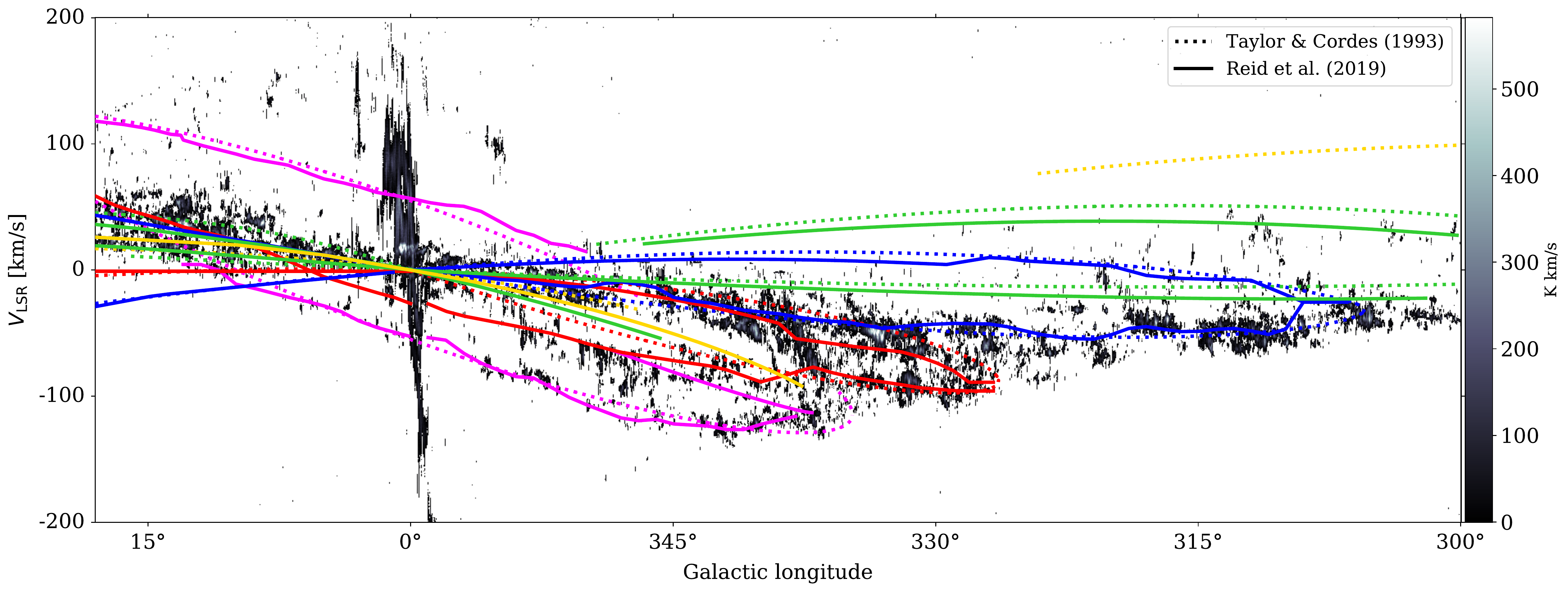}
   \caption{Spiral-arm models from \cite{taylor_cordes1993} (dotted line) and \cite{reid2019} (full line) overlaid on the \lvmap{} from the full data set. The 3\,kpc arms are shown in magenta, the Norma-Outer arm in red, the Scutum-Centaurus arm in blue, the Sagittarius-Carina arm in green, and the Perseus arm in yellow. Other symbols follow the convention of Fig.~\ref{F:sparm_lvmaps}.}
    \label{F:lvmap_tc_reid}
\end{figure*}

\begin{figure}
        \includegraphics[width=0.45\textwidth]{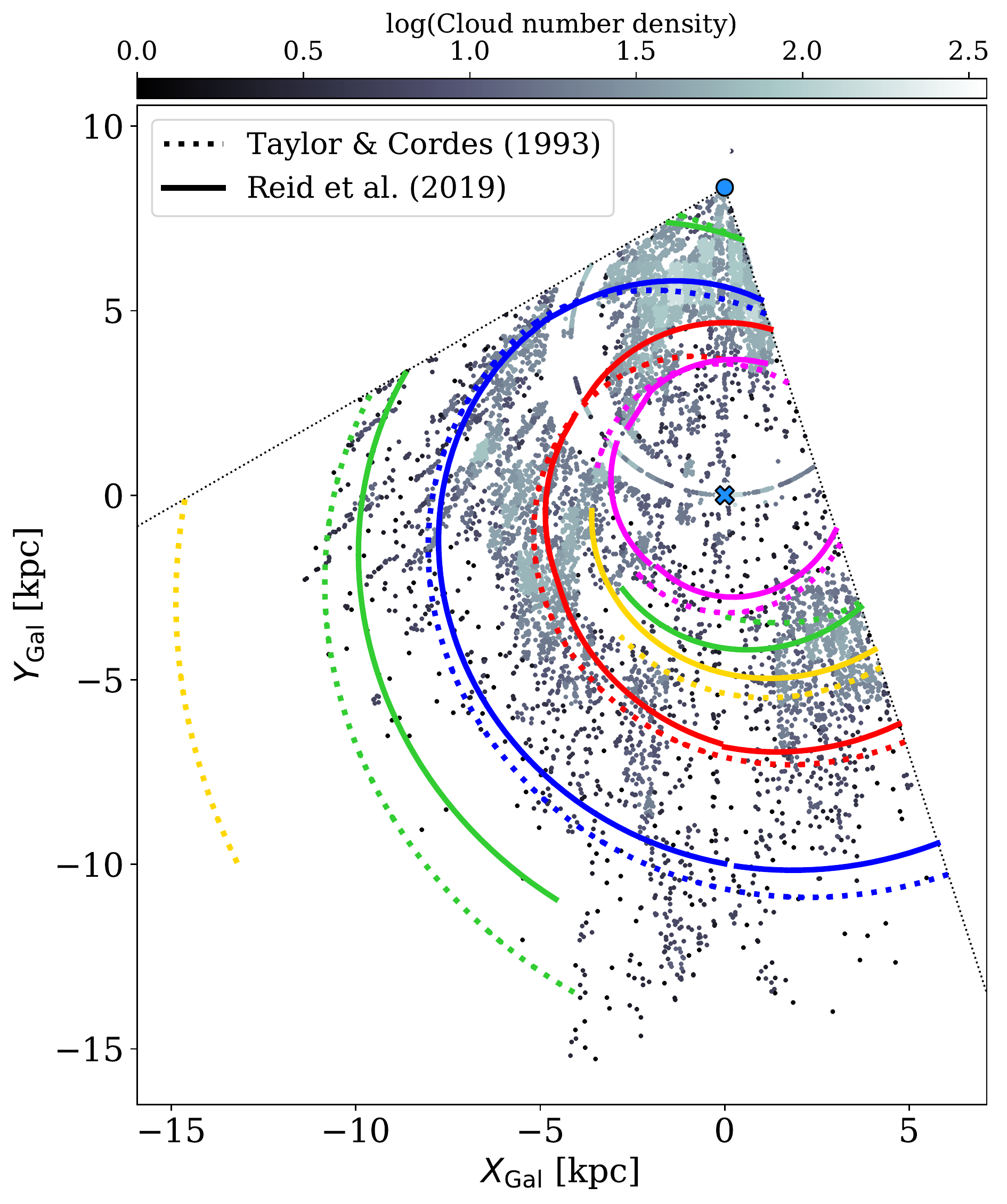}
   \caption{Spiral-arm models from \cite{taylor_cordes1993} (dotted line) and \cite{reid2019} (full line) overlaid on the cloud distribution. The 3\,kpc arms are shown in magenta, the Norma-Outer arm in red, the Scutum-Centaurus arm in blue, the Sagittarius-Carina arm in green, and the Perseus arm in yellow. Other symbols follow the convention of Fig.~\ref{F:sparm_gmc_match}.}
    \label{F:xymap_tc_reid}
\end{figure}

To test how our results and conclusions are dependent on the spiral arm model adopted, in this section, we repeated some key analyses of the paper using the models from \citep{reid2019} (hereafter R19). These models are included in the Bar And Spiral Structure Legacy (BeSSeL) Survey parallax-based distance calculator v2.4 bundle\footnote{\url{http://bessel.vlbi-astrometry.org/node/378}}. The bundle provides a table where the longitude ($l$), latitude ($b$), velocity ($V_{\rm lsr}$), Galactocentric radius ($R$), heliocentric distance ($d$), and azimuth ($\beta$) are defined at each longitudinal degree, which we interpolate on the SEDIGISM data cube grids following the method in Section~\ref{S:sparm_models}.  

Figure \ref{F:lvmap_tc_reid} shows the longitude--velocity tracks for the main spiral arms considered in this work drawn from the TC93 and R19 models. Generally, the two model sets are consistent with each other. However, several differences can be noted. Tracks from TC93 models are smooth, while the ones from R19 models, which follow the $^{12}$CO(1-0), show peaks and troughs. The 3\,kpc arms are the most compatible between the two models. The Sagittarius-Carina models are also largely compatible at large velocities and longitudes and the Scutum-Centaurus arms defined by TC93 and by R19 are almost equivalent. The Norma-Outer arm tracks are, instead, the ones that show the largest differences between the two models across the $lv$ space, as well as the Perseus arm. For the latter, in particular, the TC93 model considers also a segment between $300^{\circ}\leq l \leq 330^{\circ}$ and at large velocities which is not present in R19 models. Figure~\ref{F:xymap_tc_reid} shows the equivalent comparison for the top-down view of the Milky Way. Here the differences between the models are more evident. Spiral arm locations are similar, but start and end points generally do not coincide. Additionally, it appears that TC93 models are shifted towards larger Galactocentric radii with respect to R19 models. Therefore, we expect major discrepancies to occur on the analysis of the cloud distribution with respect to the assumed spiral arm model. 

The IDI and LDI calculated from flux and luminosity PDFs (respectively), and built through TC93 and R19 models, are collected in Table~\ref{T:lidi_reid2019}. The values have been obtained with the same thresholds as indicated in Section~\ref{SS:pdf}. The table clearly shows that indexes calculated with the two models are almost indistinguishable, except for the LDI of the inter-arm region, which are slightly different. Nevertheless, those differences are marginal and do not influence our conclusions on the moderate prominence of spiral arm PDFs over inter-arm region PDFs. 

\begin{table}
\centering
\caption{Distribution indexes for given models and regions.}
\begin{tabular}{c|cc|cc}
\hline
& \multicolumn{2}{|c|}{TC93} & \multicolumn{2}{|c}{R19} \\
Region & IDI & LDI & IDI & LDI\\
\hline
All & -0.50 & -0.63 & -0.50 & -0.63 \\
Spiral arms & -1.58 & -0.58 & -1.50 & -0.55 \\
Inter-arm & -2.02 & -0.85 & -2.01 & -0.97 \\
Gal. centre & 0.66 & - & 0.66 & - \\        
\hline
\end{tabular}
\tablefoot{Integrated intensity distribution index (IDI) and luminosity distribution index (LDI) calculated from flux and luminosity PDFs, respectively. The PDFs are built through \cite{taylor_cordes1993} (TC93) and \cite{reid2019} (R19) models.}
\label{T:lidi_reid2019}
\end{table}

This prominence can also be checked through the spiral-arm to inter-arm mass contrast. Using R19 models, we calculated a spiral arm mass of $10^{7.18}$\,M$_{\odot}$ and an inter-arm mass of $10^{7.03}$\,M$_{\odot}$, for a spiral-arm to inter-arm contrast equal to 1.4, which is remarkably close to the value inferred from TC93 models (1.5; see Section~\ref{SS:sparm_linmass}). In particular, considering the discussion in Section~\ref{S:disc}, this value classifies the Milky Way as an average spiral galaxy, similar to others in the nearby Universe.

The median and inter-quartile range of the cloud property distributions of spiral arms and inter-arm regions, allocated through R19 models, are collected in Table~\ref{T:med_iqr_reid2019} for both the full science and complete distance-limited samples, and their respective subsamples containing ATLASGAL or HMSF sources. The values in the table do not show significant differences compared to the corresponding ones from the cloud property distributions built through TC93 models (Table~\ref{T:med_iqr_taylor-cordes2003}). As in Section~\ref{SS:gmc_props}, we assessed the property differences through the KS test; the results are summarised in Table~\ref{T:pvalues_reid2019}. Interestingly, for the full science sample the test for all properties (except $\Sigma_{\rm mol}$) returned significant $p-$values (well below $10^{-4}$), while through TC93 models, we observed that $\sigma_{\rm v}$ and $AR$ were giving high $p-$values. Nevertheless, the test on the two model cloud associations returned results of equivalent meaning. considering the ATLASGAL and HMSF subsample. Additionally, as for the cloud association through TC93 models, we did not observe significant differences between the distributions that involve the distance-limited complete sample (and relative subsamples). 

Given these tests, we conclude that our results are fairly robust regardless of the spiral arm model set used to perform the analyses. We observe that only some cloud property distributions show more significant differences when using the R19 model set with respect to the TC93 model set. However, this does not change our conclusion on the nature of the spiral arm structure of the Milky Way.  

\begin{table*}
\centering
\caption{Cloud property distribution statistics across regions and subsamples with cloud-to-spiral-arm association performed with R19 models}
\begin{tabular}{c|c|cc|cc|cc|cc}
\hline
& & \multicolumn{8}{|c}{Full sample} \\
& & \multicolumn{4}{|c|}{Science sample} & \multicolumn{4}{|c}{Dist. lim. sample} \\
& & \multicolumn{2}{|c|}{SA} & \multicolumn{2}{|c|}{IA} & \multicolumn{2}{|c|}{SA} & \multicolumn{2}{|c}{IA} \\
Property & Units & $\mu$ & $IQR$ & $\mu$ & $IQR$ & $\mu$ & $IQR$ & $\mu$ & $IQR$ \\
\hline
$\log(R_{\rm eff})$ & pc & 0.34 & 0.54 & 0.39 & 0.37 & 0.38 & 0.36 & 0.38 & 0.35 \\
$\log(\sigma_{\rm v})$ & km~s$^{-1}$ & -0.11 & 0.32 & -0.12 & 0.26 & 0.07 & 0.33 & 0.04 & 0.28 \\
$\log(M_{\rm mol})$ & M$_{\odot}$ & 3.06 & 1.18 & 3.17 & 0.81 & 3.19 & 0.82 & 3.29 & 0.87 \\
$\log(\Sigma_{\rm mol})$ & M$_{\odot}$~pc$^{-2}$ & 1.87 & 0.25 & 1.87 & 0.22 & 1.97 & 0.28 & 2.02 & 0.32 \\
$\log(\alpha_{\rm vir})$ & & 0.16 & 0.45 & 0.04 & 0.38 & 0.27 & 0.47 & 0.21 & 0.42 \\
$\log(AR)$ & & 1.01 & 0.32 & 0.96 & 0.32 & 1.19 & 0.26 & 1.15 & 0.30 \\
\hline
\hline
& & \multicolumn{8}{|c}{ATLASGAL sample} \\
& & \multicolumn{4}{|c|}{Science sample} & \multicolumn{4}{|c}{Dist. lim. sample} \\
& & \multicolumn{2}{|c|}{SA} & \multicolumn{2}{|c|}{IA} & \multicolumn{2}{|c|}{SA} & \multicolumn{2}{|c}{IA} \\
Property & Units & $\mu$ & $IQR$ & $\mu$ & $IQR$ & $\mu$ & $IQR$ & $\mu$ & $IQR$ \\
\hline
$\log(R_{\rm eff})$ & pc & 0.38 & 0.55 & 0.51 & 0.33 & 0.36 & 0.35 & 0.42 & 0.30 \\
$\log(\sigma_{\rm v})$ & km~s$^{-1}$ & 0.03 & 0.20 & 0.03 & 0.22 & 0.08 & 0.22 & 0.06 & 0.19 \\
$\log(M_{\rm mol})$ & M$_{\odot}$ & 3.26 & 1.07 & 3.57 & 0.69 & 3.19 & 0.62 & 3.37 & 0.56 \\
$\log(\Sigma_{\rm mol})$ & M$_{\odot}$~pc$^{-2}$ & 2.05 & 0.22 & 2.08 & 0.20 & 2.03 & 0.22 & 2.07 & 0.21 \\
$\log(\alpha_{\rm vir})$ & & 0.23 & 0.55 & 0.05 & 0.40 & 0.29 & 0.45 & 0.25 & 0.37 \\
$\log(AR)$ & & 1.06 & 0.31 & 1.01 & 0.33 & 1.17 & 0.29 & 1.09 & 0.30 \\
\hline
\hline
& & \multicolumn{8}{|c}{HMSF sample} \\
& & \multicolumn{4}{|c|}{Science sample} & \multicolumn{4}{|c}{Dist. lim. sample} \\
& & \multicolumn{2}{|c|}{SA} & \multicolumn{2}{|c|}{IA} & \multicolumn{2}{|c|}{SA} & \multicolumn{2}{|c}{IA} \\
Property & Units & $\mu$ & $IQR$ & $\mu$ & $IQR$ & $\mu$ & $IQR$ & $\mu$ & $IQR$ \\
\hline
$\log(R_{\rm eff})$ & pc & 0.61 & 0.49 & 0.64 & 0.34 & 0.62 & 0.31 & 0.59 & 0.30 \\
$\log(\sigma_{\rm v})$ & km~s$^{-1}$ & 0.23 & 0.24 & 0.16 & 0.25 & 0.23 & 0.23 & 0.21 & 0.24 \\
$\log(M_{\rm mol})$ & M$_{\odot}$ & 3.96 & 1.07 & 4.05 & 0.78 & 3.99 & 0.70 & 4.02 & 0.83 \\
$\log(\Sigma_{\rm mol})$ & M$_{\odot}$~pc$^{-2}$ & 2.18 & 0.28 & 2.23 & 0.26 & 2.23 & 0.28 & 2.29 & 0.26 \\
$\log(\alpha_{\rm vir})$ & & 0.21 & 0.61 & 0.05 & 0.47 & 0.23 & 0.42 & 0.11 & 0.48 \\
$\log(AR)$ & & 1.18 & 0.30 & 1.20 & 0.34 & 1.26 & 0.27 & 1.26 & 0.26 \\
\hline
\hline
\end{tabular}
\tablefoot{Median ($\mu$) and inter-quartile range ($IQR$) of cloud property distributions (from top to bottom effective radius $R_{\rm eff}$, velocity dispersion $\sigma_{\rm v}$, molecular gas mass $M_{\rm mol}$, molecular gas mass surface density $\Sigma_{\rm mol}$, virial parameter $\alpha_{\rm vir}$, and aspect ratio $AR$) in the spiral arms (SA) with respect to the inter-arm regions (IA). The analysis is performed separately for the full sample, for the cloud sample that contains at least one ATLASGAL source, and for the cloud sample that contains clouds with a HMSF signpost; and their respective science and complete distance-limited subsamples.}
\label{T:med_iqr_reid2019}
\end{table*}

\begin{table*}
\centering
\caption{KS test results on cloud properties for spiral arm and inter-arm distributions with cloud-to-spiral-arm association performed with R19 models}
\begin{tabular}{c|cc|cc|cc}
\hline
Property & \multicolumn{2}{|c|}{Full} & \multicolumn{2}{|c|}{ATLASGAL} & \multicolumn{2}{|c}{HMSF} \\
& $p_{\rm val}$ & $D_{\rm stat}$ & $p_{\rm val}$ & $D_{\rm stat}$ & $p_{\rm val}$ & $D_{\rm stat}$ \\
\hline
\hline
\multicolumn{7}{c}{Science sample} \\
\hline
$R_{\rm eff}$ & <~0.0001 & 0.16 & <~0.0001 & 0.24 & 0.1500 & 0.14 \\
$\sigma_{\rm v}$ & <~0.0001 & 0.06 & 0.6524 & 0.06 & 0.0177 & 0.19 \\
$M_{\rm mol}$ & <~0.0001 & 0.15 & <~0.0001 & 0.28 & 0.0549 & 0.16 \\
$\Sigma_{\rm mol}$ & 0.0969 & 0.03 & 0.0453 & 0.12 & 0.1667 & 0.14 \\
$\alpha_{\rm vir}$ & <~0.0001 & 0.16 & <~0.0001 & 0.23 & 0.0187 & 0.19 \\
$AR$ & <~0.0001 & 0.08 & 0.1364 & 0.10 & 0.8472 & 0.07 \\
\hline
\hline
\multicolumn{7}{c}{Complete distance-limited sample} \\
\hline
$R_{\rm eff}$ & 0.8195 & 0.04 & 0.1435 & 0.17 & 0.4913 & 0.14 \\
$\sigma_{\rm v}$ & 0.1559 & 0.08 & 0.9045 & 0.08 & 0.3692 & 0.16 \\
$M_{\rm mol}$ & 0.1484 & 0.08 & 0.0390 & 0.20 & 0.6764 & 0.12 \\
$\Sigma_{\rm mol}$ & 0.0217 & 0.10 & 0.0594 & 0.19 & 0.7078 & 0.12 \\
$\alpha_{\rm vir}$ & 0.0851 & 0.08 & 0.0333 & 0.21 & 0.2217 & 0.18 \\
$AR$ & 0.0008 & 0.13 & 0.0524 & 0.19 & 0.9925 & 0.07 \\
\hline
\end{tabular}
\tablefoot{$p-$values ($p_{\rm val}$) and statistics ($D_{\rm stat}$) from the KS test comparing the distributions of cloud properties (from top to bottom: effective radius R$_{\rm eff}$, velocity dispersion $\sigma_{\rm v}$, molecular gas mass M$_{\rm mol}$, molecular gas mass surface density $\Sigma_{\rm mol}$, virial parameter $\alpha_{\rm vir}$, and aspect ratio $AR$) in the spiral arms (SA) with respect to the inter-arm regions (IA). The analysis is performed separately for the full sample, for the cloud sample that contains at least one ATLASGAL source, and for the cloud sample
that contains clouds with a HMSF signpost.}
\label{T:pvalues_reid2019}
\end{table*}

\section{Assessing the influence of the chosen velocity offset}
\label{A:voff}

\begin{figure*}
        \includegraphics[width=1\textwidth]{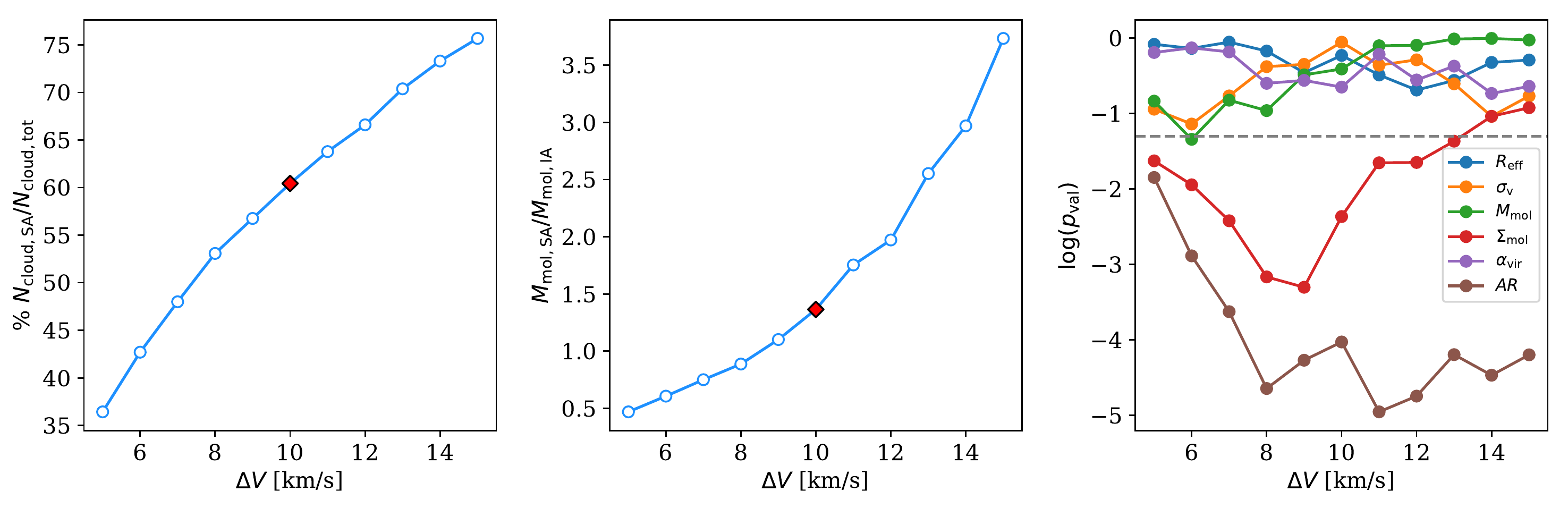}
   \caption{Number of clouds in the spiral arms ($N_{\rm cloud, SA}$) as a percentage of the total number of clouds ($N_{\rm cloud, tot}$) in the complete distance-limited sample (left panel), and the integrated cloud mass within spiral arms ($M_{\rm mol, SA}$) over integrated cloud mass within the inter-arm region ($M_{\rm mol, IA}$; middle panel) versus different velocity offsets with respect to the spiral arm model ridge line. In the panels, the red diamonds show the values for our chosen velocity offset $\Delta V=10$\,\kms{}. Also shown is the KS test $p-$value ($p_{\rm val}$, right panel) variation with respect to $\Delta V$ calculated from the property distributions of the clouds in the spiral arms and inter-arm region. Considered properties are effective radius R$_{\rm eff}$, velocity dispersion $\sigma_{\rm v}$, molecular gas mass M$_{\rm mol}$, molecular gas mass surface density $\Sigma_{\rm mol}$, virial parameter $\alpha_{\rm vir}$, and aspect ratio $AR$. The dashed line indicates $p-$value = 0.05.}
    \label{F:voff}
\end{figure*}

To associate clouds to spiral arms we have chosen a particular velocity offset ($\Delta$V) with respect to the arm ridge line of 10\,\kms{}. This choice was motivated by the magnitude of spiral arm streaming motion measured in several studies (see Section~\ref{SSS:gmc_match}). Here we tested how this choice might influence the properties of the spiral arms versus the inter-arm region inferred through cloud measurements. We used the complete distance-limited sample in order to remove distance biases that might influence our interpretation of the results. Varying this threshold increases the number of clouds associated with spiral-arm
regions  steeply: between $35\%-75\%$ of the clouds in the complete distance-limited sample are found within spiral arms considering $\Delta V=5-15$\,\kms{} (Fig.~\ref{F:voff}, left). The same is true for the spiral-arm--inter-arm region contrast (measured through the ratio between the total cloud mass within spiral arms and the total cloud mass in the inter-arm region; Fig.~\ref{F:voff}, middle), which also steeply increases with $\Delta V$. In addition, we checked how the property distributions of the clouds in the two considered regions compare by evaluating the KS test $p-$value at different $\Delta V$. For R$_{\rm eff}$, $\sigma_{\rm v}$, M$_{\rm mol}$, and $\alpha_{\rm vir}$ we did not observe a significant variation of $p-$value, which also indicates that spiral-arm and inter-arm region distributions can be drawn from the same parental distribution regardless of the velocity offset. The $p-$value for $\Sigma_{\rm mol}$ distributions showed larger variations, mostly below the significance level. The $p-$value for $AR$ distributions instead goes well below 0.05 for each $\Delta V$.  

We conclude that the choice of $\Delta V$ has a potentially large impact on spiral-arms and inter-arm region differences inferred from cloud measurements. However, this conclusion also indicates that there are no clear discontinuities between the molecular gas across velocity space, reinforcing our general conclusion on the flocculent nature of the Milky Way spiral structure. 

\section{Possible biases introduced by the spiral arm KDA solver}
\label{A:dflag13}

In Section~\ref{SS:cxya}, we solved the KDA of 139 clouds with unreliable distances using the spiral arm association: a cloud uniquely associated with given spiral arm region assumes the distance of that region if the cloud distance was originally tagged as 'unreliable' in the DC21 catalogue. This explicitly supposed that spiral arms favour the formation of molecular clouds. In addition, this operation increased the spiral arm cloud population in our sample. Those clouds have been tagged with $d_{\rm flag}=13$ in the catalogue. In this section, we check whether this changes the main conclusions of our analysis. In Section~\ref{SS:sparm_linmass}, we observed that 54\% of the clouds with reliable distances are in the spiral arms; without the inclusion of $d_{\rm flag}=13$ clouds, the percentage decreases to 53\%. Similar decrements are observed considering the well-resolved subsamples of clouds used in Section~\ref{SS:gmc_props}. The addition of the $d_{\rm flag}=13$ clouds did not change the property distributions in Section~\ref{SS:gmc_props} or the relationships between the distributions, as given by the $p-$values, which show similar significance. The clouds with $d_{\rm flag}=13$ add a mass of $10^{5.3}$\,M$_{\odot}$ to the spiral arms. Removing them would not significantly affect the results presented in Table~\ref{T:densities} as spiral arms show a total molecular gas mass in clouds of $\sim10^{7.2}$\,M$_{\odot}$.

\section{Distance bias in the cloud science sample}
\label{A:distbias}

\begin{figure}
        \includegraphics[width=0.45\textwidth]{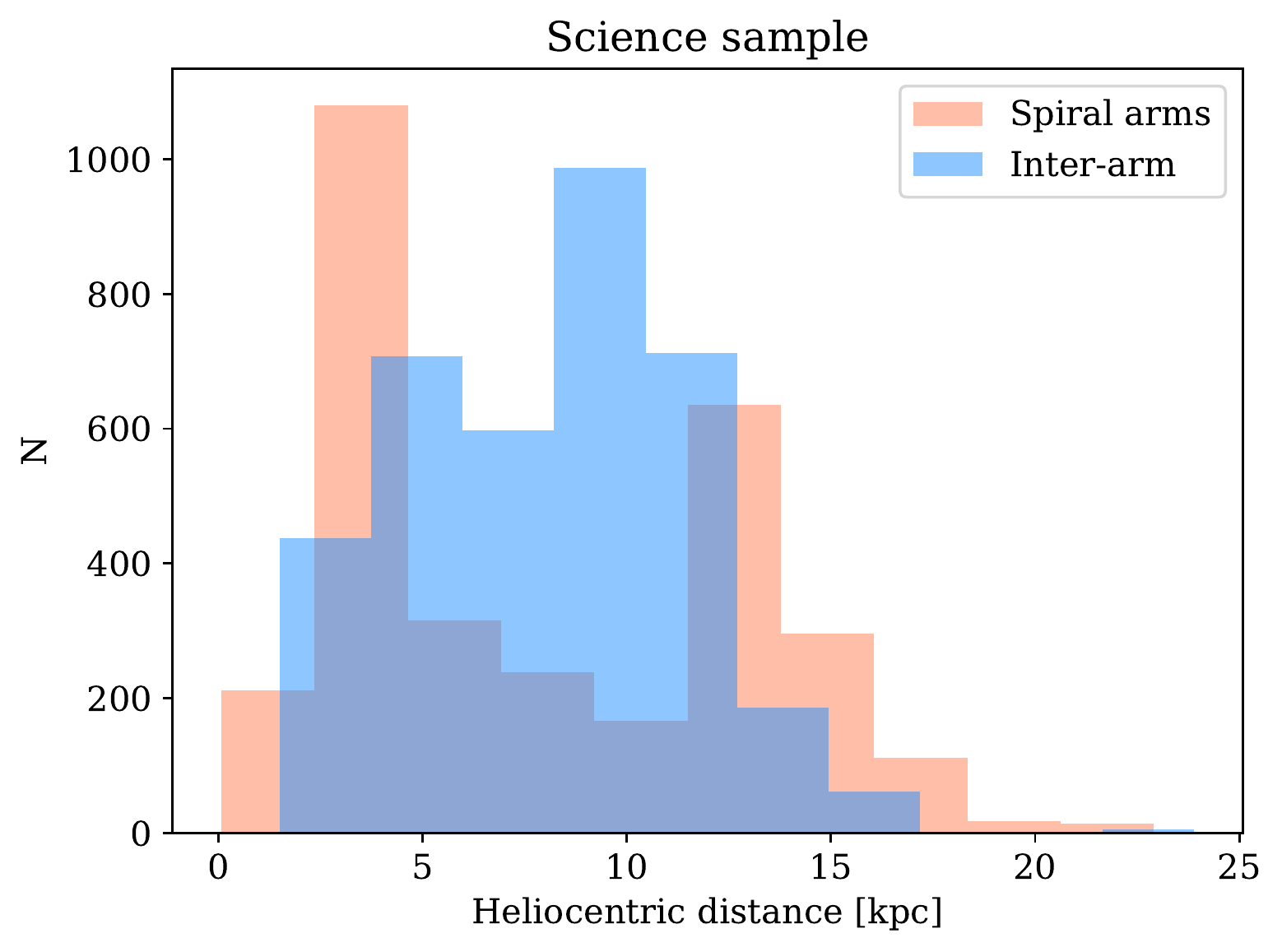}
   \caption{Distribution of distances from spiral arm and inter-arm region clouds in the science sample.}
    \label{F:distbias}
\end{figure}

In Section~\ref{SS:gmc_props} we compared the cloud property distributions in the spiral arms and inter-arm regions. Besides the science sample, we also used a complete distance-limited sample to access whether the distribution differences might be driven by a distance bias in the science sample. In Fig.~\ref{F:distbias}, we show the distribution of heliocentric distance for clouds in the spiral arms and inter-arm regions in the science sample. The spiral arm histogram shows two peaks at distances of $\sim4$\,kpc and $\sim12$\,kpc, while inter-arm clouds populate the distance region between these two peaks. In addition, spiral arm clouds show distances that extend up to $\sim20$\,kpc. This discrepancy might indeed be responsible for the low $p-$values observed for some property distributions in the science sample. As such, a KS test performed on the distance distributions considered here gives a very low $p-$value (on the order of $10^{-96}$).

\section{Full SEDIGISM survey integrated intensity map multi-coloured images}\label{A:multi_colored}
Here we collect all multi-coloured integrated intensity maps of the $^{13}$CO\,(2-1) emission across the full SEDIGISM data. Clouds attributed to a given spiral arm are colour coded as follows: magenta (3~kpc), red (Norma-Outer), blue (Scutum-Centaurus), green (Sagittarius-Carina), yellow (Perseus), and grey (inter-arm or unreliable distance clouds).

\begin{figure*}
        \includegraphics[width=0.85\textwidth]{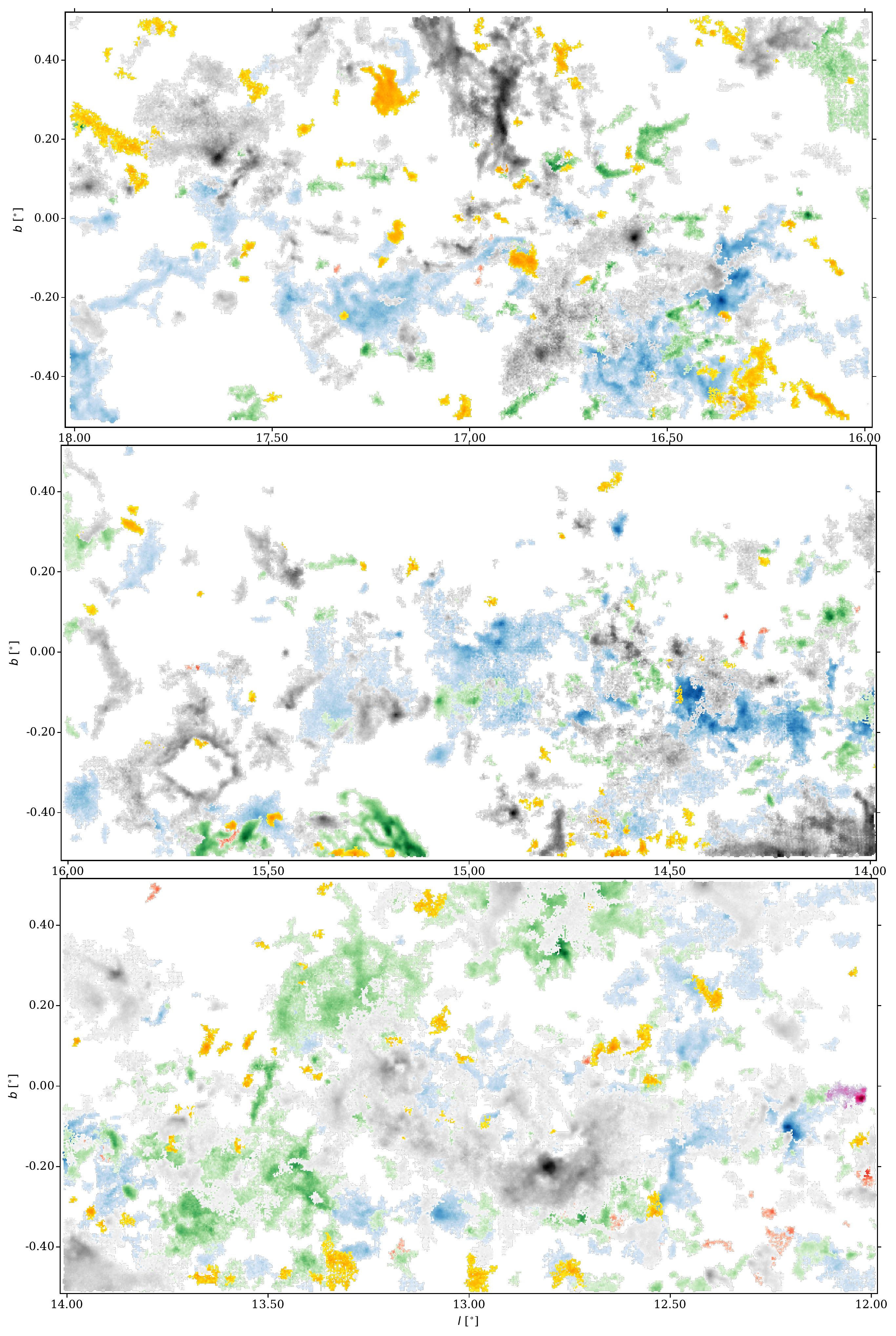}
   \caption{Multi-coloured integrated intensity maps of the $^{13}$CO\,(2-1) emission in the SEDIGISM field across $2^{\circ} \leq l \leq 18^{\circ}$.}
    \label{F:sed_sparms_013}
\end{figure*}

\begin{figure*}
        \includegraphics[width=0.85\textwidth]{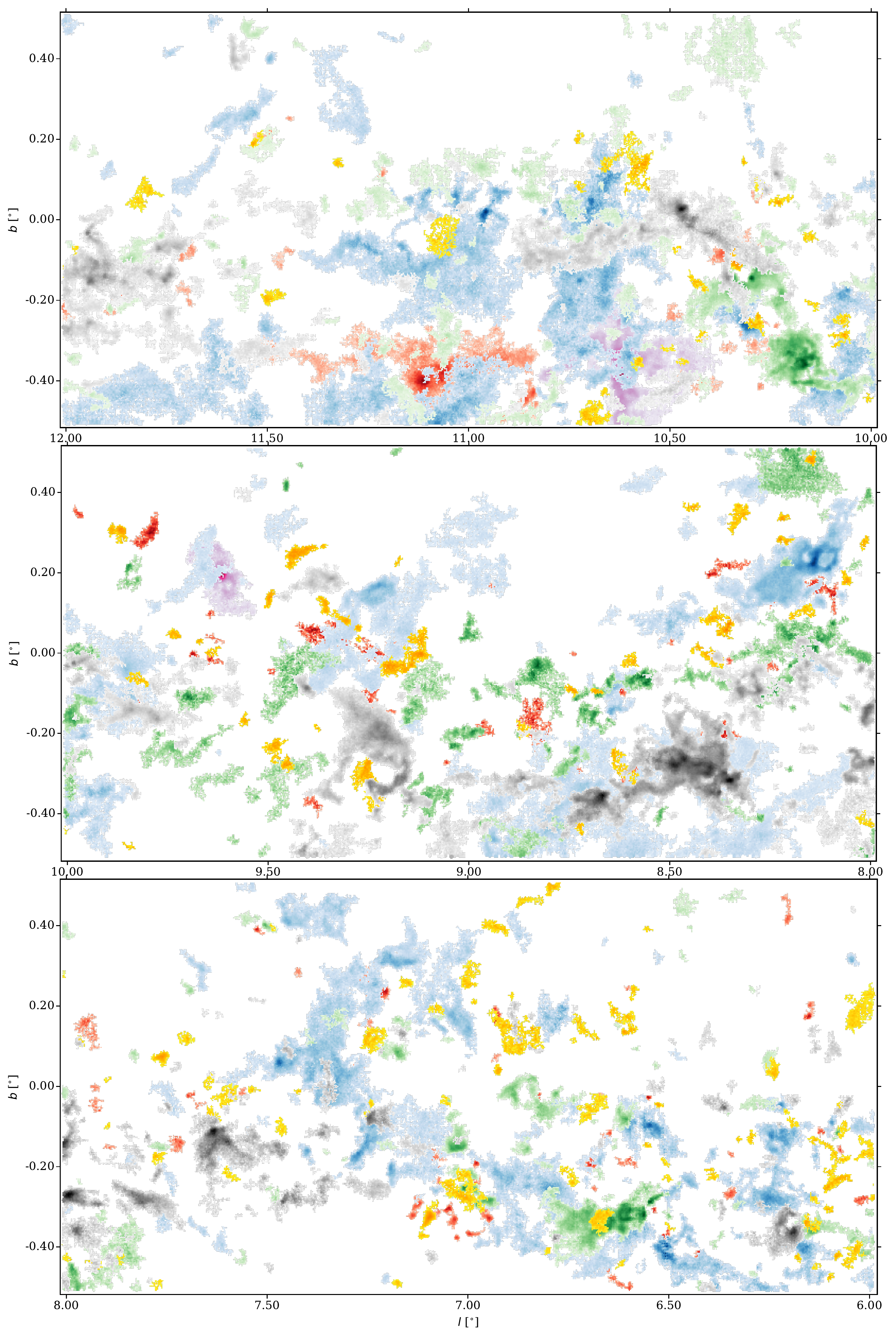}
   \caption{Multi-coloured integrated intensity maps of the $^{13}$CO\,(2-1) emission in the SEDIGISM field across $0^{\circ} \leq l \leq 2^{\circ}$ and  $346^{\circ} \leq l \leq 360^{\circ}$.}
    \label{F:sed_sparms_007}
\end{figure*}

\begin{figure*}
        \includegraphics[width=0.85\textwidth]{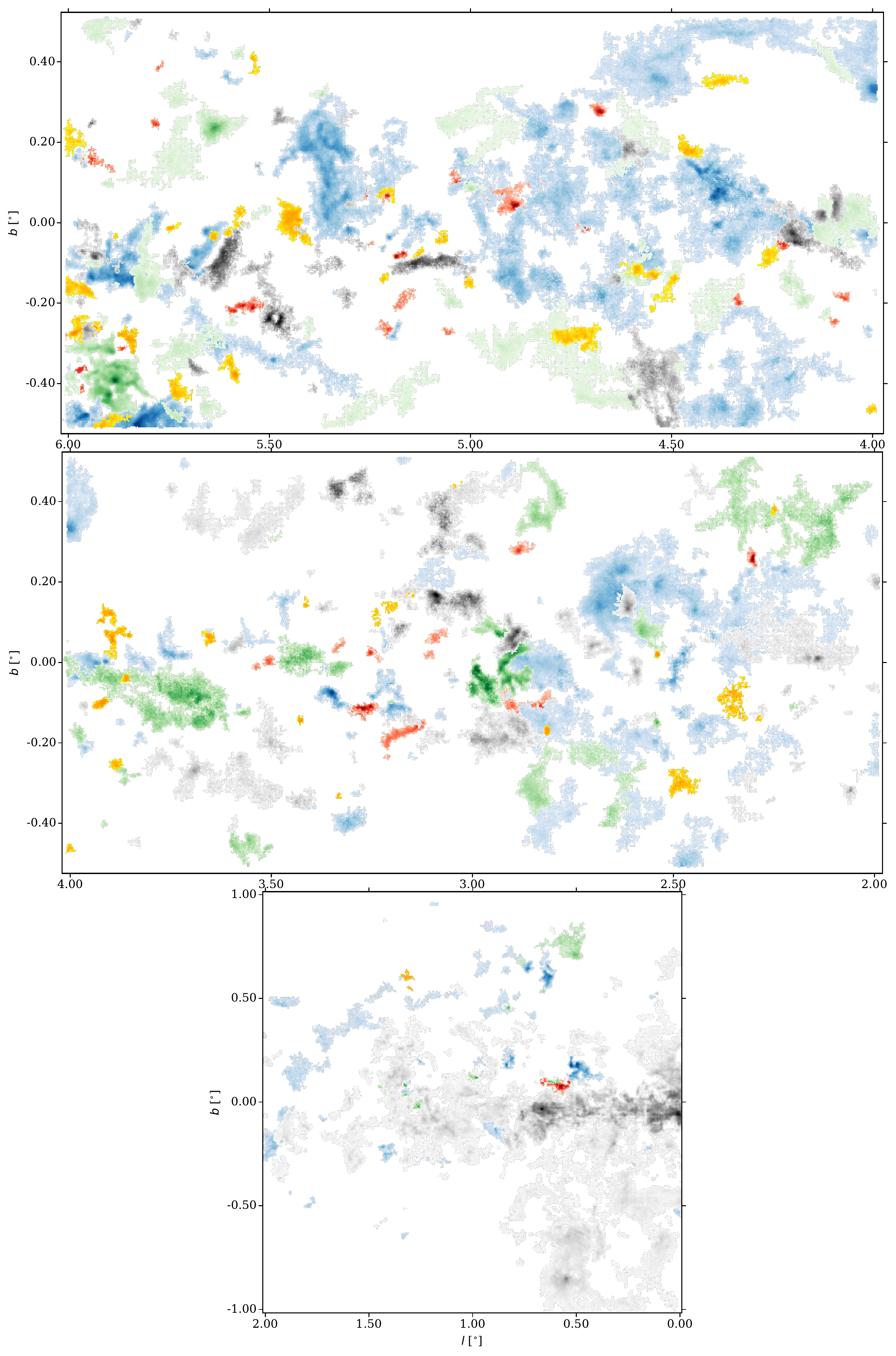}
   \caption{Multi-coloured integrated intensity maps of the $^{13}$CO\,(2-1) emission in the SEDIGISM field across $331^{\circ} \leq l \leq 346^{\circ}$.}
    \label{F:sed_sparms_001}
\end{figure*}

\begin{figure*}
        \includegraphics[width=0.85\textwidth]{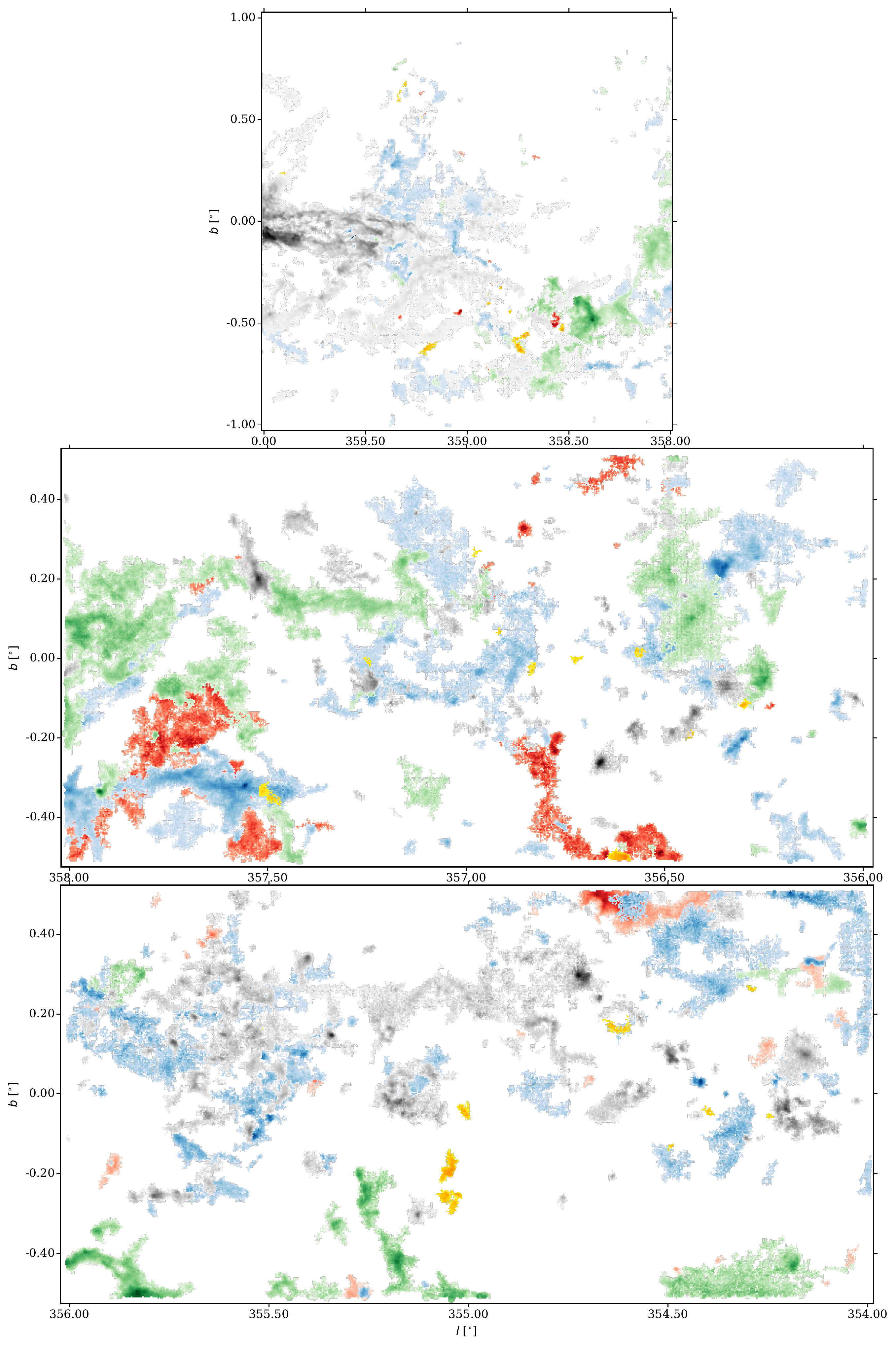}
   \caption{Multi-coloured integrated intensity maps of the $^{13}$CO\,(2-1) emission in the SEDIGISM field across $314^{\circ} \leq l \leq 331^{\circ}$.}
    \label{F:sed_sparms_355}
\end{figure*}

\begin{figure*}
        \includegraphics[width=0.85\textwidth]{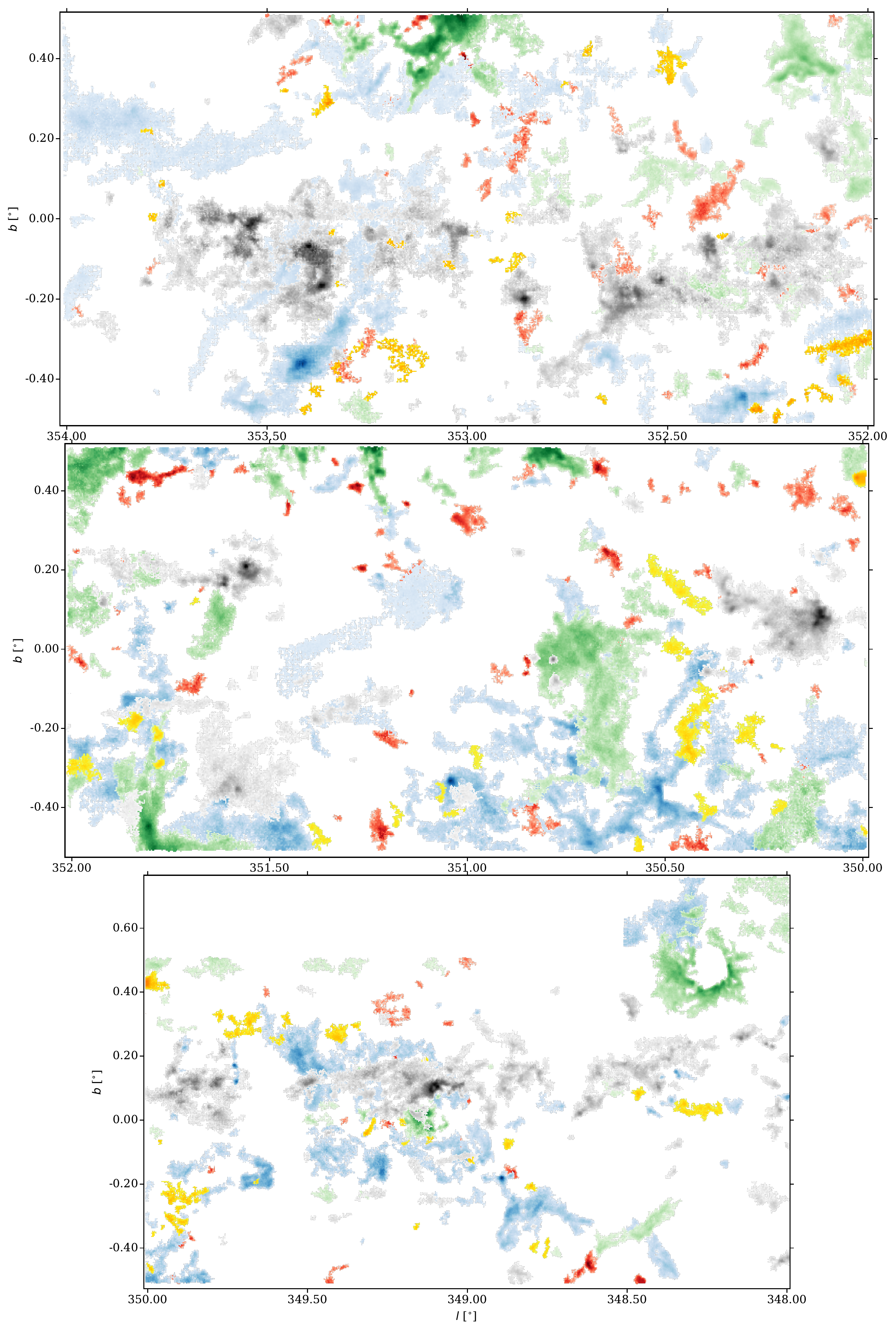}
   \caption{Multi-coloured integrated intensity maps of the $^{13}$CO\,(2-1) emission in the SEDIGISM field across $301^{\circ} \leq l \leq 314^{\circ}$.}
    \label{F:sed_sparms_349}
\end{figure*}

\begin{figure*}
        \includegraphics[width=0.85\textwidth]{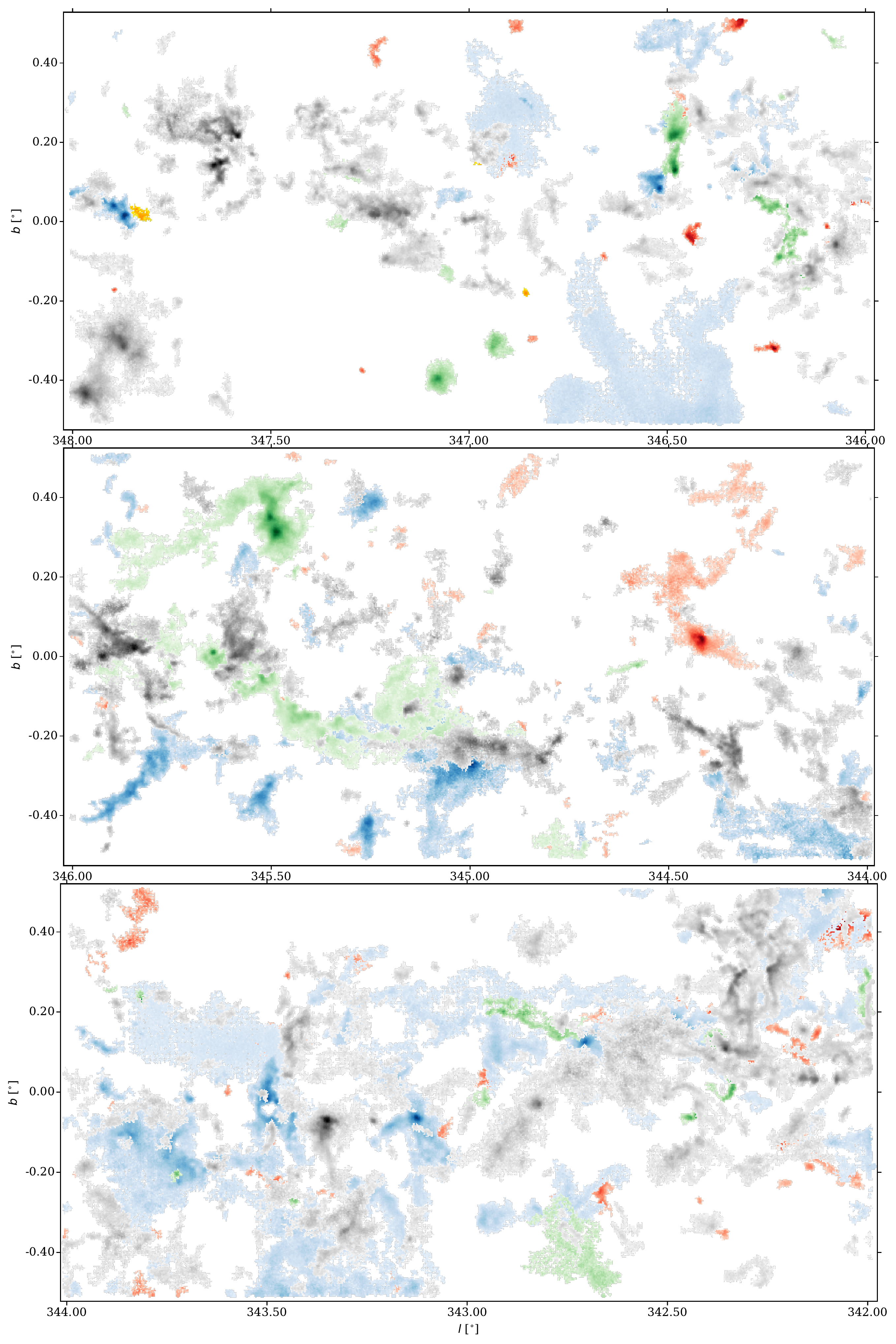}
   \caption{Multi-coloured integrated intensity maps of the $^{13}$CO\,(2-1) emission in the SEDIGISM field across $301^{\circ} \leq l \leq 314^{\circ}$.}
    \label{F:sed_sparms_343}
\end{figure*}

\begin{figure*}
        \includegraphics[width=0.85\textwidth]{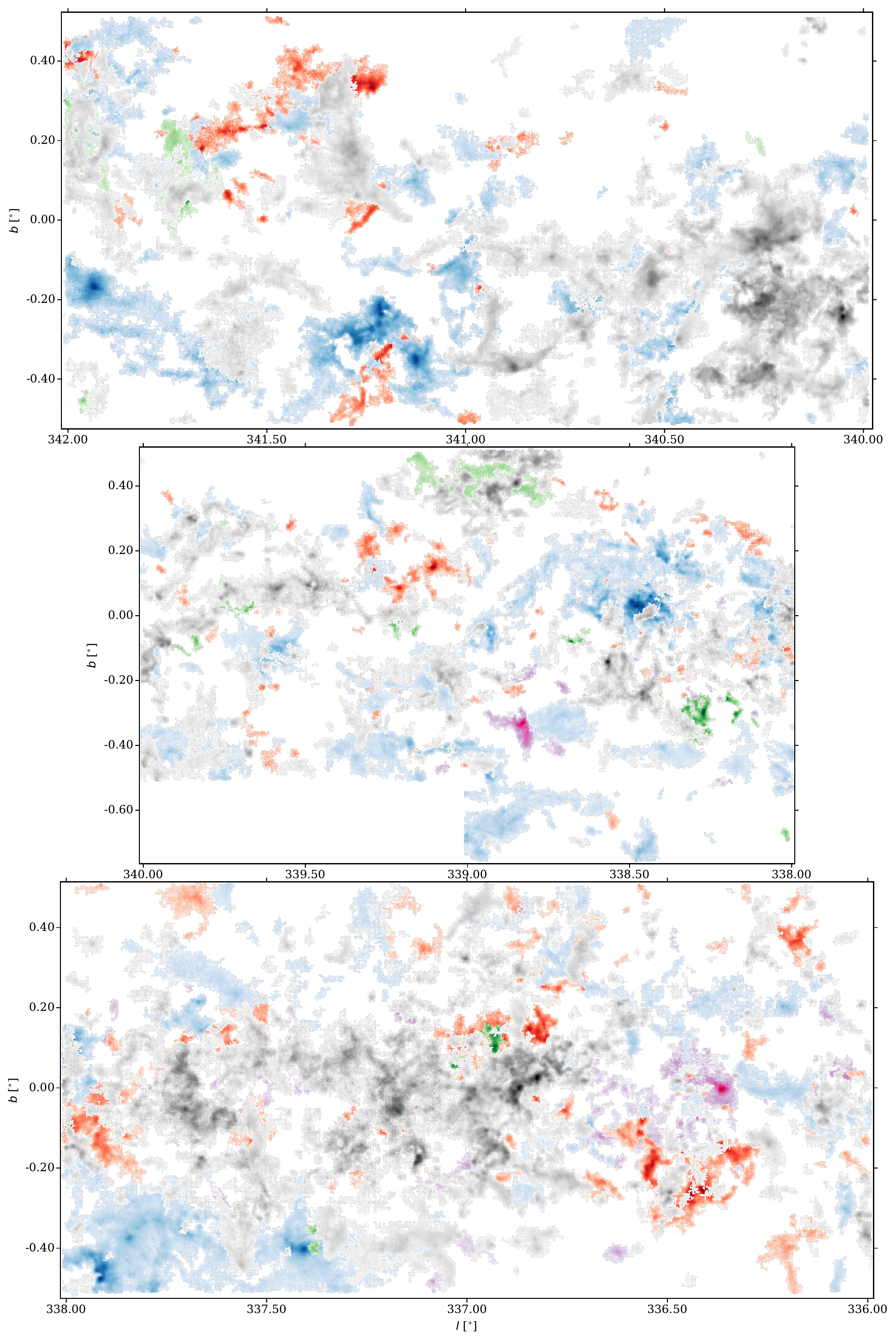}
   \caption{Multi-coloured integrated intensity maps of the $^{13}$CO\,(2-1) emission in the SEDIGISM field across $301^{\circ} \leq l \leq 314^{\circ}$.}
    \label{F:sed_sparms_337}
\end{figure*}

\begin{figure*}
        \includegraphics[width=0.85\textwidth]{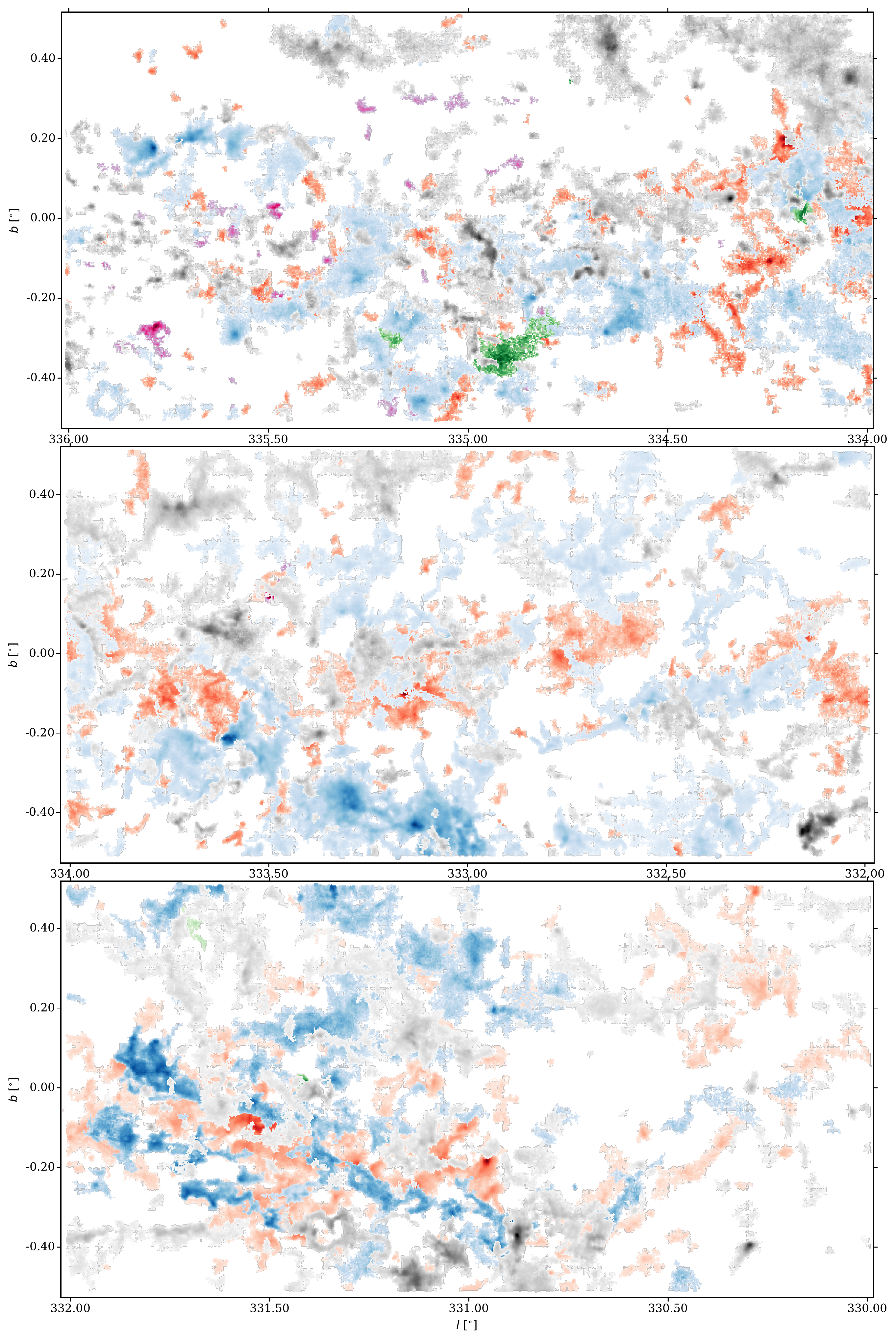}
   \caption{Multi-coloured integrated intensity maps of the $^{13}$CO\,(2-1) emission in the SEDIGISM field across $301^{\circ} \leq l \leq 314^{\circ}$.}
    \label{F:sed_sparms_331}
\end{figure*}

\begin{figure*}
        \includegraphics[width=0.85\textwidth]{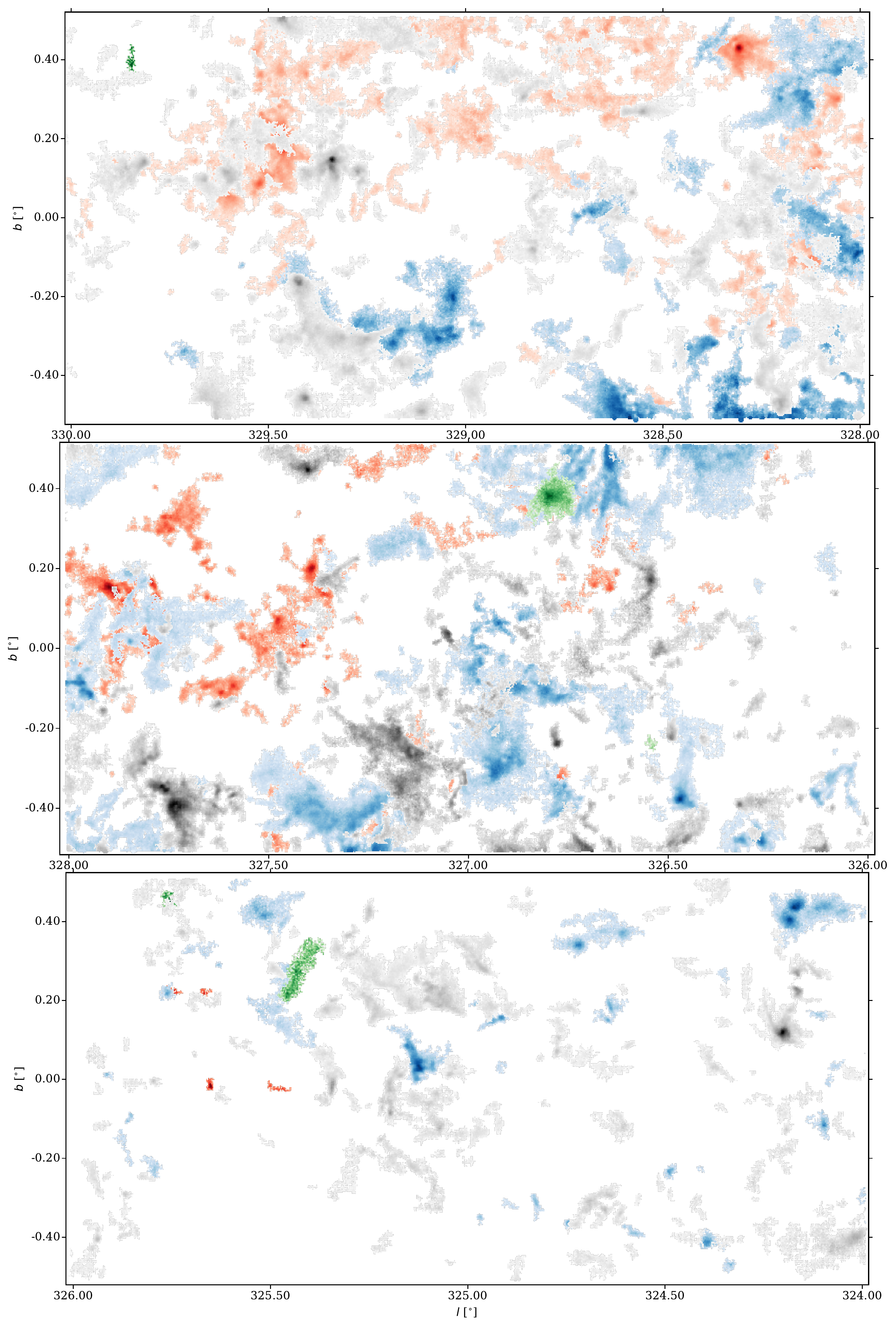}
   \caption{Multi-coloured integrated intensity maps of the $^{13}$CO\,(2-1) emission in the SEDIGISM field across $301^{\circ} \leq l \leq 314^{\circ}$.}
    \label{F:sed_sparms_325}
\end{figure*}

\begin{figure*}
        \includegraphics[width=0.85\textwidth]{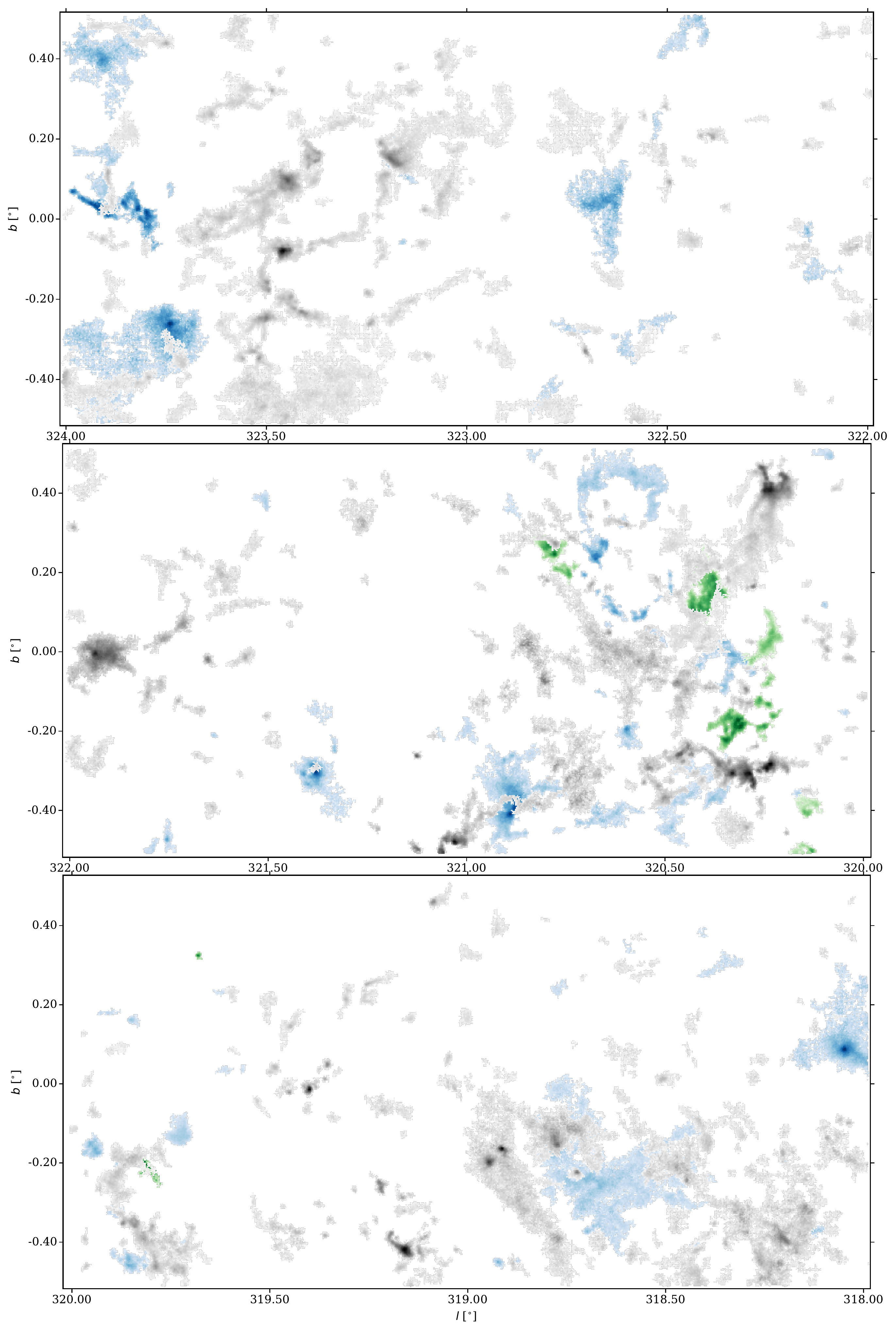}
   \caption{Multi-coloured integrated intensity maps of the $^{13}$CO\,(2-1) emission in the SEDIGISM field across $301^{\circ} \leq l \leq 314^{\circ}$.}
    \label{F:sed_sparms_319}
\end{figure*}

\begin{figure*}
        \includegraphics[width=0.85\textwidth]{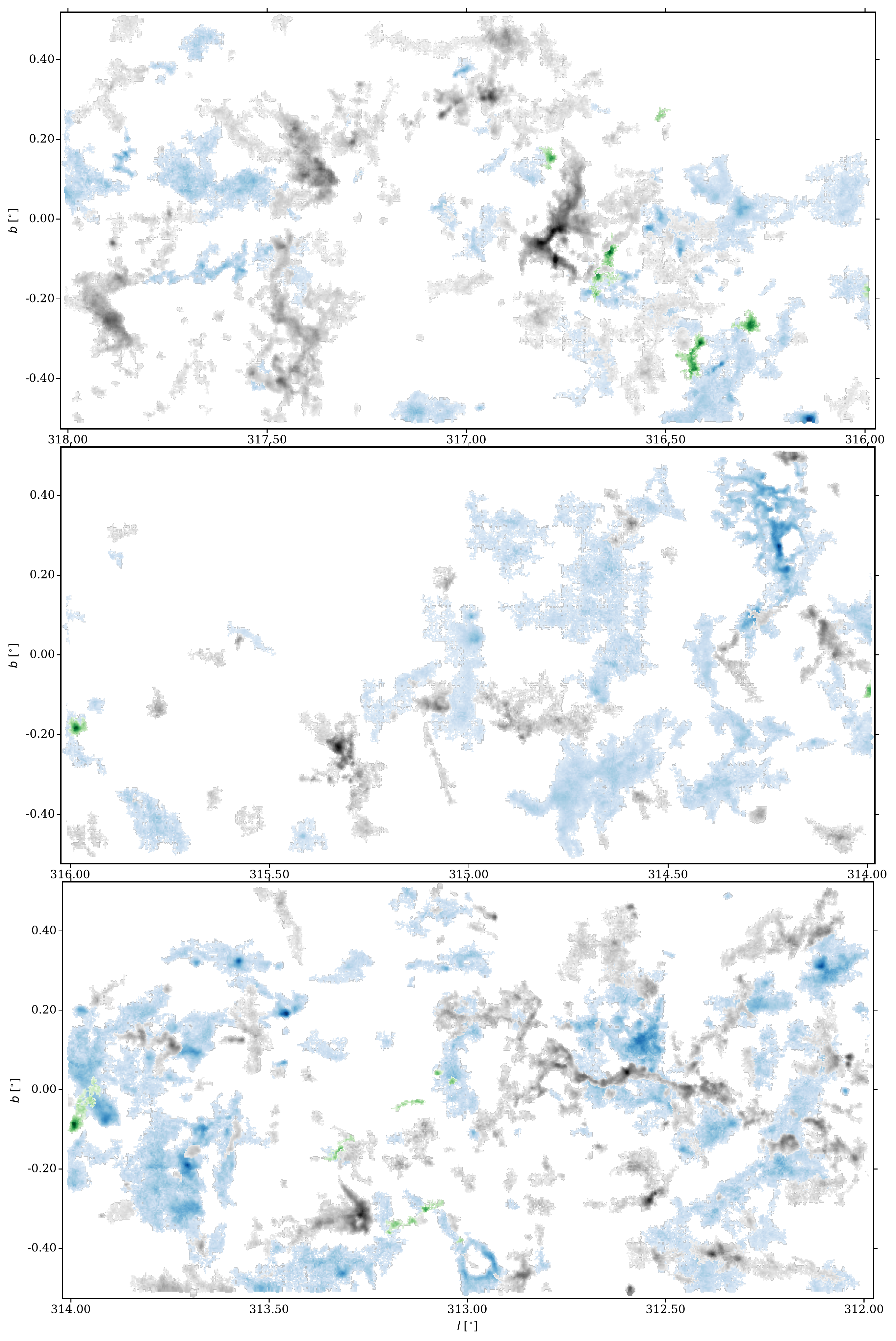}
   \caption{Multi-coloured integrated intensity maps of the $^{13}$CO\,(2-1) emission in the SEDIGISM field across $301^{\circ} \leq l \leq 314^{\circ}$.}
    \label{F:sed_sparms_313}
\end{figure*}

\begin{figure*}
        \includegraphics[width=0.85\textwidth]{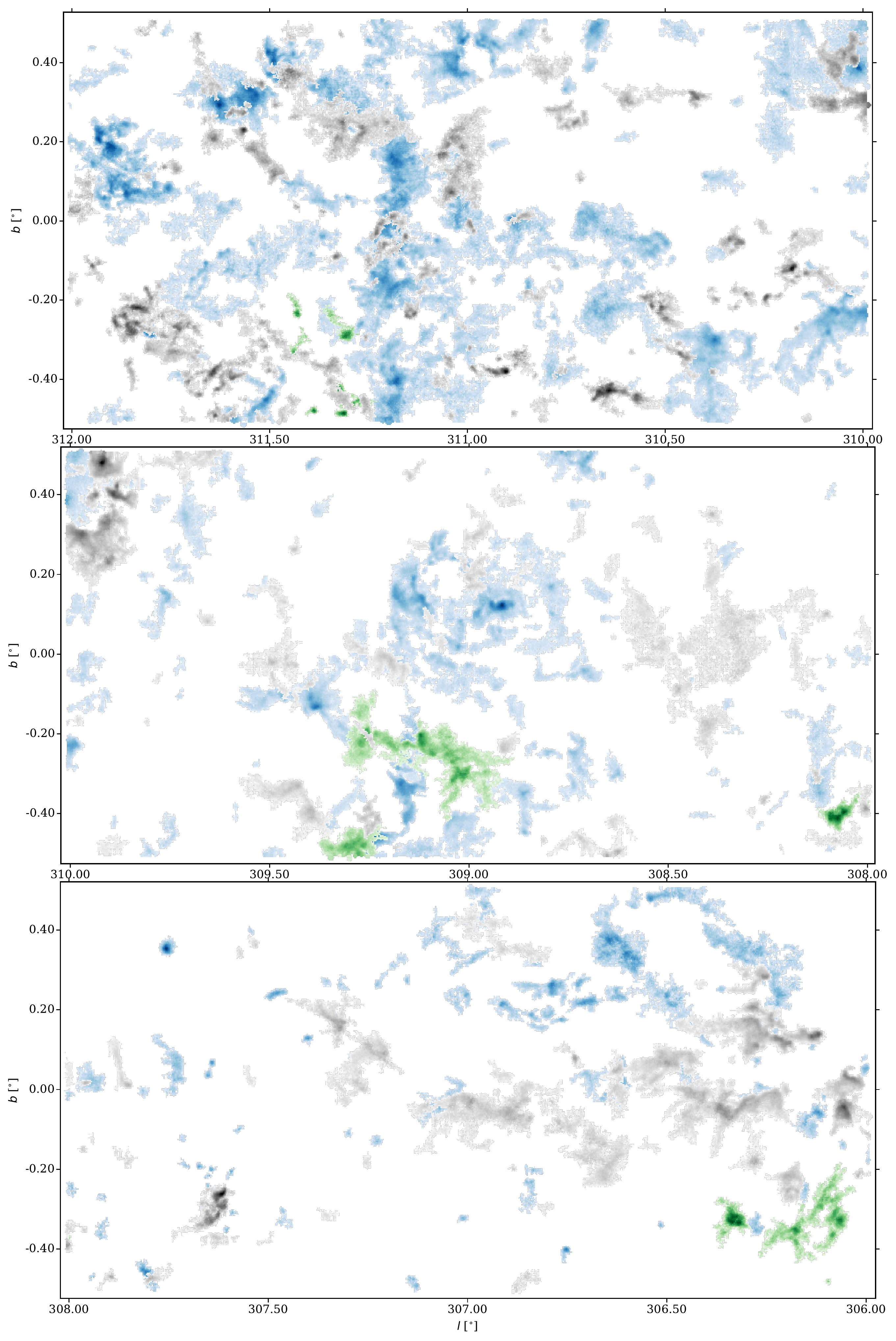}
   \caption{Multi-coloured integrated intensity maps of the $^{13}$CO\,(2-1) emission in the SEDIGISM field across $301^{\circ} \leq l \leq 314^{\circ}$.}
    \label{F:sed_sparms_307}
\end{figure*}

\begin{figure*}
        \includegraphics[width=0.85\textwidth]{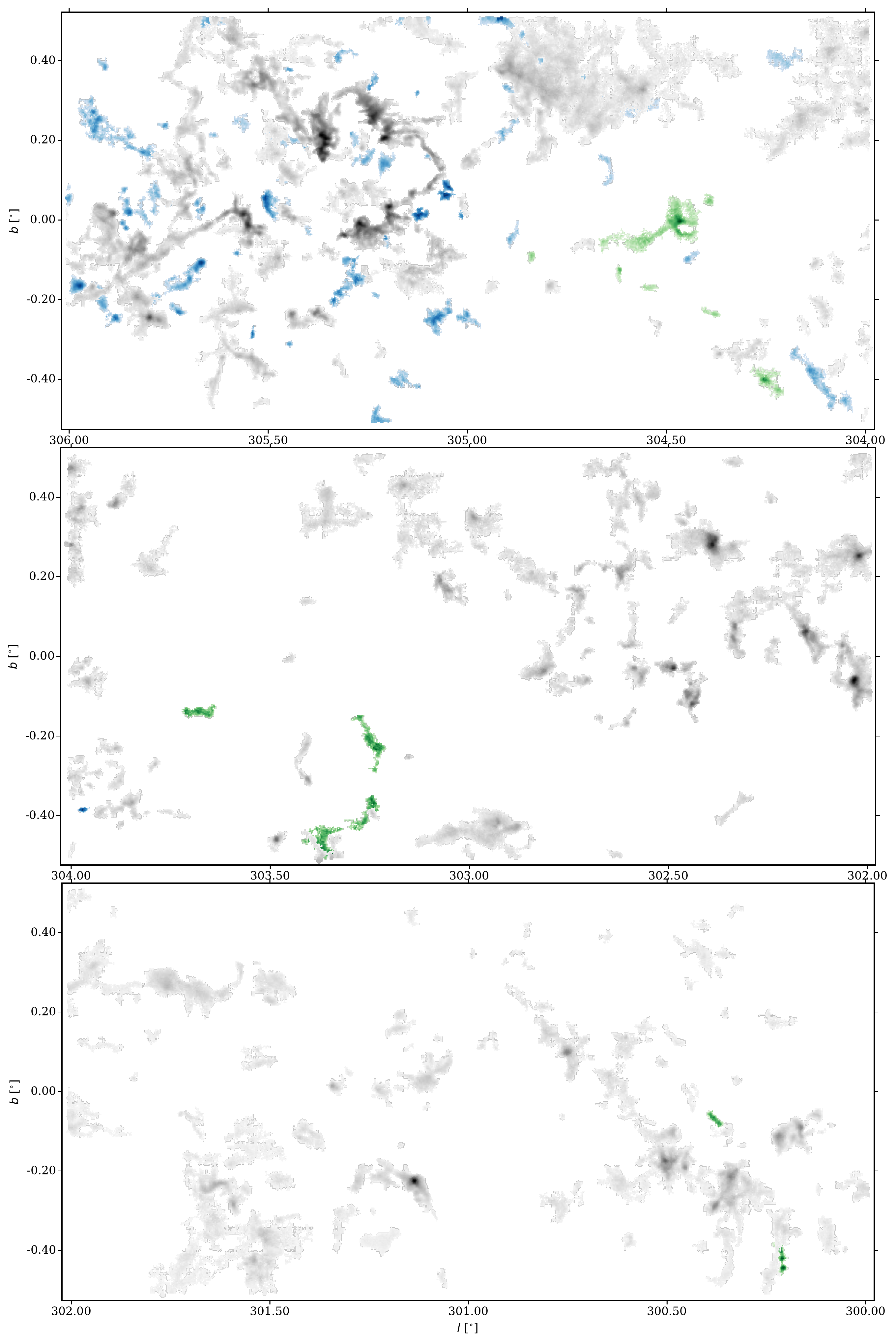}
   \caption{Multi-coloured integrated intensity maps of the $^{13}$CO\,(2-1) emission in the SEDIGISM field across $301^{\circ} \leq l \leq 314^{\circ}$.}
    \label{F:sed_sparms_301}
\end{figure*}


\end{document}